\documentclass[twocolumn]{aastex62}

\usepackage{amsmath,amssymb,amstext}

\usepackage{natbib}
\shorttitle{WISE-2MASS Red Quasars}
\shortauthors{Glikman et al.}

\begin{document}

\title{The WISE-2MASS Survey: Red Quasars Into the Radio Quiet Regime}

\author[0000-0003-0489-3750]{E. Glikman}
\affiliation{Department of Physics, Middlebury College, Middlebury, VT 05753, USA }

\author[0000-0002-3032-1783]{M. Lacy}
\affiliation{National Radio Astronomy Observatory, Charlottesville, VA, USA}

\author[0000-0002-5907-3330]{S. LaMassa}
\affiliation{Space Telescope Science Institute, 3700 San Martin Drive, Baltimore MD, 21218, USA}

\author{C. Bradley}
\affiliation{Department of Physics, Middlebury College, Middlebury, VT 05753, USA }

\author[0000-0002-0603-3087]{S.~G. Djorgovski}
\affiliation{California Institute of Technology, Pasadena, CA 91125, USA}

\author[0000-0001-6746-9936]{T. Urrutia} 
\affiliation{Leibniz Institut f\"{u}r Astrophysik, An der Sternwarte 16, D-14482 Potsdam, Germany}

\author[0000-0002-3739-0423]{E.~L. Gates} 
\affiliation{UCO/Lick Observatory, P.O. Box 85, Mount Hamilton, CA 95140, USA}

\author[0000-0002-3168-0139]{M.~J. Graham}
\affiliation{California Institute of Technology, Pasadena, CA 91125, USA}

\author[0000-0002-0745-9792]{M. Urry}
\affiliation{Yale Center for Astronomy \& Astrophysics, Physics Department, P.O. Box 208120, New Haven, CT 06520, USA}
\affiliation{Department of Physics, Yale University, P.O. Box 208121, New Haven, CT 06520, USA}

\author[0000-0001-9163-0064]{I. Yoon}
\affiliation{National Radio Astronomy Observatory, Charlottesville, VA, USA}

\begin{abstract}
We present a highly complete sample of broad-line (Type~1) QSOs out to $z\sim3$ selected by their mid-infrared colors, a method that is minimally affected by dust reddening. 
We remove host galaxy emission from the spectra and fit for excess reddening in the residual QSOs, resulting in a Gaussian distribution of colors for unreddened (blue) QSOs, with a tail extending toward heavily reddened (red) QSOs, defined as having $E(B-V)>0.25$.  
This radio-independent selection method enables us to compare red and blue QSO radio properties in both the FIRST (1.4~GHz) and VLASS ($2-4$~GHz) surveys. 
Consistent with recent results from optically-selected QSOs from SDSS,
we find that red QSOs have a significantly higher detection fraction and a higher fraction of compact radio morphologies at both frequencies. 
We employ radio stacking to investigate the median radio properties of the QSOs including those that are undetected in FIRST and VLASS, finding that red QSOs have significantly brighter radio emission and steeper radio spectral slopes compared with blue QSOs. 
Finally, we find that the incidence of red QSOs is strongly luminosity dependent, where red QSOs make up $>40\%$ of all QSOs at the highest luminosities. Overall, red QSOs comprise $\sim40\%$ of higher luminosity QSOs, dropping to only a few percent at lower luminosities. 
Furthermore, red QSOs make up a larger percentage of the radio-detected QSO population. 
We argue that dusty AGN-driven winds are responsible for both the obscuration as well as excess radio emission seen in red QSOs.
\end{abstract}


\section{Introduction}

Our incomplete understanding of  the relationship between supermassive black hole (SMBH) growth and the growth of galaxies in the Universe remains an outstanding problem in astrophysics. There is ample evidence that SMBHs are linked to their host galaxies through scaling relations, such as the $M_{BH} - \sigma$ relationship \citep{Gebhardt00}, which tell us that galaxies and their nuclear black holes likely grew in tandem. In order for galaxies to build up stars while growing a nuclear SMBH, a galaxy-scale energy exchange, or ``feedback'', is required to regulate this process and tie the two systems together. This feedback is still very poorly understood, and may come in the form of radiation, winds, outflows, and/or jets \citep[c.f.,][]{Fabian12}. 

A population of sources that may help elucidate the nature of feedback are dust-reddened quasars, which appear to represent an important evolutionary phase linking galaxy mergers to black hole growth. 
Luminous quasars are thought to be triggered by major galaxy mergers \citep{Sanders88a,Treister12} and simulations of major, gas-rich mergers are able to reproduce many of the aforementioned correlations and galaxy properties \citep{DiMatteo05,Hopkins05}.  
During the merger, some gas loses angular momentum and feeds the black hole while shocks trigger a starburst. As the SMBH grows, it starts out in a heavily-obscured state followed by a relatively brief transitional phase during which the dust is cleared via feedback mechanisms. Subsequently, an unobscured, blue quasar emerges and dominates the radiation output for the system. The objects in the brief transitional phase are moderately reddened (or, red) quasars and can serve as laboratories for studying how quasar feedback impacts their host galaxies.

Red quasars can be elusive, because their optical and near-infrared colors resemble those of low-mass stars, which are far more abundant at these wavelengths.  
Early work used radio selection to find red quasars and avoid contamination from red stars, which are  weak radio sources \citep{Webster95,White03b}.
Results from these studies suggested that red quasars make up a large fraction (up to $\sim80\%$) of the overall quasar population but had been missed by optical selection methods. 
 \citet{Glikman04} combined the Faint Images of the Radio Sky at Twenty-centimeters \citep[FIRST;][]{Becker95} radio survey and 2 Micron All Sky Survey \citep[2MASS;][]{Skrutskie06} to develop an efficient selection method for finding these missed red quasars. Subsequent work identified $\gtrsim130$ dust-reddened quasars via the same method \citep[hereafter referred to as F2M quasars][]{Glikman07,Glikman12,Urrutia09} that have broad emission lines and are moderately obscured by $A_V \sim 1 - 4 ~ (0.1 < E(B - V) \lesssim1.5$) across a broad range of redshifts ($0.1 < z \lesssim 3$). 
Follow up studies of F2M quasars showed that they are accreting with very high Eddington rates \citep{Urrutia12,Kim15}, are overwhelmingly in merger-dominated systems \citep{Urrutia08,Glikman15}, and often have broad absorption lines that are typically associated with outflows and feedback \citep[LoBALs and FeLoBALs;][]{Urrutia09,Farrah12,Glikman12}. 
This body of evidence suggests that red quasars are merger-induced systems, in a transitional phase, emerging from their shrouded environments, as predicted by the galaxy merger simulations.

Careful comparison with blue quasar samples found that red quasars comprise $\sim20-30\%$ of the overall quasar population \citep{Glikman12,Glikman18}, at least at the highest luminosities.  When interpreted as an evolutionary phase, this fraction implies that the duration of this transitional phase is $\sim20-30\%$ as long as the unobscured phase, consistent with theoretical models of quasar ignition and evolution triggered by a major galaxy merger \citep[e.g.,][]{Hopkins05}.

However, because the F2M survey used radio selection, those quasars belong to the rarer radio-loud and radio-intermediate populations that make up $\sim10\%$ of the overall quasar population. Assuming that radio emission from these quasars is unrelated to their surrounding dust, we could extend the F2M results to the entire quasar population. However, if the radio emission and reddening are not independent, then any conclusions about the red quasar population derived from the F2M sample, such as the duration of the transitional phase, could be biased and does not apply to the full quasar population.

Recent results indeed suggest a correlation between reddening and radio emission. \citet{White07} used stacking of FIRST images of known (mostly radio-quiet) quasars and found that redder quasars have higher median radio fluxes: objects that are 0.8 mag redder than average have radio fluxes that are $\sim3$ times higher than average. Another study of the brightest red quasars ($K < 14.5$) by \citet{Georgakakis09}, using only a $J - K > 1.5$ color selection and no radio constraint, found that 6 out of their 10 objects were detected in the radio. In a sample of extremely red quasars (ERQs) found in the Sloan Digital Sky Survey (SDSS) without a radio criterion, all of the mid-infrared-brightest and reddest sources are detected in FIRST \citep{Ross15,Hamann17}.  

More recently, \citet{Klindt19} and \citet{Fawcett20} found distinct differences between the radio properties of blue and red SDSS quasars. They find that the redder quasars have a significantly higher detection fraction in FIRST. When stacked, radio quiet red quasars have higher median radio fluxes than an unreddened sample. In addition, red quasars' radio morphologies are more compact compared with blue quasars. 
However, the optical selection of SDSS quasars misses the more heavily reddened sources like those found in the F2M survey because their optical colors place the sources atop the stellar locus \citep{Urrutia09}; most of the red QSOs in the SDSS sample have $E(B-V) \lesssim 0.2$.
To avoid any biases in the SDSS QSO selection algorithm that misses heavily reddened quasars, a selection method is needed at wavelengths that are minimally impacted by dust extinction and is also radio independent. Such a method should have sufficient depth and coverage area to enable a robust comparison between the red and blue populations.

In this paper, we invoke mid-infrared selection, as it has been shown to successfully identify broader populations of QSOs\footnote{Because the focus of this paper is on the differences between radio detected and undetected systems, we adopt the canonical nomenclature that distinguishes quasars -- radio-detected luminous AGN whose radio emission is essential to their selection -- from QSOs -- the overall class of luminous AGN.  } that were less affected by dust extinction  \citep[e.g.,][]{Lacy04,Stern05,Donley12,Jarrett11}. 
The Wide-Field Infrared Space Explorer \citep[{\em WISE};][]{Wright10}, scanned the sky at 3.4, 4.6, 12, and 22 $\mu$m down to flux densities of 0.08, 0.11, 1, and 6 mJy, respectively\footnote{The sensitivity limit varies with latitude due to the scanning cadence of the satellite, with the deepest fields near the poles.}, providing the wide-area coverage needed for identifying large numbers of luminous QSOs. 
In {\it WISE} color-color space QSOs can be isolated from stars and other extragalactic sources, making the mid-IR an excellent wavelength region for our purposes \citep[e.g.,][]{Mateos12,Jarrett11,Stern12,Assef13,Assef18}.  

In a pilot study, \citet{Glikman18} identified a complete sample of QSOs using near-to-mid-infrared color selection over $\sim260$ deg$^2$ that overlap the SDSS Stripe 82 legacy field \citep{Frieman08}. 
Here we expand upon that work over a $7.5\times$ larger area reaching fainter mid-infrared fluxes. We identify a sample of QSO candidates according to their mid-infrared colors and obtain spectroscopy of sources missed by SDSS and other optical QSO surveys. 
Because mid-infrared selection identifies both blue and red luminous Type 1 QSOs\footnote{However, see \citet{LaMassa19} and Anicetti et al. (in prep) for examples of luminous X-ray selected QSOs that are undetected in {\em WISE}.}, we can compare sources that are drawn from the same mid-infrared selection criteria.

In this paper, we present a sample of red QSOs \citep[defined as having $E(B-V) > 0.25$;][]{Lacy07,Glikman18} without relying on radio selection and aim to determine whether the fraction of red quasars found in the F2M survey ($\sim$20\%) holds for the full red QSO population, including radio quiet sources.  
In addition, we explore the differences between radio-detected and radio-undetected QSOs in the red and blue populations, as well as their average radio properties through stacking, to explore possible differences in the mechanisms giving rise to their radio emission, informing notions of jet formation, dusty winds, or other physical processes.

The surveys employed in this paper use a mix of AB and Vega for their photometric zeropoints. Rather than transform to a common system, we adhere to the native systems presented in each survey. {\em WISE} and 2MASS photometry are on the Vega system, while SDSS uses AB magnitudes. When colors are derived from catalogs on mixed systems, we provide specificity as to which system we are using. 
Throughout this work, we adopt the concordance $\Lambda$CDM cosmology with $H_0 = 70$ km s$^{-1}$ Mpc$^{-1}$, $\Omega_M = 0.3$, and $\Omega_\Lambda = 0.7$ when computing cosmology-dependent values \citep{Bennett13}.

\section{Sample Selection} \label{sec:sel}

Our survey area covers an equatorial region overlapping the SDSS over two fields: a region spanning a range in right ascension of $\alpha = 8^h - 16^h$ and in declination of $\delta = 0.5^\circ - 17^\circ$, available for follow-up in the spring months, and the region over Stripe 82 identical to that in \citet{Glikman18}, $\alpha = 20^h 40^m - 3^h 56^m$ and $\delta = -1.25^\circ$ to $+1.25^\circ$ (excluding the region $00^h < \alpha < 00^h 15^m$).  This covers a total area of 2213  deg$^2$ (1950 deg$^2$ and 263$^2$ in the two regions, respectively). 

Our aim is to define a sample of luminous Type 1 QSOs in order to compare the red and blue sub-populations with minimal reddening and radio biases. 
We begin by selecting sources with {\em WISE} colors consistent with QSO emission, focusing on sources brighter than $K=14.7$ mag to enable near-infrared spectroscopy with 3-m class telescopes (\S \ref{sec:obs}) over the chosen survey area. 
We include all spectroscopically confirmed broad-line QSOs from SDSS and identify red QSO candidates among the sources lacking a spectrum in SDSS, selected by their optical through near-infrared colors. 
We perform follow-up spectroscopy of all such candidates and keep all broad-line QSOs to construct a complete sample of QSOs that obey uniform mid-to-near infrared selection criteria. 
To construct the blue and red QSO subsamples, we fit a reddened QSO template to each spectrum, subtracting host galaxy emission when necessary, and define red QSOs as having $E(B-V) > 0.25$. 
Finally, we identify a luminosity-restricted subsample that excludes the blue quasars that would not have been detected if they were reddened by $E(B-V) \ge 0.25$ to enable a valid comparison between the red and blue QSOs. 

We note that due to the $K< 14.7$ mag limit of our survey, the red QSO population is incomplete, as there are likely to be red QSOs with $E(B-V) > 0.25$ with lower intrinsic luminosities that when reddened, fall below the flux limit. 
On the other hand, our selection does not miss significant numbers of blue QSOs. Therefore, we can compare the blue and red populations to arrive at a red quasar fraction, understanding that it is a lower limit. 

\begin{figure*}
\epsscale{1}
\plotone{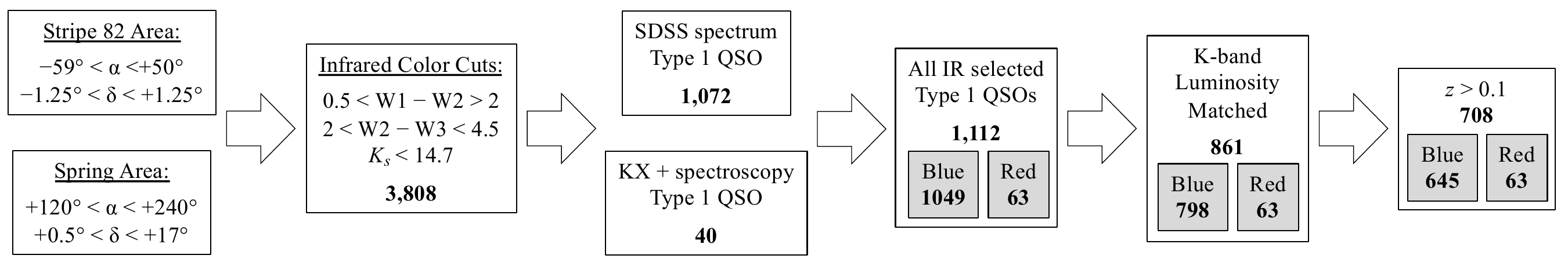}
\caption{Flowchart showing the process for selecting the full QSO sample in this work. Each box reports a major selection stage, with the number of sources that passed each stage shown in boldface.  Table \ref{tab:selection} complements the flowchart with references to the relevant sections in the text that elaborate on each step in the process.}\label{fig:flowchart}
\end{figure*}

Figure \ref{fig:flowchart} shows a flowchart following the selection process as described in the steps below and Table \ref{tab:selection} presents an overview of the selection with references to the sections in the text that elaborate on each step in the process.

\begin{deluxetable*}{cllc}




\tablecaption{Overview of sample selection steps \label{tab:selection}}

\tablenum{1}

\tablehead{\colhead{Step} & \colhead{Definition} & \colhead{Relevant Section Reference} & \colhead{Number of sources} } 

\startdata
I      & Mid-Infrared color selection &  \S \ref{sec:mid-ir} + Eqn. \ref{eqn:gator_sel} &  3808 \\
IIa    & SDSS Type 1 QSOs             &  \S \ref{sec:sdss}  &  1072 \\
IIb    & KX selected Type 1 QSOs      &  \S \ref{sec:kx}  + Eqn. \ref{eqn:diag} + Eqn. \ref{eqn:kx} + \S \ref{sec:spec} &  40 \\
III    & all W2M QSOs                 &  \ldots &  1112 \\
\ldots & red QSOs                     &  \S \ref{sec:reddening} + $E(B-V) > 0.25$ &  63 \\
\ldots & blue QSOs                    &  \S \ref{sec:reddening} + $E(B-V) < 0.25$ &  1049 \\
IV     & luminosity restricted blue QSOs &  \S \ref{sec:luminosity} &  798 \\
\ldots & $z> 0.1$                     &  \ldots &  645 \\
\enddata
%




\end{deluxetable*}

\subsection{Mid-infrared selection}\label{sec:mid-ir}

Since the effects of dust reddening diminish with longer wavelengths, we expect red QSOs to have nearly the same mid-infrared spectral shape as unreddened QSOs\footnote{At far infrared wavelengths, we expect the spectra to deviate once again, as the scattered and absorbed UV photons are reprocessed into thermal radiation from heated dust.  However, the QSO continuum dominates at the {\em WISE} wavelengths.}.  \citet[Figure 1]{Glikman18} showed that the F2M red quasars lie in the same region of {\em WISE} color space as blue QSOs. We therefore inform our selection of red QSOs by studying the regions that normal, blue QSOs occupy in {\em WISE} color-color space.  

\begin{figure}
\epsscale{1}
\plotone{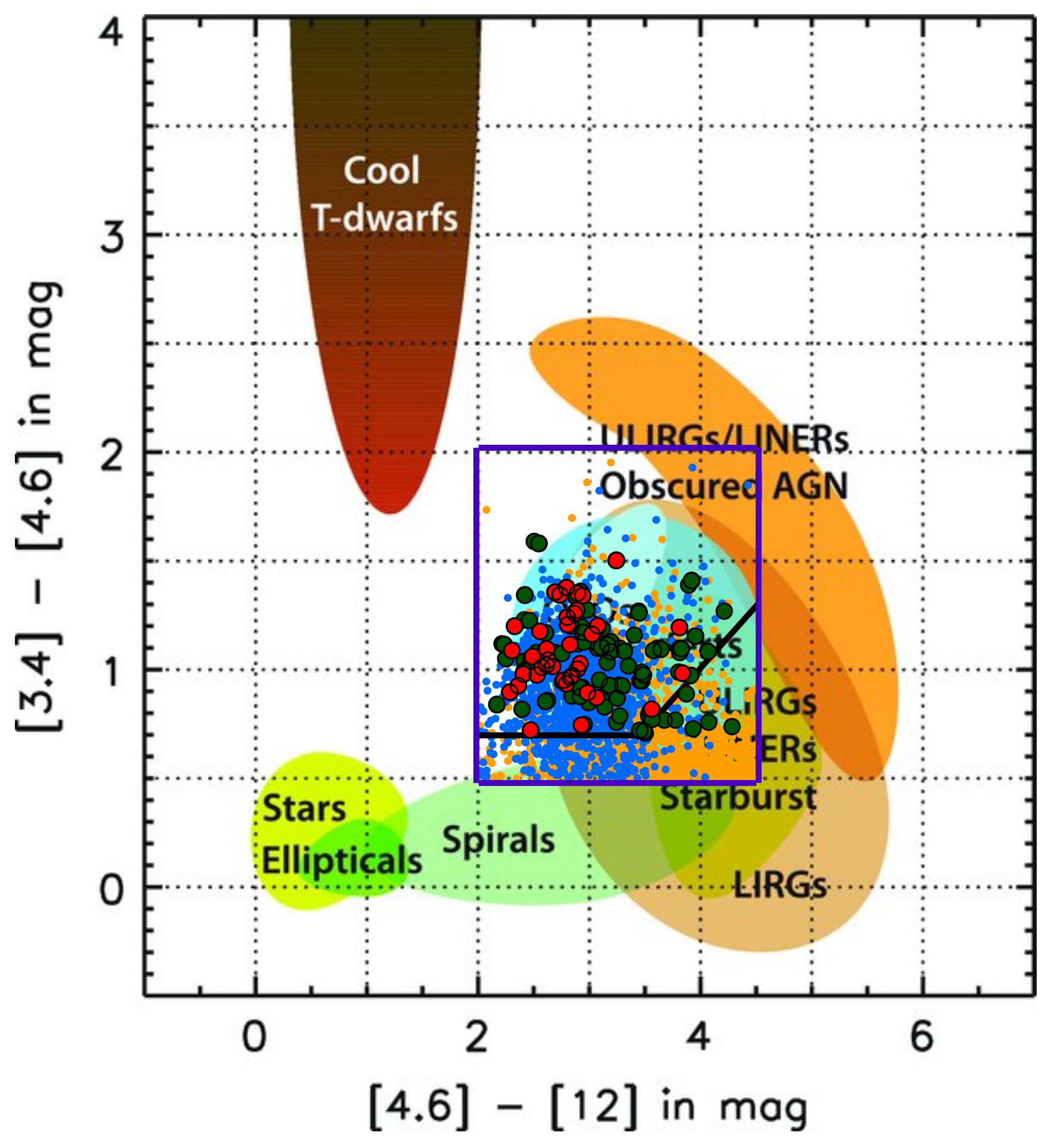}
\caption{We reproduce Figure 12 from \citet{Wright10} showing the location of various astrophysical objects in {\em WISE} color-color space.  We overplot a box that outlines our initial selection region (Eqn \ref{eqn:gator_sel}) with thick purple solid lines and plot within the box sources with SDSS spectra identified as {\tt QSO} (blue circles) and {\tt GALAXY} (orange circles). 
We also plot our selected candidates with large outlined circles. Newly confirmed QSOs are filled red and sources that are not obviously QSOs are filled green. The black solid line shows our refined selection criterion (Eqn \ref{eqn:diag}) that avoids significant contamination from non-QSOs (\S \ref{sec:kx}). 
}\label{fig:wise_colors}
\end{figure}
 
We began by selecting sources from the {\em WISE} catalog, which presents its photometry in the Vega system. We select sources with $0.5 < W1 - W2 < 2$ and $2 < W2 - W3 < 4.5$, which is a liberal cut around the location of QSO colors in these bands, as shown in Figure 12 of \citet{Wright10}, restricting our sample to sources with $K<14.7$ mag. 
To generate our sample, we use the IRSA GATOR catalog selection tool to identify all sources in the AllWISE Source Catalog obeying the following search criteria via an SQL query:
\begin{equation}
\begin{array}{l}
\tt  WHERE ~(~(w1mpro - w2mpro) > 0.5 ~ and \\ 
\tt (w1mpro - w2mpro) < 2 ~ and \\
\tt (w2mpro - w3mpro) > 2 ~ and \\
\tt (w2mpro - w3mpro) < 4.5~) ~ and \\ 
\tt k\_m\_2mass < 14.7, \\
\end{array}\label{eqn:gator_sel}
\end{equation}
applied to the two sky regions. The GATOR query returned 3,808 sources (3,330 in the northern region and 478 over Stripe 82).  
 
We match the {\em WISE} sources to SDSS DR9 photometric catalog \citep{Ahn12} within 2.0\arcsec\ using the Centre de Donn\'{e}es astronomiques de Strasbourg (CDS) Upload X-Match service through the TOPCAT tool \citep{Taylor05} and excluded the handful of sources with $K < 10$ mag.  This resulted in 3,741 matches. Of these, 2,779 have spectroscopic identifications in SDSS (1398 {\tt QSO}s, 1372 {\tt GALAXY}s; 9 {\tt STAR}s). Another seventeen sources were observed with the BOSS spectrograph on SDSS as part of the DR 14 campaign \citep{Abolfathi18}.  Figure \ref{fig:wise_colors} plots these sources in the {\em WISE} $[3.4]-[4.6]$ vs. $[4.6]-[12]$ color-color space defined by \citet{Wright10}, which shows the location of different astrophysical populations in this space.  The blue circles are SDSS-identified {\tt QSO}s and they overlap their expected location (cyan oval) on this diagram. The orange circles are SDSS-identified {\tt GALAXY}s and they are concentrated toward lower $[3.4]-[4.6]$ values (i.e, {\tt GALAXY}s are bluer at these wavelengths).  

\subsection{Type 1 QSOs}

\begin{figure}
\epsscale{1}
\plotone{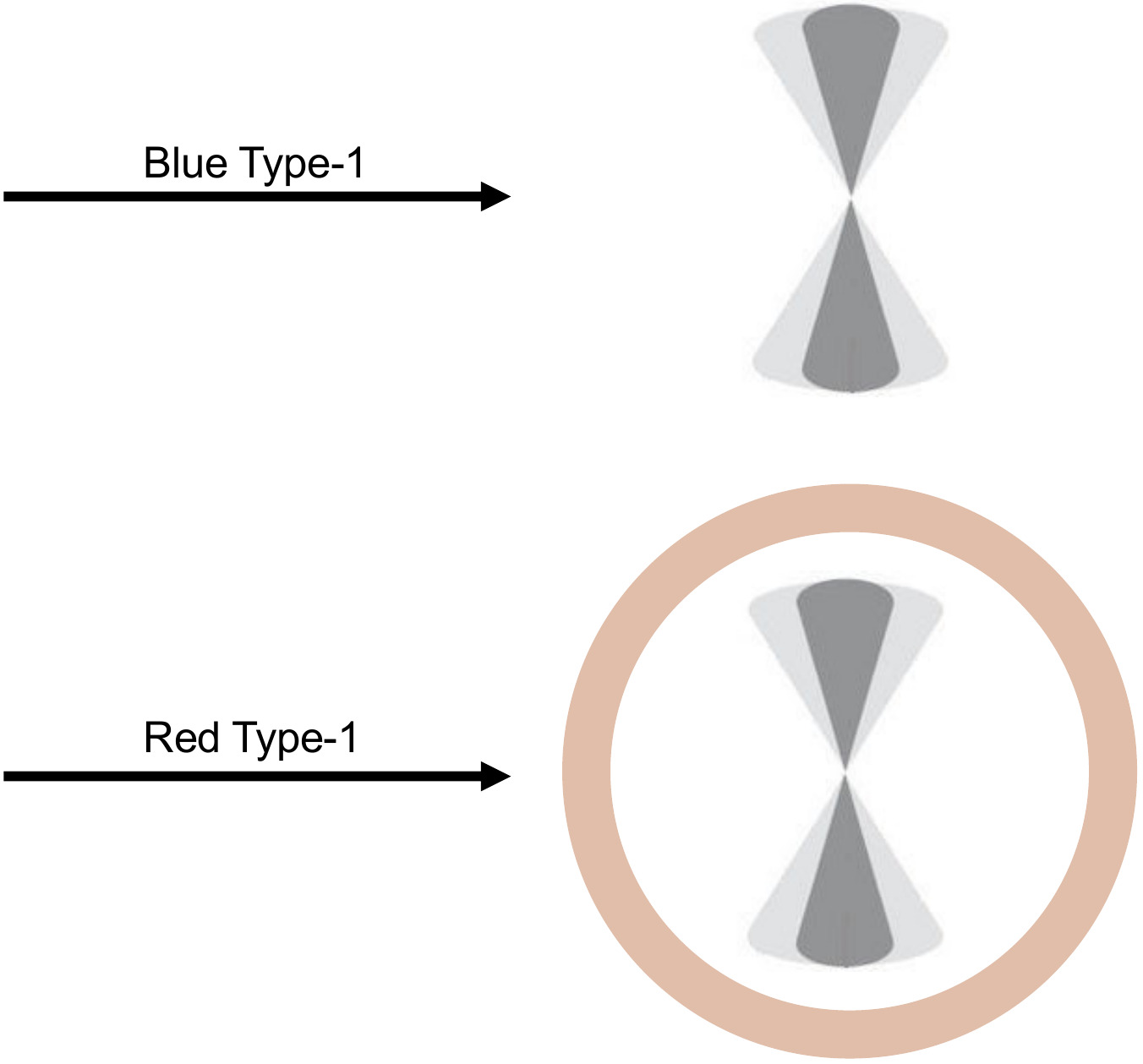}
\caption{
This Figure shows the purported obscuration geometry of high-Eddington-ratio AGN viewed along an unobscured line-of-sight.  In both cases, broad lines are seen from this viewing angle. Therefore, by focusing on only broad-line (Type 1) QSOs, under the assumption that red QSOs are not reddened by the nuclear obscuration (i.e., by the torus, gray regions) but are rather embedded in a dusty environment  (bottom figure) where the dust may arise as the the consequence of a merger, we can translate the fraction of red QSOs to the duration of   the red QSO phase. }\label{fig:cartoon}
\end{figure}

The unification model for AGN \citep{Urry95} says that the difference in viewing angle to the central engine of the accreting-black-hole system determines the observed spectral shape, including emission line widths. Type 1 (broad-line) sources are understood through this model to be seen at orientation angles nearer to the pole. Beyond a certain range of viewing angles, the line-of-sight to the broad line region is blocked by close-in high-column-density gas and dust (i.e., the so-called `torus'). In this study, we wish to focus on Type 1 sources in both the blue and red samples so as to compare sources with the same approximate distribution of viewing angles, knowing that our line-of-sight is not intersecting the dense clouds of the torus (see Figure \ref{fig:cartoon} for an illustration of this argument). 
If we include Type 2 (narrow-line) sources in the sample, it becomes more challenging to determine the location and nature of the obscuring material. 
Therefore, focusing exclusively on broad-line sources allows us to directly compare the blue and red populations.  
As the canonical definition of a broad-line quasar requires having line widths $>1000$ km s$^{-1}$ \citep{Glikman04,Glikman12,Gregg96,Schneider03}, our sample will focus on all objects with line widths broader than 1000 km s$^{-1}$.  

\subsection{QSOs in SDSS}\label{sec:sdss}

The bulk of the QSOs in our sample comes from the SDSS, seen among the blue dots in Figure \ref{fig:wise_colors}. 
We selected from the SDSS DR9 spectroscopic catalog all sources with a {\tt class} of {\tt QSO}, including all {\tt QSO}s in the {\em WISE} color selection box (Eqn.~\ref{eqn:gator_sel}; Figure \ref{fig:wise_colors}; 1398 {\tt QSO}s).

As noted in \citet{Glikman18}, among the spectra that SDSS classifies as {\tt QSO} there are sources that only show narrow lines. To eliminate the Type 2 QSOs, we utilize the ALPAKA catalog \citep{Mullaney13} which provides detailed line analysis for 25,670 AGN with spectra in SDSS DR7.  The spectra of these sources were fitted with multi-component Gaussians to study their kinematics and Eddington ratios. 
Line fitting was performed on [\ion{O}{3}]~4959\AA,  [\ion{O}{3}]~5007\AA, [\ion{N}{2}]~6548\AA, H$\alpha$, and [\ion{N}{2}]~6584\AA, including a broad component for forbidden species and an additional broad component for permitted species, if warranted. The ALPAKA sample is limited to $z<0.4$ in order not to lose H$\alpha$ beyond the SDSS spectroscopic wavelength limit of 9000\AA.
We matched the QSO sample to the ALPAKA catalog and found 1016 matches which we use to examine line-widths and select broad-line sources; 382 sources are left to be dealt with separately.
The ALPAKA catalog classifies sources as Type 1 when the broad component of their H$\alpha$ line constitutes $\ge 50\%$ of the total line flux and a 600 km~s$^{-1}$ threshold \citep{Mullaney13}. However, we impose the additional requirement that the broad component have a Full Width at Half Maximum (FWHM) velocity, $v_{FWHM} > 1000$ km~s$^{-1}$.  We also exclude any source not classified as Type 1 in the ALPAKA catalog; 733 spectra obey these criteria.  

\begin{figure}
\epsscale{1}
\plotone{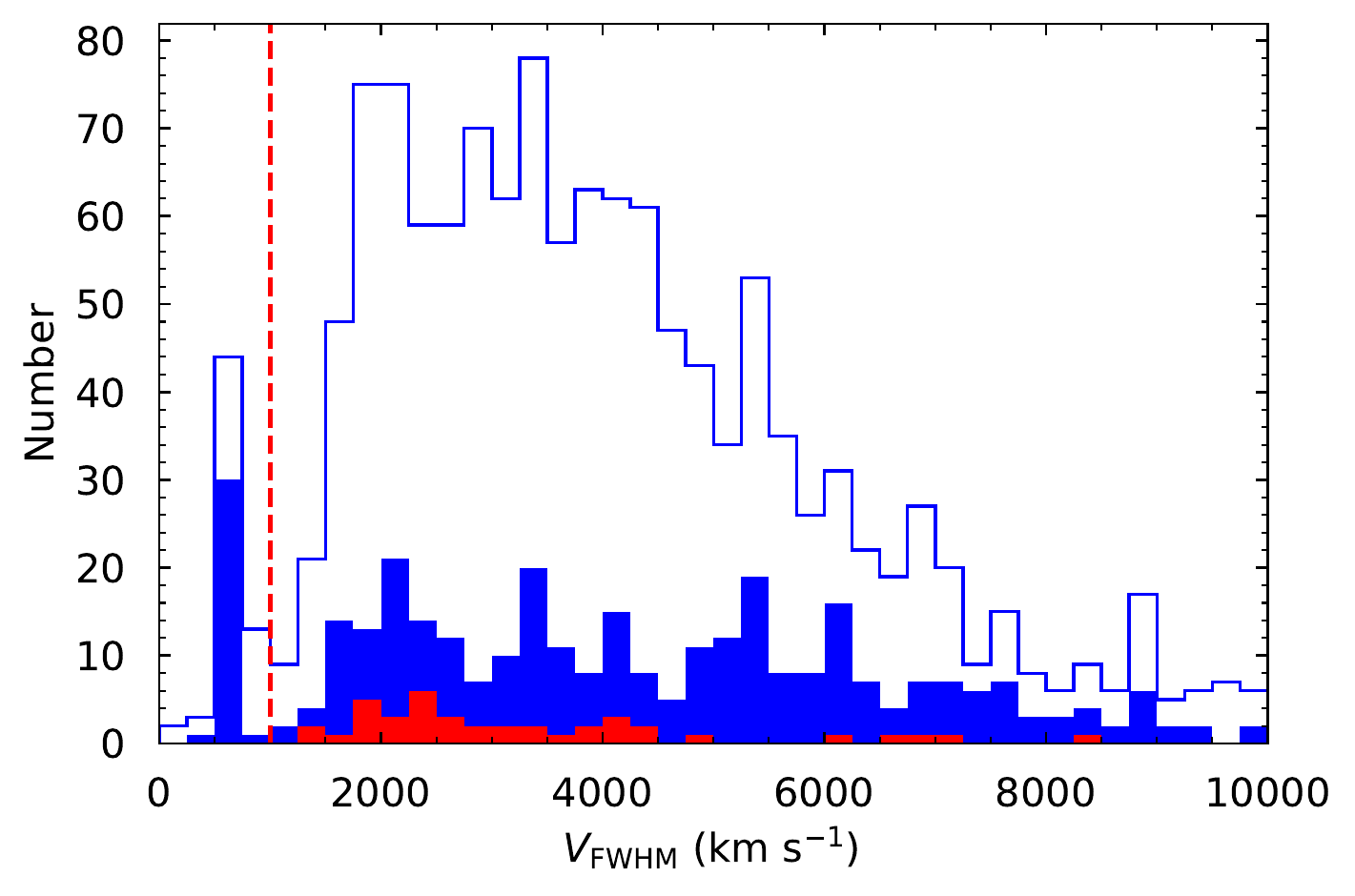}
\caption{We show the distribution of the velocity width, $v_{FWHM}$, for the QSOs derived from both the ALPAKA and \citet{Bolton12} analyses. The solid line represents all 1398 SDSS {\tt QSO}s found within the selection box defined by Eqn \ref{eqn:gator_sel}.  The filled blue histogram shows the line width distribution for QSOs not analyzed by ALPAKA, whose line widths were derived in \citet{Bolton12}.
We also show the same for the newly identified W2M red QSO spectra (red filled histogram) which we derived by fitting a single Gaussian profile to the strongest broad line in each QSO spectrum (\S \ref{sec:rqlines}). The FWHM of the red QSO spectra range from 1000 km s$^{-1}$ to 9000 km s$^{-1}$. The red vertical line shows the $v_{\rm FWHM} = 1000$ km s$^{-1}$ cutoff. }\label{fig:fwhm}
\end{figure}

For the 382 remaining AGN without line analysis in ALPAKA, we batch-downloaded the spectra via the Science Archive Server (SAS) web interface\footnote{https://dr15.sdss.org/optical/spectrum/search} and used the value-added measurements provided in their multi-extension FITS headers to further reduce the size of the SDSS QSO sample by examining the distribution of their maximum emission-line-width. 
We examined the emission-line fits performed on the SDSS spectra through DR9 \citep{Bolton12}.  Single-component Gaussians were fitted to common UV and optical emission lines for all the spectra, tying the Balmer line widths to each other. Forbidden line widths are also tied to each other and fitted with a separate Gaussian. These are provided in the third extension of the FITS tables \citep[see][for details\footnote{This link provides details on how to extract the line information from the FITS tables {\tt https://data.sdss.org/datamodel/files/BOSS\_SPECTRO\_REDUX/ RUN2D/spectra/PLATE4/spec.html}}]{Bolton12}.
The line widths are reported in terms of the Gaussian $\sigma$ parameter, which we convert to a FWHM velocity, $v_{\rm FWHM} = 2.355\sigma$.  Figure \ref{fig:fwhm} shows the FWHM distribution of the broadest line component in each spectrum (blue line) where $v_{\rm FWHM} = 1000$ km s$^{-1}$ is shown with a vertical red line.  Thirty-two sources fail this criterion. %

We visually examined all the spectra that obeyed the broad-line criterion and identified several sources with erroneous redshifts.  We corrected these using the catalog of \citet{Hewett10} which provides improved redshifts for SDSS QSOs.
We also identified and removed another 11 sources whose spectra were featureless, with no discernible emission lines, suggesting an error in the automated line analysis for these spectra. This leaves 339 (382 - 32 -11) additional broad-line QSOs that we add to the QSO sample.  

\begin{figure}
\epsscale{1}
\plotone{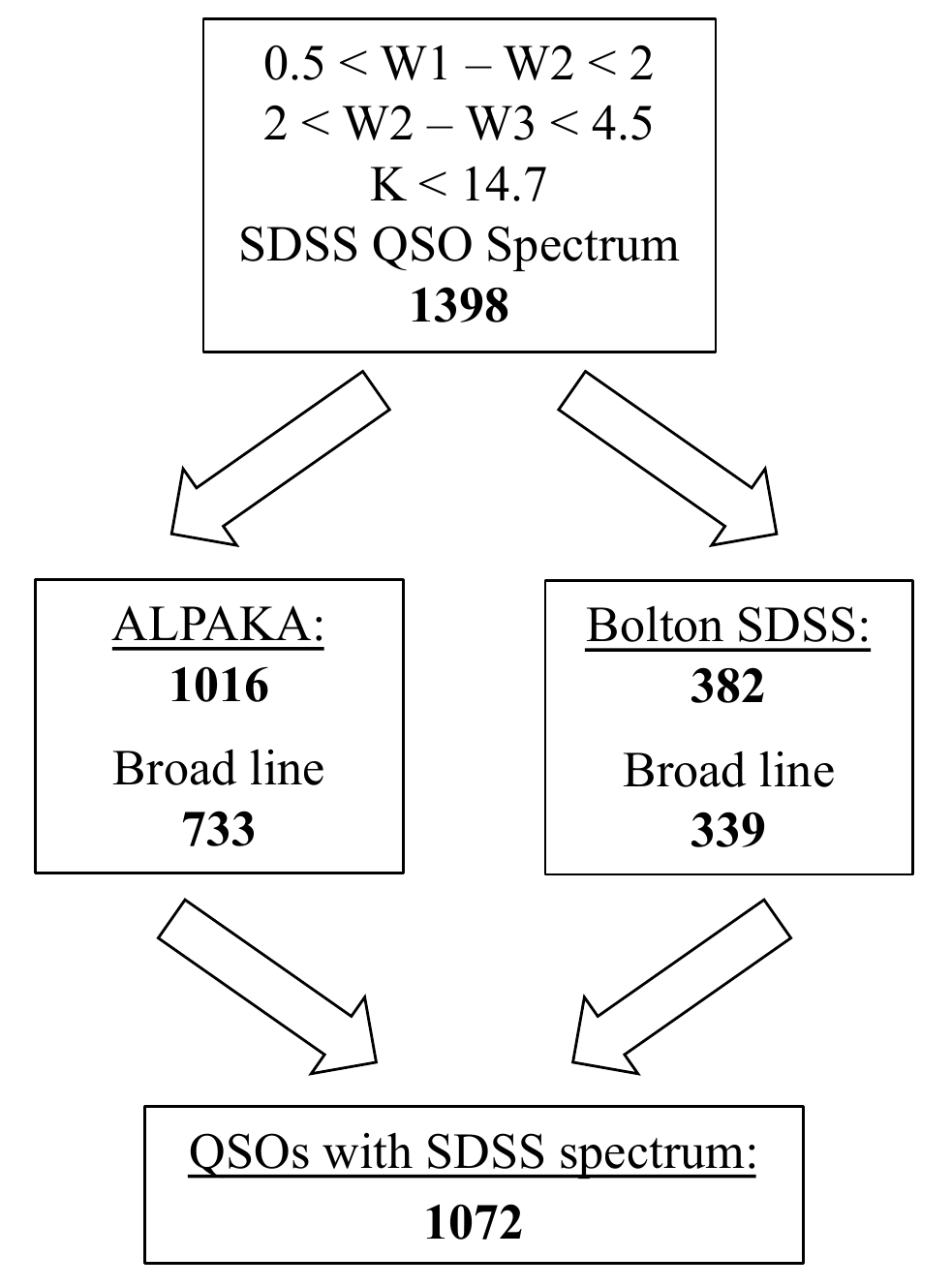}
\caption{Flowchart showing the process for identifying the sample of broad-line QSOs with SDSS spectra. }\label{fig:comp_flow}
\end{figure}

Therefore, together with the ALPAKA-line-width-selected sources, our QSO sample contains 1,072 ($733 + 339$) broad-line QSOs with SDSS spectra.  Figure \ref{fig:comp_flow} shows a flowchart of the process.  
As we we show in Section \ref{sec:reddening}, some of these SDSS-identified QSOs show significant amounts of reddening in their spectra and will be part of the red QSO subsample. 

\subsection{KX selection of red QSOs}\label{sec:kx}

We are interested in recovering likely reddened QSOs missed by the SDSS and other QSO selection algorithms by performing spectroscopy  on sources lacking a spectral classification in SDSS.
Initially, our follow-up spectroscopy spanned the full {\em WISE} color selection box (Eqn. \ref{eqn:gator_sel}; Figure \ref{fig:wise_colors}).  
Although incomplete, we recovered no new QSOs (red circles) in the lower right region of the diagram.  
We therefore aimed to refine our color selection to increase the efficiency of the follow-up spectroscopy of unidentified sources. 

\begin{figure*}
\epsscale{1}
\plottwo{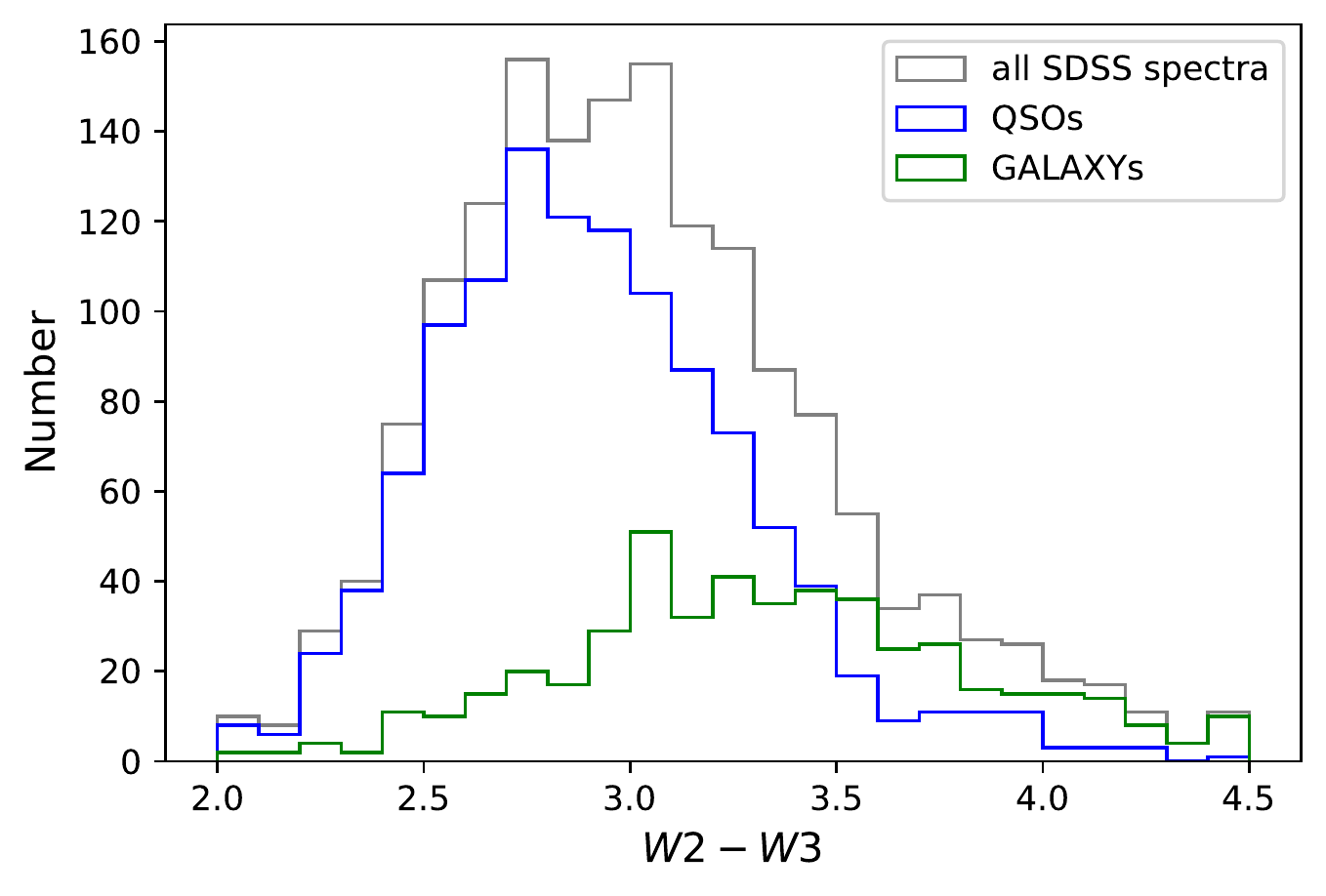}{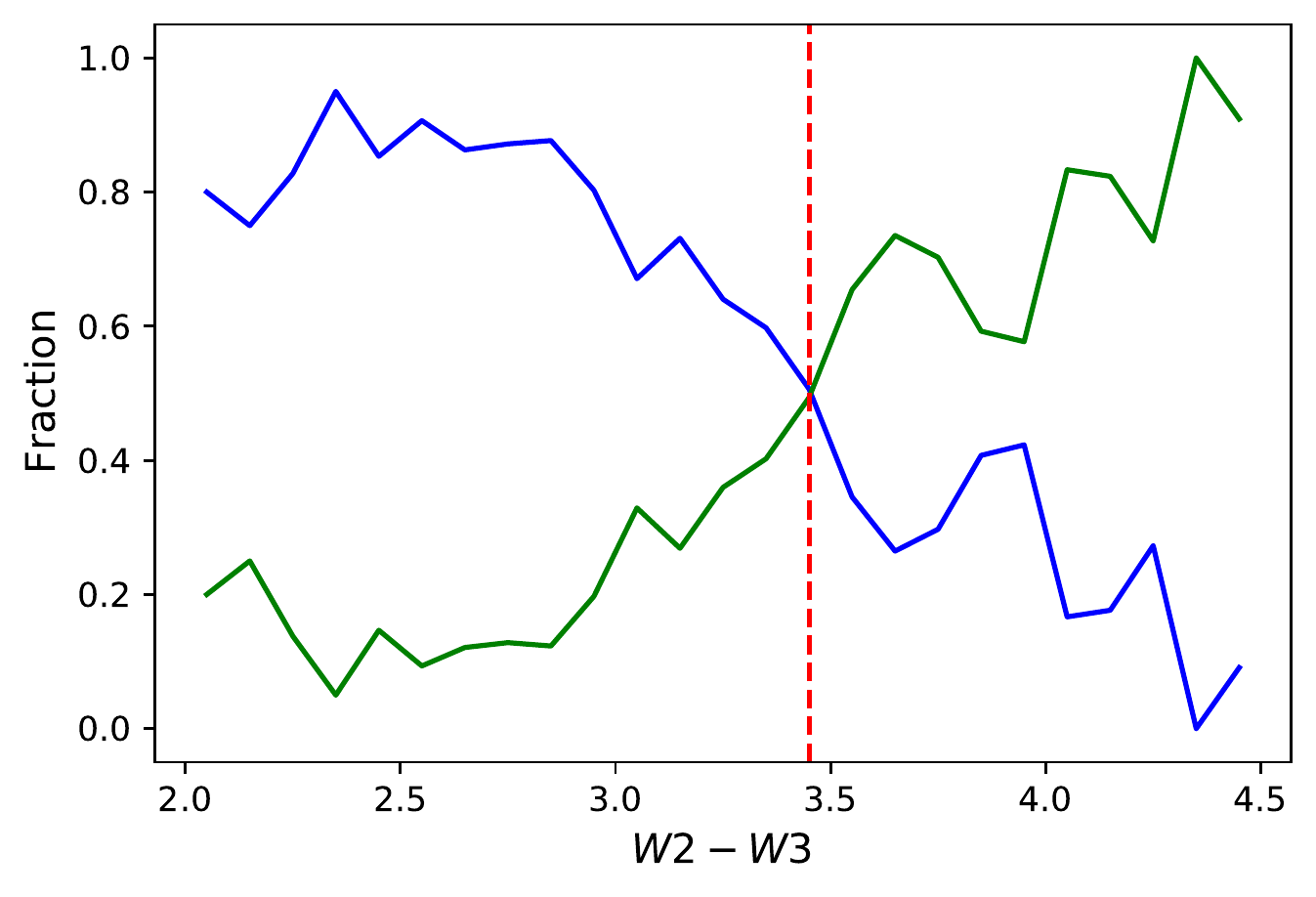}
\caption{These figures demonstrate the reasoning behind applying a diagonal color cut at large $[4.6]-[12]$ (i.e., $W2 - W3$) colors (Eqn. \ref{eqn:diag}). {\em Left -- } Histogram of SDSS-identified objects defined by Eqn. \ref{eqn:gator_sel} and $[3.4]-[4.6]>0.7$ (grey) and divided by classification: {\tt QSO}s (blue) and {\tt GALAXY}s (green). We see that beyond $[4.6]-[12] \gtrsim 3.5$, {\tt GALAXY}s begin to overtake {\tt QSO}s thereby reducing the efficiency of QSO selection at these colors. 
{\em Right --} Fraction of SDSS-identified objects classified as {\tt QSO}s (blue) and {\tt GALAXY}s (green). The red line shows the point, $[4.6]-[12] = 3.45$, at which the two populations are equal in proportion. }\label{fig:w2w3_colors}
\end{figure*}

\begin{figure}
\epsscale{1}
\plotone{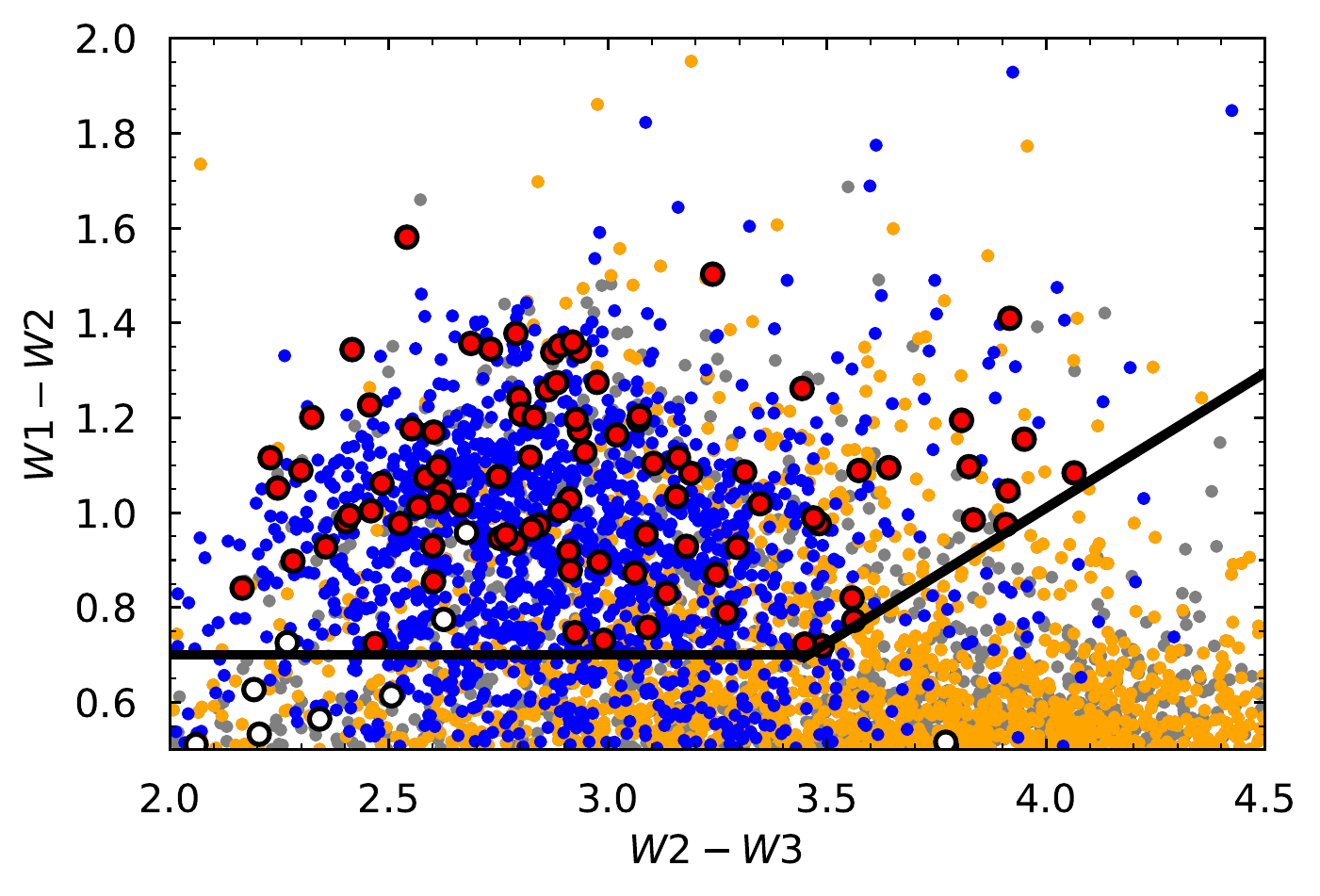}
\caption{Here we focus on the sources that obey the selection criteria defined in Equation \ref{eqn:gator_sel} with SDSS counterpart. 
Gray circles are sources without a spectrum in SDSS. Blue circles are identified as {\tt QSO} in SDSS, orange circles are identified as {\tt GALAXY}, and open white circles are {\tt STAR}s.  Red circles are the objects that meet the selection criteria in Equations \ref{eqn:diag} and \ref{eqn:kx} that amount to the sample that we follow up spectroscopically.
Although we defined a generous selection box around the QSO region, most of our candidates overlap the SDSS-identified QSOs and the solid black line represents our refined selection cuts.}\label{fig:wise_colors_zoom}
\end{figure}

We explore the occurrence of SDSS-identified {\tt GALAXY}s and {\tt QSO}s as a function of $W2 - W3$ (i.e., $[4.6]-[12]$) color in Figure \ref{fig:w2w3_colors}.  On the left, we plot a histogram of the number of SDSS spectra (gray line), {\tt QSO}s (blue line), and {\tt GALAXY}s (green line) and see a sharp decline of {\tt QSO} number beyond $W2-W3\sim3.5$.  On the right we show the fraction of {\tt QSO}-classified spectra in blue, compared with the fraction of {\tt GALAXY}-classified spectra in green.  The vertical red dashed line at $W2-W3=3.45$ shows the point where the fraction of {\tt GALAXY}s exceeds the {\tt QSO} fraction. 
To maximize success of identifying QSOs, we impose an additional cut:
\begin{equation}
\begin{array}{l}
    W1 - W2 > 0.7  \\ 
   \qquad {\rm ~for~} \quad 2 < W2-W3 < 3.45 \\
    W1 - W2  >  0.57~(W2 - W3) - 1.27   \\ 
    \qquad {\rm ~for~} \quad 3.45 < W2-W3 < 4.5, \label{eqn:diag}
\end{array}
\end{equation}
shown by the thick black line in Figures \ref{fig:wise_colors} and  \ref{fig:wise_colors_zoom}.  

We note that this cut remains liberal compared to other studies of infrared-selected AGN \citep[e.g.,][who used $W1-W2 > 0.8$]{Stern12,Assef13} and recovers all but four of the 120 F2M red quasars \citep[as seen in Figure 1 of ][]{Glikman18}.  The shorter wavelengths of the $W1$ and $W2$ bands are more affected by reddening than the longer {\em WISE} bands, so it is not surprising that reddened QSOs have slightly redder $W1 - W2$ colors. Applying the color cut in Eqn. \ref{eqn:diag} leaves 393 objects with no spectrum in SDSS.

Although only 889 Type 1 QSOs with SDSS spectra obey this stricter cut, we keep the full sample of 1072 QSOs established in \S \ref{sec:sdss} for our subsequent study. In Section \ref{sec:rad_stack} we demonstrate that the radio properties, which are the interest of this paper, are indistinguishable between the full blue QSO sample and the blue QSOs obeying Eqn. \ref{eqn:diag} (above the `diagonal' cut). 

The SDSS QSO selection algorithm \citep{Richards02,Ross12} is very successful at finding blue, broad-line QSOs whose colors naturally deviate from the color locus produced by Galactic stars.  
However, dust-reddened QSOs often overlap the stellar locus in the optical, making them hard to find by their SDSS colors alone \citep[Figure 5]{Urrutia09}.  As a remedy, \citet{Warren00} showed that in optical-to-near-infrared (so-called KX) color space, reddened QSOs can be cleanly separated from stars.  
We apply the following optical-to-near-infrared color cuts (in AB magnitudes),
\begin{equation}
J_{AB}-K_{AB} > 0.5 \quad {\rm and} \quad g_{AB}-J_{AB} > 2.0,  \label{eqn:kx}
\end{equation}
to the remaining 393 sources using their 2MASS $J$ and $K$ magnitudes combined with their SDSS $g$ magnitude. 
These color cuts correspond to $ J_{V}-K_{V} > 1.462$ and $g_{AB}-J_{V} > 2.938$, which are in line with the cuts used by \citet[$J_{V}-K_{V} > 1.3$]{Urrutia09} and \citet[$J_{V}-K_{V} > 1.5$]{Glikman13}.

\begin{figure}
\epsscale{1}
\plotone{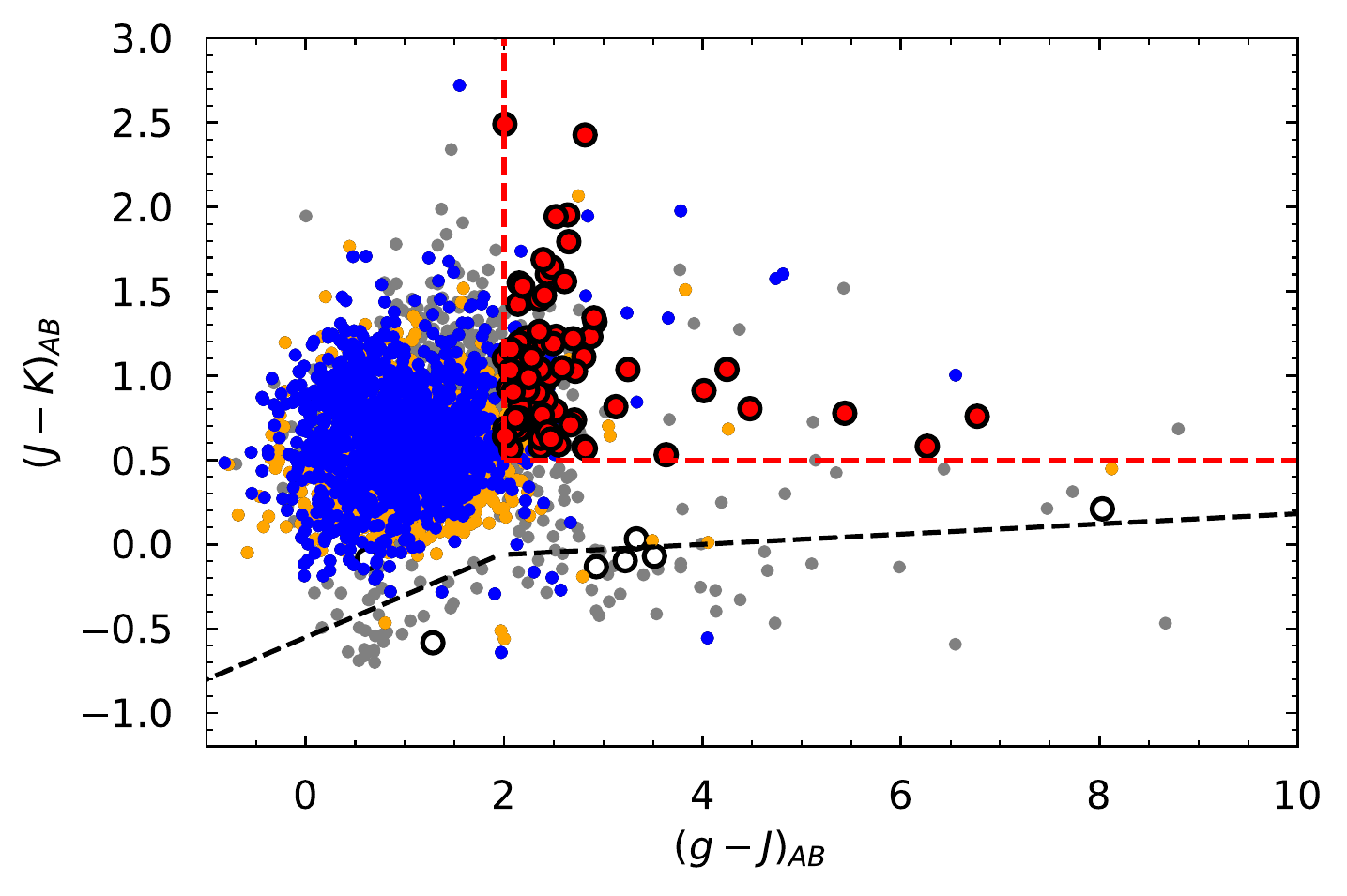}
\caption{Optical-to-near-infrared colors used in the KX selection method defined by \citet{Warren00} and refined by \citet{Maddox08}. The KX color cut separating stars from QSOs is shown by the black dashed line.  The red dashed line is the cut we impose (Eqn. \ref{eqn:kx}) in this space to identify red QSOs. The colors of sources are the same as in Figure \ref{fig:wise_colors_zoom}.}\label{fig:kx_colors}
\end{figure}

Figure \ref{fig:kx_colors} shows the $J-K$ vs.~$g - J$ KX-selection colors, using the same color scheme as in Figure \ref{fig:wise_colors_zoom}.  
In this space, the black dashed line separates stars from QSOs \citep{Maddox08}.  
Our candidate sources appear to complete the cloud of points around $g-J\simeq 2$ but also extend to very red tails in both $g-J$ and $J-K$.  Applying these criteria to the 393 sources results in 91 candidates. Seven of these are in the Stripe 82 region and, of those, three were originally identified in \citet{Glikman18}.  Another source in Stripe 82 was originally identified in the FIRST-2MASS survey \citep{Glikman07} and was recovered in \citet{Glikman18}. 
The 91 selected sources are shown with large outlined red circles in Figures \ref{fig:wise_colors_zoom}, and \ref{fig:kx_colors}.  
They obey the criteria outlined in Equation \ref{eqn:diag} at {\em WISE} wavelengths and have the optical through near-infrared colors in Eqn. \ref{eqn:kx}.  

\begin{figure*}
\epsscale{1}
\plotone{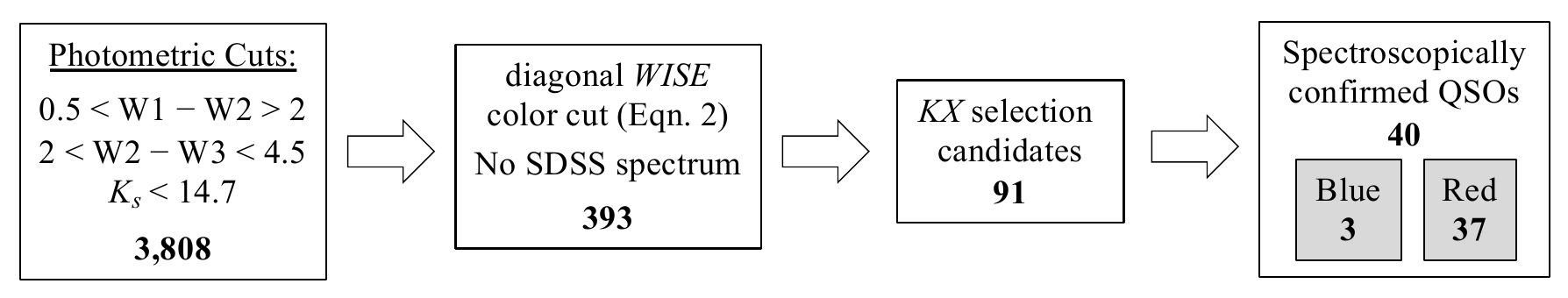}
\caption{Flowchart showing the process for selecting the red QSO candidate sample in this work. Each box reports the candidate selection step, with the number of sources that passed each stage shown in boldface. The final box reports the confirmed QSOs that are added to the Type 1 QSOs with SDSS spectra.} \label{fig:kx_sel_flowchart}
\end{figure*}

Figure \ref{fig:kx_sel_flowchart} shows a flowchart of the process to select these redder QSO candidates missed by SDSS. 
We list these candidates in Table \ref{tab:candidates}, including their positions, optical through mid-infrared magnitudes, as well as classification and redshift based on spectroscopic followup (\S \ref{sec:spec}).

\subsection{Spectroscopic Classification} \label{sec:spec} 

\subsubsection{Archival Spectroscopy}

Among the 91 QSO candidates, we recover twelve objects that were identified in red QSOs surveys. 
Four were in the pilot study of \citet{Glikman18} over Stripe 82: W2M J0030$-$0027 is a Type 2 AGN at $z=0.242$ originally identified in \citet{Glikman12}; W2M J0306+0108 is a Type 2 AGN at $z=0.189$; and W2M J0349+0054 is a Type 2 AGN at $z=0.109$. F2M J2216$-$0054 is a red quasar at $z=0.2$ originally identified in \citet{Glikman07}. 

The other eight objects are in the spring sky region and were all identified in the F2M survey \citep{Glikman12}. Four of them are red quasars.  F2M J1232+1112 is at $z=0.25$, F2M J1248+0531 is at $z=0.749$, F2M J1439+1136 is at $z=0.296$, and F2M J1554+0714, which was unclassified in  \citet{Glikman12}, has had new spectroscopy (\S \ref{sec:obs}) revealing broad H$\alpha$ and corresponding, narrower, H$\beta$ at $z=1.64$.  

\subsubsection{Spectroscopic Observations}\label{sec:obs}

We obtained spectroscopic classifications of all but four out of the 91 candidates in our sample ($96\%$ spectroscopic completeness).
We also obtained near-infrared spectra for QSOs whose SDSS spectrum revealed strong reddening (\S \ref{sec:reddening}) to broaden their wavelength coverage.
These observations were conducted over six observing runs at three different telescope facilities.  We used the SpeX spectrograph \citep{Rayner03} at the NASA InfraRed Telescope Facility (IRTF), TripeSpec \citep{Herter08} at the Palomar Observatory's 200 inch Hale telescope, and TripleSpec \citep{Wilson04} on the 3.5 meter telescope at the Apache Point Observatory. The data were reduced using the Spextool software package \citep{Cushing04}, which was originally written for the SpeX instrument but has been modified to also reduce data from TripleSpec.  We followed the procedure outlined in \citet{Vacca03} to correct for telluric absorption using spectra of nearby A0V stars obtained immediately before or after our target exposures. 

We also obtained optical spectroscopy of twenty-seven sources. 
Nine sources were observed with the MODS1B Spectrograph on the Large Binocular Telescope (LBT) observatory, with the red and blue arms simultaneously, with a 0\farcs6-wide slit on UT 2013 March 14, covering the wavelength range $3300 - 10100$\AA. 
We also obtained 18 optical spectra with the KAST spectrograph at the Lick Observatory.  All data were reduced using standard IRAF routines and flux calibrated using Feige 34.

Examination of the spectra for the presence of a broad emission line (see \S \ref{sec:rqlines}) results in 40 new QSOs; 37 of them have $E(B-V) > 0.25$ deeming them as red QSOs (see \S \ref{sec:reddening}) and label them with the prefix ``W2M'', which is an abbreviation of WISE-2MASS, consistent with our definition established in \citet{Glikman18}.  
The three newly-discovered blue QSOs are included with the blue sample. 

\begin{figure*}
\epsscale{1.2}
\figurenum{10a}
\plotone{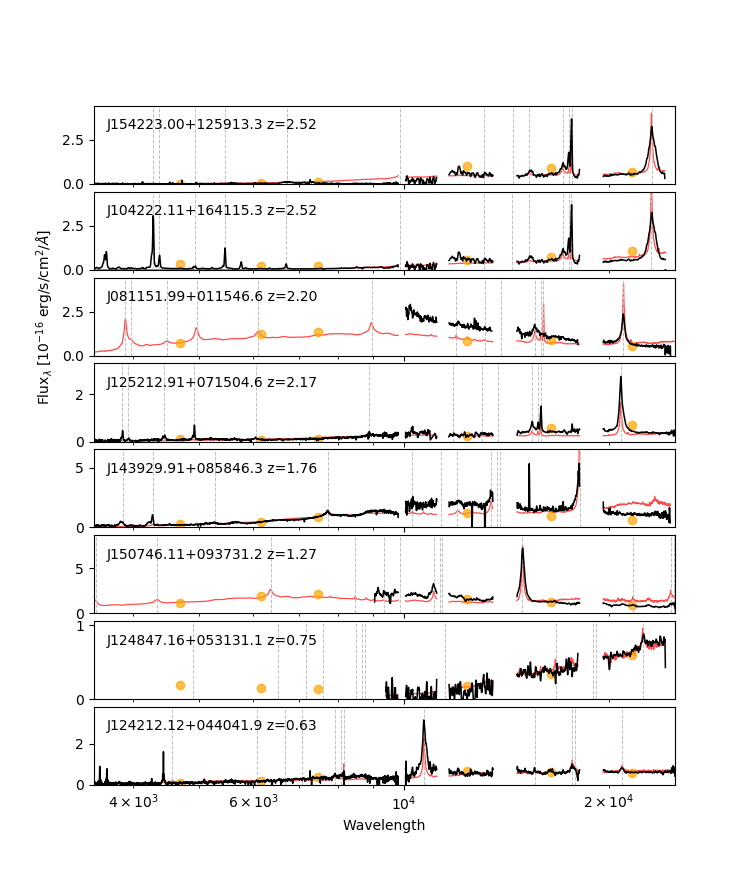}
\caption{Optical through near-infrared spectral atlas of newly identified QSOs, in decreasing redshift order. Typical QSO emission lines are marked with vertical dashed lines.  Orange circles represent the photometry-based fluxes in the $g$, $r$, and $i$ bands from SDSS, to which the optical spectrum was scaled, and $J$, $H$, and $K_s$ bands from 2MASS, to which the near-infrared spectrum has been scaled. The red line is the best-fit reddened QSO template from which $E(B-V)$ is derived.  }\label{fig:qso_atlas1}
\end{figure*}

Figures \ref{fig:qso_atlas1} - \ref{fig:qso_atlas5} show a spectral atlas of the optical through near-infrared spectra of the 40 objects classified as QSOs in decreasing redshift order.  
The last five columns of Table \ref{tab:candidates} provides the details on the objects' spectroscopic observations, source classification, and assigned redshift. 

\begin{figure*}
\epsscale{1.2}
\figurenum{10b}
\plotone{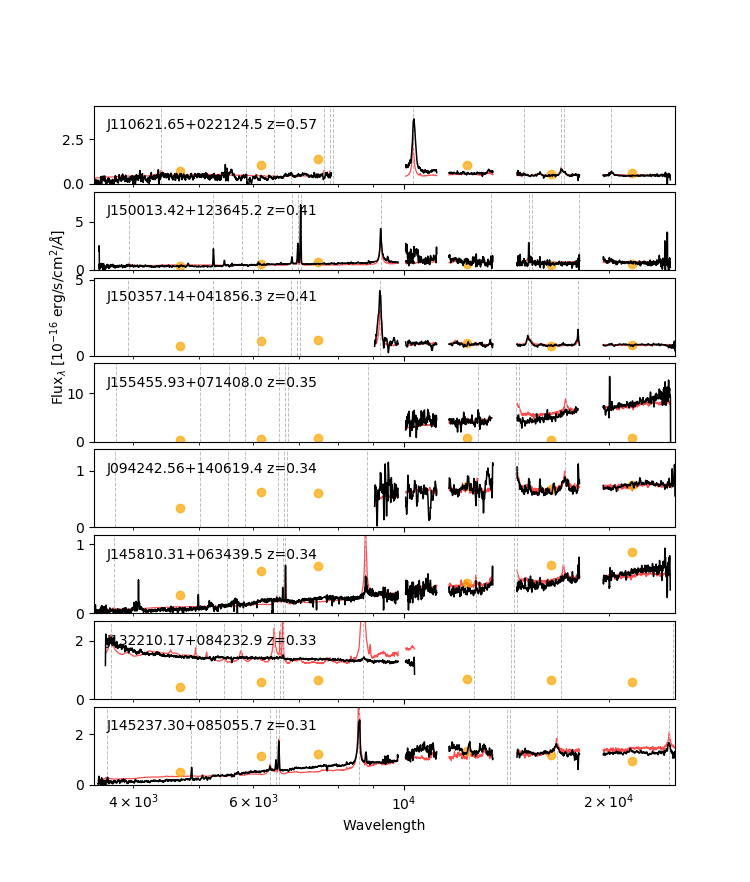}
\caption{Same as Figure \ref{fig:qso_atlas1}. }
\end{figure*}

\begin{figure*}
\epsscale{1.2}
\figurenum{10c}
\plotone{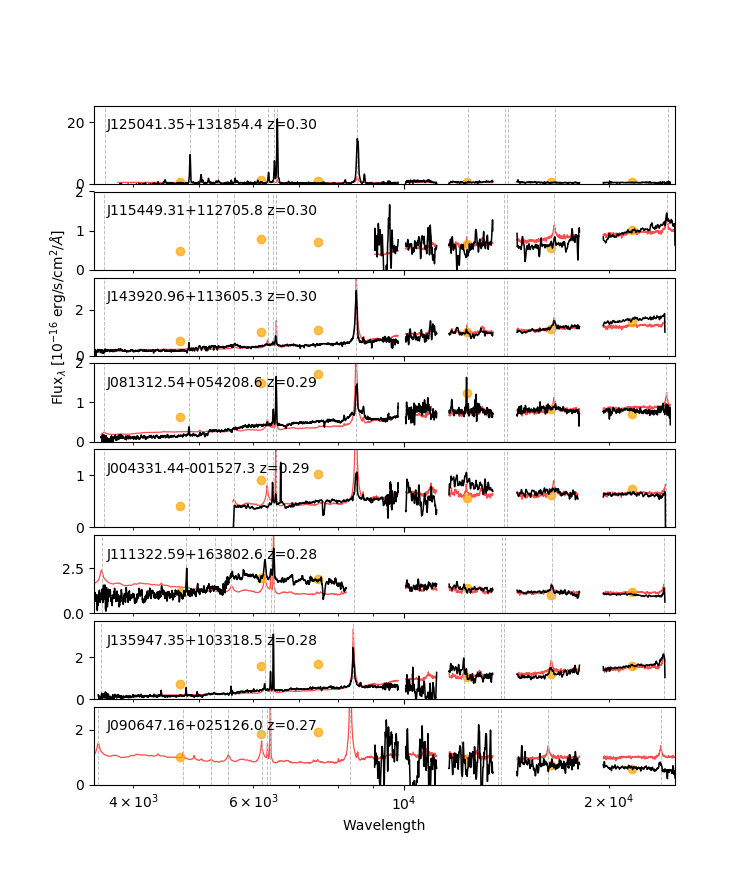}
\caption{Same as Figure \ref{fig:qso_atlas1}.}
\end{figure*}

\begin{figure*}
\epsscale{1.2}
\figurenum{10d}
\plotone{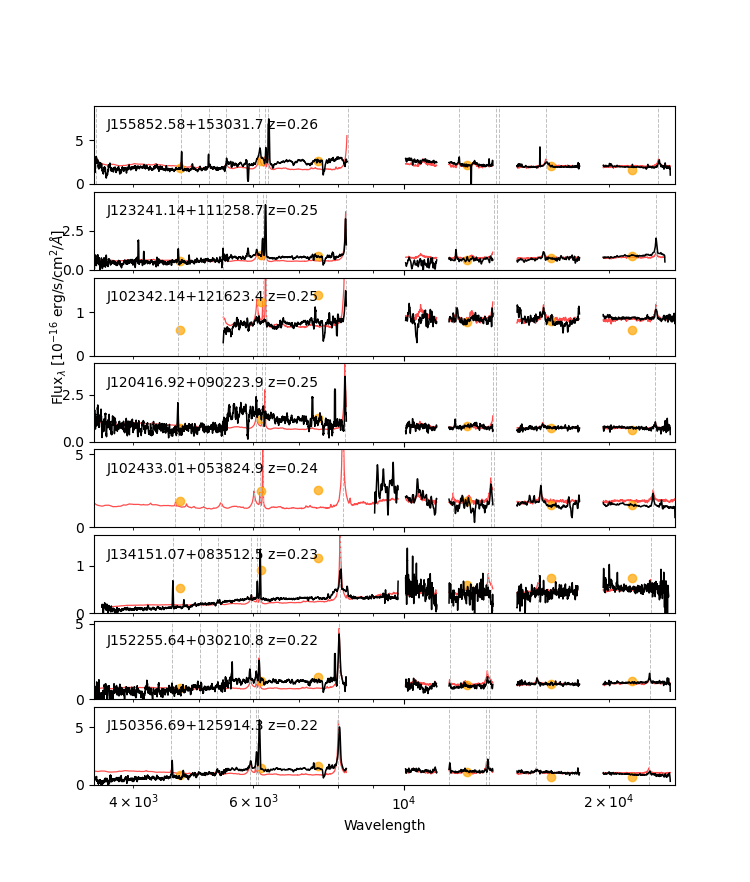}
\caption{Same as Figure \ref{fig:qso_atlas1}.}
\end{figure*}

\begin{figure*}
\epsscale{1.2}
\figurenum{10e}
\plotone{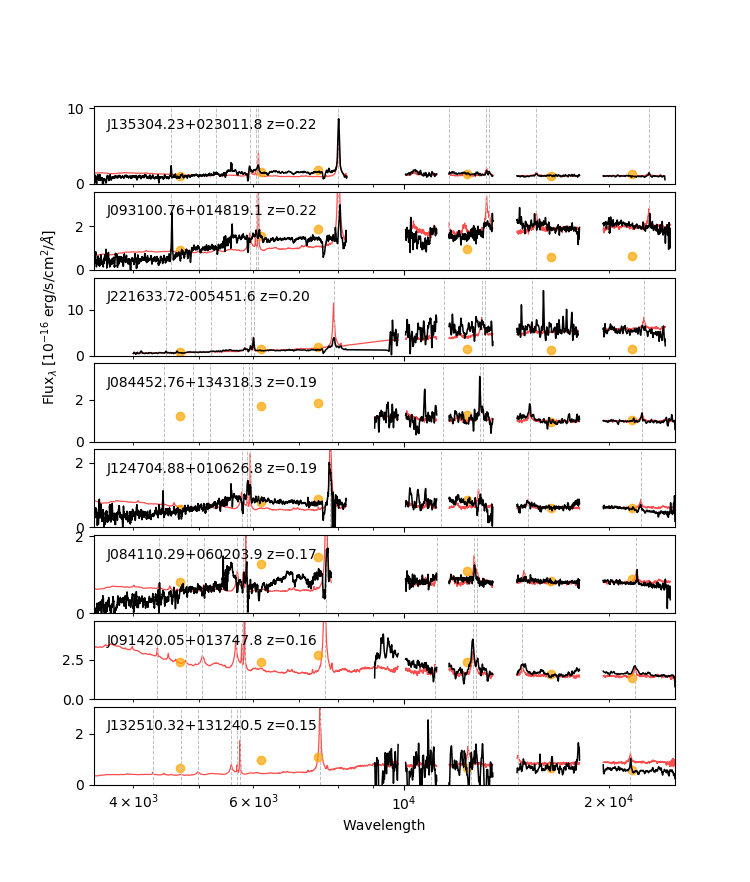}
\caption{Same as Figure \ref{fig:qso_atlas1}.}\label{fig:qso_atlas5}
\end{figure*}




\begin{deluxetable*}{cccccccccccccc}

\rotate



\tablecaption{WISE-2MASS red QSO candidates \label{tab:candidates}}

\tablenum{2}

\tablehead{\colhead{Name} & \colhead{$g$} & \colhead{$r$} & \colhead{$i$} & \colhead{$J$} & \colhead{$K$} & \colhead{$W1$} & \colhead{$W2$} & \colhead{$W3$} & \colhead{$W4$} & \colhead{Redshift} & \colhead{Class} & \colhead{Spectrum} & \colhead{References} \\ 
\colhead{} & \colhead{(mag)} & \colhead{(mag)} & \colhead{(mag)} & \colhead{(mag)} & \colhead{(mag)} & \colhead{(mag)} & \colhead{(mag)} & \colhead{(mag)} & \colhead{(mag)} & \colhead{} & \colhead{} & \colhead{} & \colhead{} } 

\startdata
J003009.08$-$002744.2 & 20.35 & 19.18 & 18.44 & 16.50 & 14.20 & 13.86 & 12.81 & 8.90 & 6.48 & 0.242 & NLAGN & LRIS & G18 \\
J004331.44$-$001527.3 & 20.20 & 18.74 & 18.19 & 16.86 & 14.42 & 13.27 & 12.34 & 9.74 & 7.39 & 0.294 &  QSO   &  LRIS/IRTF &\\
J030654.88+010833.6   & 19.39 & 18.05 & 17.53 & 16.36 & 14.49 & 13.22 & 12.02 & 9.19 & 6.30 & 0.189 & NLAGN &  LICK/SpeX  & G18 \\
J034902.65+005430.5   & 18.97 & 17.92 & 17.14 & 15.84 & 13.35 & 11.59 & 10.23 & 7.31 & 4.86 & 0.109 & NLAGN & LICK/SpeX  & G18 \\
J080121.75+004028.4   & 19.10 & 18.10 & 17.56 & 16.16 & 14.51 & 13.66 & 12.79 & 9.54 & 7.23 & \ldots &  ?          &  IRTF &  \\
J080155.37+132827.7   & 22.73 & 18.14 & 17.55 & 16.36 & 14.62 & 13.47 & 12.72 & 9.63 & 7.17 & 0.110 & Galaxy & LICK   &  \\
\enddata

\tablerefs{G18 = \citet{Glikman18}}

\tablerefs{Only a portion of this table is shown here to demonstrate its form and content. A machine-readable version of the full table is available.}

\end{deluxetable*}

\subsubsection{Red QSO Line Properties}\label{sec:rqlines} 

All the red QSOs for which we obtained spectra have at least one broad emission line to which we fit a single Gaussian profile. We convert the best-fit $\sigma$, in wavelength units, to a FWHM in velocity units through the expression,
\begin{equation}
v_{\rm FWHM} = c \times \frac{2.355\sigma}{\lambda_{\rm 0}},
\end{equation}
where $c$ is the speed of light and $\lambda_{\rm 0}$ is the rest-wavelength of the line being fit.  The red filled histogram in Figure \ref{fig:fwhm} shows the distribution of $v_{FWHM}$ for the newly-discovered red QSOs.  

Our master sample of mid-infrared color selected Type 1 QSOs now contains 1112 sources, with 1072 objects coming from the SDSS spectroscopic sample and 40 newly-identified QSOs selected by their red colors in KX color space. 
Table \ref{tab:w2m} lists the full QSO sample -- referred to as the W2M sample going forward -- listing their coordinates, SDSS, 2MASS, and {\em WISE} magnitudes as well as peak flux densities in the FIRST (1.4 GHz) and Very Large Array Sky Survey \citep[VLASS;$2-4$ GHz;][]{Lacy20} surveys and the spectroscopic redshift. The table also indicates whether the object obeyed the stricter selection criteria of Eqn \ref{eqn:diag}, whether the object qualifies as a red QSO (\S \ref{sec:reddening}) with its best-fit $E(B-V)$ value. 




\begin{deluxetable*}{ccccccccccccccccccc}

\rotate

\tabletypesize{\scriptsize}


\tablecaption{WISE-2MASS QSOs \label{tab:w2m}}

\tablenum{3}

\tablehead{\colhead{Name} & \colhead{$g$} & \colhead{$r$} & \colhead{$i$} & \colhead{$J$} & \colhead{$K$} & \colhead{$W1$} & \colhead{$W2$} & \colhead{$W3$} & \colhead{$W4$} & \colhead{$F_{pk}$\tablenotemark{a}} & \colhead{$V_{pk}$\tablenotemark{b}} & \colhead{Redshift} & \colhead{Diag\tablenotemark{c}} & \colhead{Class} & \colhead{$E(B-V)$} & \colhead{$M_K$\tablenotemark{d}} & \colhead{Lum\tablenotemark{e}} & \colhead{$\log L_{\rm bol}$} \\ 
\colhead{} & \colhead{(mag)} & \colhead{(mag)} & \colhead{(mag)} & \colhead{(mag)} & \colhead{(mag)} & \colhead{(mag)} & \colhead{(mag)} & \colhead{(mag)} & \colhead{(mag)} & \colhead{(mJy)} & \colhead{(mJy)} & \colhead{} & \colhead{} & \colhead{} & \colhead{(mag)} & \colhead{(mag)} & \colhead{Sel} & \colhead{(erg s$^{-1}$) }  } 

\startdata
W2M~J003147.21$-$004359.2 &  17.35 &  16.47 &  15.95 &  15.50 &  14.24 &  13.22 &  12.68 &   9.88 &   8.20 &          &          &  0.067 & F &       QSO &         & $-23.15$ & T & 43.90 \\ 
W2M~J003238.19$-$010035.2 &  17.58 &  17.14 &  16.76 &  15.80 &  14.34 &  13.41 &  12.61 &   9.19 &   6.78 &          &          &  0.092 & T &       QSO &         & $-23.77$ & T & 44.36 \\ 
W2M~J003328.05$-$001912.8 &  17.37 &  16.54 &  16.01 &  15.29 &  13.92 &  12.72 &  11.96 &   9.41 &   7.07 &   1.02 &  2.02 &  0.107 & T &       QSO &         & $-24.55$ & T & 44.59 \\ 
W2M~J003659.78$+$010544.3 &  18.19 &  17.68 &  17.20 &  16.00 &  14.45 &  13.57 &  12.82 &  10.02 &   8.15 &        &          &  0.121 & T &        QSO &         & $-24.31$ & T & 44.42 \\ 
W2M~J003659.82$-$011332.5 &  21.29 &  20.19 &  19.66 &  16.57 &  13.63 &  11.67 &  10.51 &   7.88 &   5.21 &   1.92 &  2.77 &  0.294 & T & red-QSO & 0.79 & $-27.67$ & T & 46.25 \\ 
W2M~J003847.97$+$003457.5 &  17.07 &  16.58 &  16.09 &  15.34 &  14.01 &  13.15 &  12.50 &   9.91 &   7.63 &  1.12 &          &  0.081 & F &       QSO &         & $-23.81$ & T & 44.12 \\ 
\enddata

\tablenotetext{a}{FIRST peak flux density.}
\tablenotetext{b}{VLASS peak flux density.}
\tablenotetext{c}{Boolean indicating whether the object obeys the stricter 'diagonal' selection cut of Equation \ref{eqn:diag}.}
\tablenotetext{d}{Absolute $K$-band magnitude, de-reddened by $E(B-V)$ for red QSOs.}
\tablenotetext{e}{Boolean indicating whether the object is part of the luminosity restricted subsamples.}
\tablecomments{Only a portion of this table is shown here to demonstrate its form and content. A machine-readable version of the full table is available.}

\end{deluxetable*}

\section{Defining Red and Blue QSO subsamples}\label{sec:reddening}

Here, we aim to construct well-defined red and blue QSO subsamples whose properties can be distinguished and compared. We study the reddening properties of the QSOs and use that information to determine the sample's de-reddened absolute magnitudes to ensure the two samples are intrinsically similar. 

\subsection{Reddening investigation}

\begin{figure*}
\figurenum{11}
\plottwo{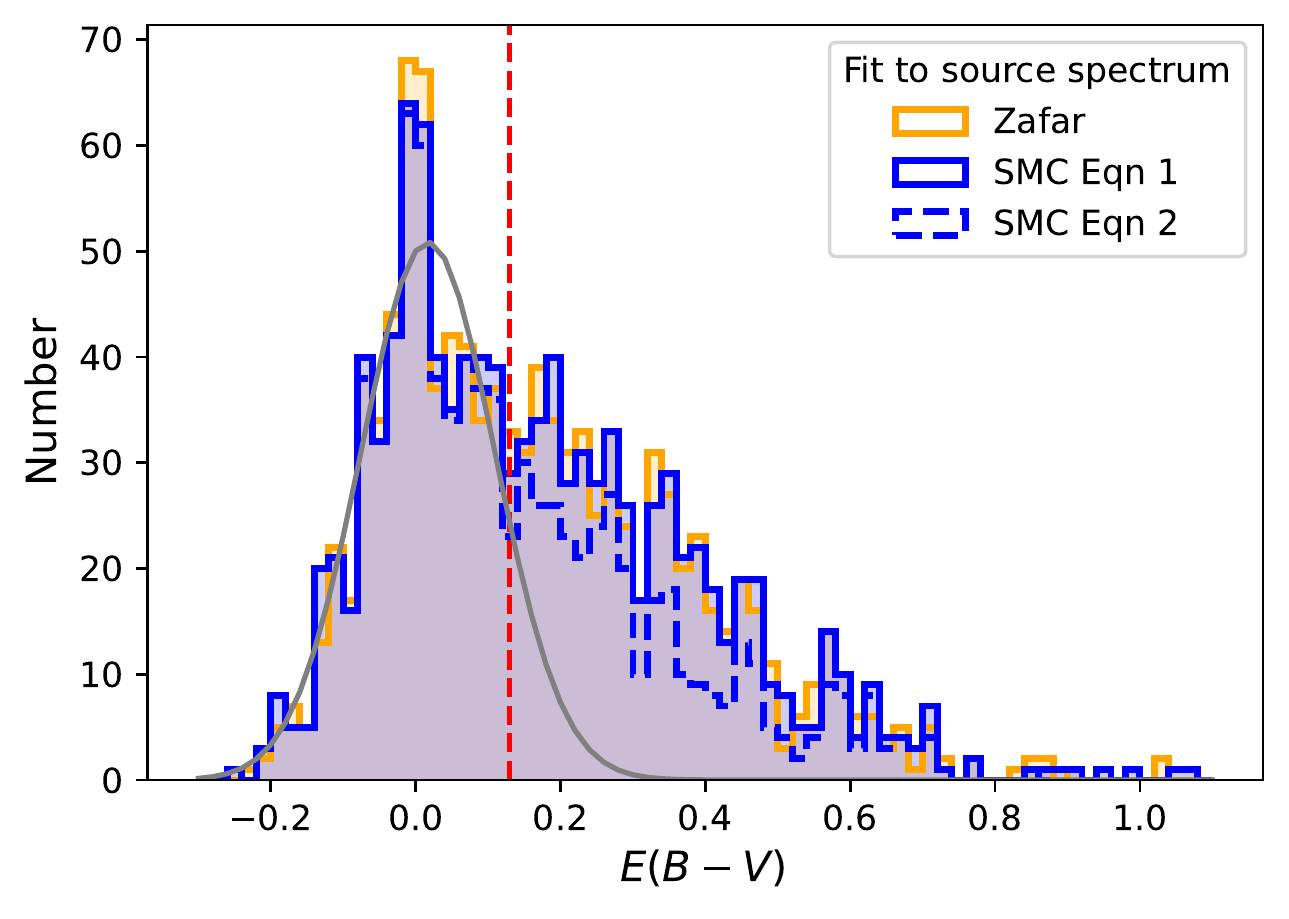}{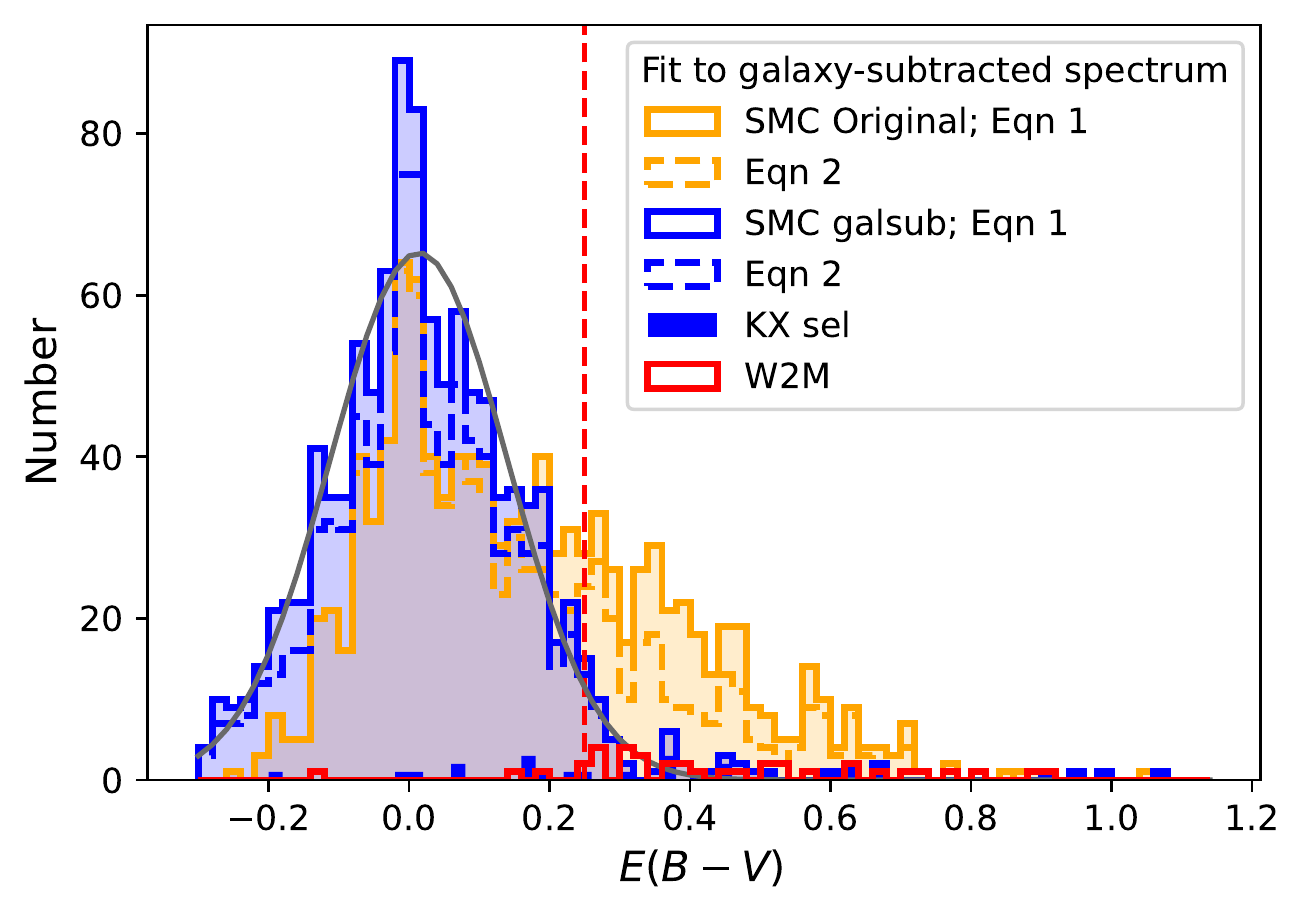}
\caption{
{\em Left --} Distribution of $E(B-V)$ for the SDSS QSOs, which we determined by fitting a reddened QSO template to each spectrum.  The solid line represents the broad line QSOs obeying Eqn \ref{eqn:gator_sel} with $E(B-V)$ derived using the SMC dust law of \citet{Gordon98} shown in blue and the \citet{Zafar15} dust law shown in orange. The distributions are nearly identical and therefore we proceed with the SMC dust law. The dashed line is the $E(B-V)$ distribution for the subsample obeying the diagonal color cut of Eqn \ref{eqn:diag}.  The Gaussian curve shown in gray is centered at $E(B-V) = 0$ with excess reddening likely due to host galaxy contamination. 
{\em Right --} Distribution of $E(B-V)$ for the SDSS QSOs after subtracting host galaxy light from each spectrum (blue). Solid lines represent the full sample, while the dashed line is for the subsample obeying the diagonal color cut. 
The orange histogram shows, for comparison, the original values plotted in the left panel. Subtracting a host galaxy shifts the $E(B-V)$ to lower values and is subsequently well-fit by a Gaussian distribution centered on $E(B-V)=0$. The vertical red dashed line at $E(B-V) = 0.25$ is our defined limit for a QSO to be considered dust-reddened.
}\label{fig:ebv_hist}
\end{figure*}

We fit a reddened QSO template to each SDSS spectrum, following the formalism described in \citet{Glikman07} and plot the distribution of $E(B-V)$ in the left panel of Figure \ref{fig:ebv_hist}. 
We use the \citet{Gordon98} SMC dust law (blue line) and, for comparison, we also used the dust law from \citet{Zafar15} that was derived directly from QSOs (orange line).  
The distributions are nearly identical and we find no systematic differences between the two dust laws.  Therefore, we choose to use the SMC dust law of \citet{Gordon98} in order to maintain consistency with previous red quasar studies.
The dashed blue line shows the $E(B-V)$ distribution for the QSOs that obey Eqn \ref{eqn:diag}, which demonstrates that including sources at lower $W1-W2$ values finds optically redder objects.  This is likely due to contamination from host galaxy emission, since dust reddening would have the effect of increasing the $W1-W2$ color. We discuss the removal of host galaxy emission to account for this in Section \ref{sec:gandalf}, below. 

The distributions appear as a Gaussian (gray curve; fitted to the data with $E(B-V) < 0.12$), peaked at $E(B-V) = 0$ with a broad tail extending toward redder colors. The Gaussian distribution is attributed to an intrinsic spread of the power-law continuum slope of un-reddened quasars. The tail extending to higher $E(B-V)$ values is due to dust-reddening as well as host galaxy contamination.  This was pointed out earlier by \citet{Richards02} when examining the relative $g^* - i^*$ colors of SDSS QSOs in an early data release of the SDSS survey. 

\subsection{Removal of host galaxy emission}\label{sec:gandalf}

\begin{figure*}
\figurenum{12}
\plottwo{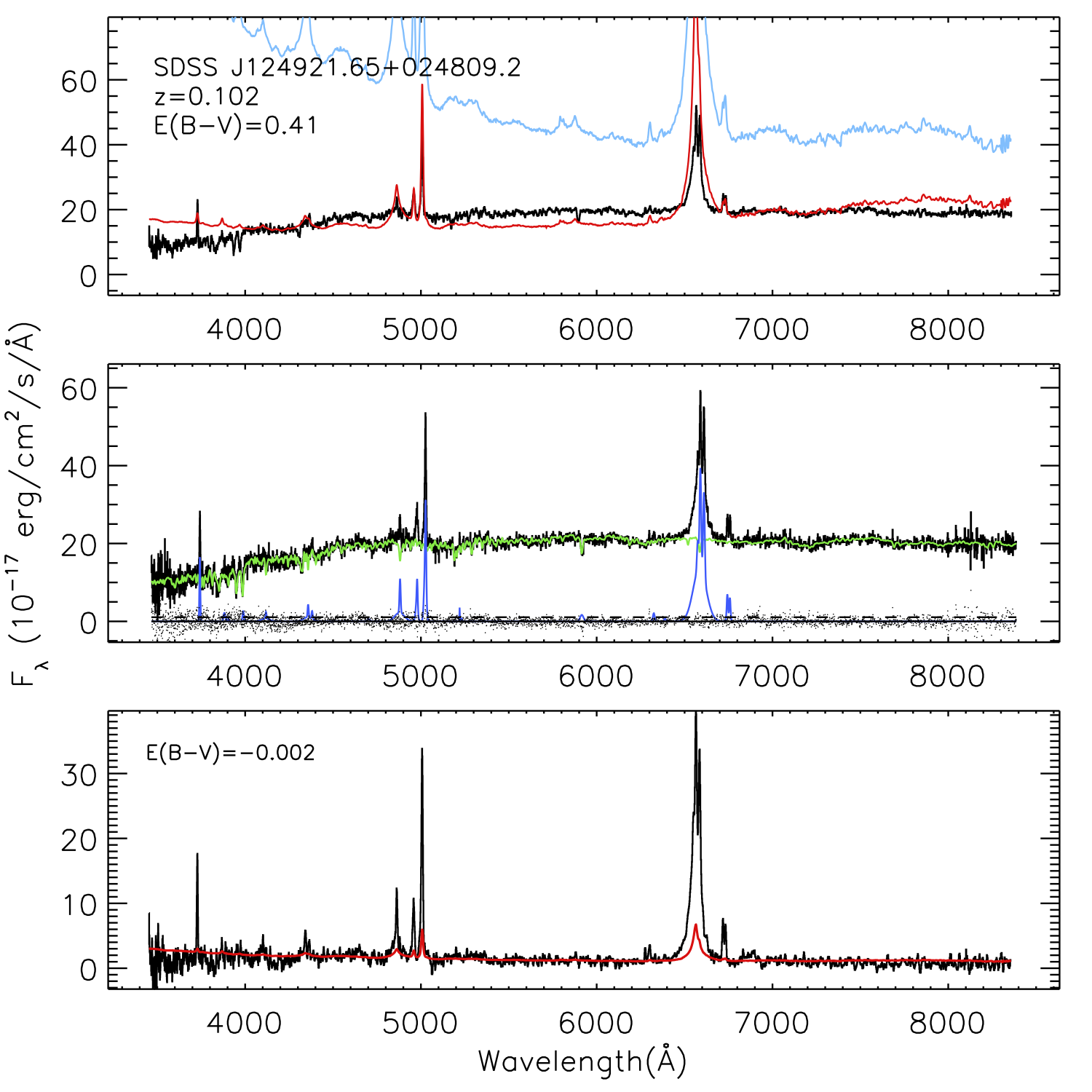}{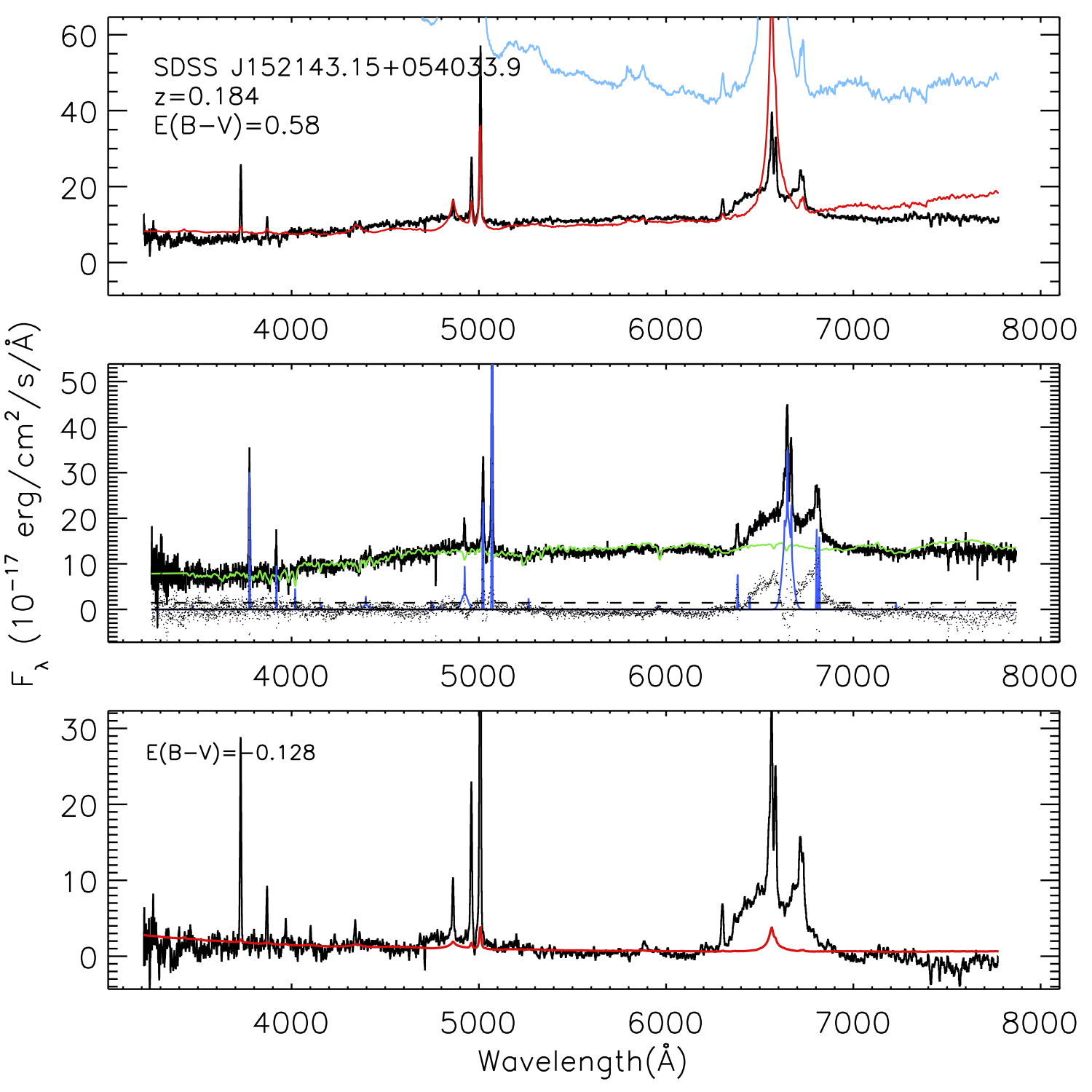}
\caption{Examples of two QSO spectra whose host galaxies were fitted with GANDALF and subtracted to improve the QSO template fitting process and better-estimate the reddening, parametrized by $E(B-V)$, experienced by the QSO. 
The top panel in each column is the original, poor, fit. The black line is the original spectrum. The red line is the best-fit, reddened QSO template and the cyan line is the unreddened QSO template.  
The middle row shows the fits produced by GANDALF, with the host galaxy spectrum shown in green, atop the original spectrum in black. 
The bottom panel shows the galaxy-subtracted spectrum in black and the newly best-fitted reddened QSO template in red. 
In both examples, and in the vast majority of spectra in our sample, host galaxy emission results in artificially large extinctions. Subtracting the host galaxy exposes predominantly unreddened QSOs.  }\label{fig:gandalf}
\end{figure*}

Because this sample contains a majority of sources at low redshift ($z<0.4$), lower luminosity AGN may have red colors as a result of added light from a host galaxy rather than reddening of the AGN continuum.  To correct for this, we used the Gas AND Absorption Line Fitting code \citep[GANDALF;][]{Sarzi06} to fit a model host galaxy simultaneously with Gaussian profiles fitted to specified emission lines to the $z<0.4$ objects with SDSS spectra.  We examined the fits and subtract the best-fit host galaxy model from the spectra when a good fit is achieved (i.e., the galaxy continuum traces stellar absorption features around 4000\AA).  Figure \ref{fig:gandalf} shows two representative examples of spectra whose reddened QSO template fits were poor and needed to have a host galaxy template subtracted.  The top panel in each column is the original, poor, fit.  The middle panel shows the fits produced by GANDALF, with the host galaxy spectrum shown in green. The bottom panel shows the galaxy-subtracted spectrum and its fitted reddened QSO template.  In both cases -- as well as in the other 368 sources that needed this treatment -- removing the galaxy results in an AGN spectrum with little to no reddening.

We then re-fit the galaxy-subtracted AGN with the reddened QSO template to minimize host galaxy effects on our $E(B-V)$ estimates. The right panel of Figure \ref{fig:ebv_hist} shows this newer distribution in blue (solid line is the full SDSS sample, the dashed line shows only sources obeying Eqn \ref{eqn:diag} with the original distribution, seen also ion the left, shown for comparison in orange).  The galaxy-subtracted AGN reddening distribution is now symmetric and well-fit by a slightly broader Gaussian, still centered at $E(B-V) = 0$.  The fact that host-galaxy removal shifts the QSO subsample that obeys the diagonal color selection (dashed blue line; Eqn \ref{eqn:diag} ) to agree with the full sample provides reassurance that the excess red color in that subsample was indeed due to host galaxy light and that the blue QSOs are otherwise similar. 
The removal of the host galaxy results in a reddening estimate that is lower by a mean of $0.42$ mag. 

\subsection{Reddening in the W2M sample}\label{sec:ebv_w2m}

The 40 newly identified W2M QSOs do not have the uniform optical spectroscopy that SDSS provides, and thus our estimates of $E(B-V)$ for these sources requires an object-by-object approach, depending on what data exist for each source. \citet{Glikman07,Glikman12,Glikman13,Glikman18} discuss the challenges of estimating $E(B-V)$ depending on the available spectroscopy and photometry. Here we use wisdom gained from those studies to determine the reddening for the W2M QSO sample. 

Twelve sources have only a near-infrared spectrum and, when fit with a reddened composite template, six of those yield $E(B-V) \ge 0.25$.  The remaining six are fit with a lower $E(B-V)$ but when the reddened template is compared with the SDSS photometry, there are significant mis-matches, suggesting that the optical emission is far more reddened than is constrained by the the near-infrared spectrum. Some of this is due to the near-infrared spectra becoming noisier toward shorter wavelength, where the shape of the spectrum is more sensitive to $E(B-V)$. For the six sources whose reddened fits do not match their optical photometry, we fit a reddened template to the effective flux in eight photometric bands, $u$,$g$,$r$,$i$,$z$ from SDSS plus $J$, $H$, $K_s$ from 2MASS, following the procedure that we describe in Section 6.1 of \citet{Glikman13}. In this process, three QSOs were well fit with $E(B-V) < 0.25$ and we do not classify them as red QSOs. 

The two highest redshift QSOs in the W2M sample, W2M J1542+1259 and W2M J1042+1641\footnote{W2M J1042+1641 is also a quadruply lensed quasar with significant flux anomaly. The system is thoroughly analyzed in \citet{Glikman18b} and Glikman et al. (2022, submitted). }, both at $z=2.52$, are heavily absorbed in their optical (rest-frame UV) spectrum due to outflowing gas that gives rise to broad absorption lines (BAL).  For these sources, we conducted the template fitting to the spectrum considering only wavelengths above $\lambda > 8000$\AA\ and $\lambda > 10000$\AA, respectively.  Details on this fitting approach are provided in \citet{Glikman18b} 

The remaining sources that had both an optical and near-infrared spectrum were combined in one of two ways. Objects that had overlapping optical and near-infrared spectral regions with sufficiently high signal-to-noise were scaled to match, and the combined spectrum was fitted to a reddened QSO template. 
In other cases, where the spectral scaling was either impossible due to no overlap between the optical and near-infrared spectrum or where the overlapping regions were so noisy that small differences in the region used for comparison yielded wildly different results, we scaled the spectra to their SDSS and 2MASS photometry, combined them into a single spectrum and fit that spectrum with a reddened QSO template. 

We note that in many of the low-redshift QSOs ($z\lesssim0.3$) the optical spectrum does not match well the reddened template, often exposing excess light from the host galaxy. We attempted to remove the galaxy as described in Section \ref{sec:gandalf}, but were then unable to reliably combine the galaxy-subtracted optical QSO spectrum with the infrared spectrum.  In one case, W2M J1250+1318 at $z=0.3$, the galaxy-subtracted optical spectrum had sufficient signal in the observed $g$-band to allow a scaling to that part of the spectrum, yielding an excellent fit to the reddened template.  However, the galaxy-subtracted spectra for the other sources were not easily scalable to the near-infrared spectrum. Therefore, since the QSO dominates at longer wavelengths, we use the near-infrared spectra alone to determine $E(B-V)$. 

We find 37 W2M QSOs with $E(B-V) > 0.25$ which we classify as red QSOs, and 3 W2M QSOs with $E(B-V) < 0.25$ which we classify as blue QSOs. This information is recorded in a column in Table \ref{tab:w2m}.

\subsection{Red QSOs in the SDSS sample} \label{sec:sdss_rq}

Red QSOs also exist among the sources with SDSS spectra and we wish to identify them and combine them with the 37 newly-identified red QSOs.  As seen in Figure \ref{fig:kx_colors}, there exist SDSS-identified {\tt QSO}s in the KX-selection box along with the newly identified W2M red QSOs. All but three of them obey the diagonal color cut (Eqn \ref{eqn:diag}) and 28 of them meet the broad-line criterion. One source, J131327.46+145338.5, whose SDSS spectrum failed the broad-line criterion is a known F2M red quasar at $z=0.584$ \citep[F2M1313+1453][]{Glikman12} where strong broad H$\alpha$ and Pa$\beta$ lines are seen in its near-infrared spectrum. The object was assigned a redshift of $z=5.513$ by SDSS based on a mis-identification of [\ion{O}{3}] $\lambda 5007$ as Ly$\alpha$, which is a common occurrence in automated redshift assignments of red QSOs \citep[see, e.g., \S 3.1 of][]{Glikman18}.  

We plot the $E(B-V)$ values for these KX-obeying sources as filled blue bars in the histogram shown in Figure \ref{fig:ebv_hist}.  After correcting for the host galaxy, many of them recover their blue colors, but 19 of them have $E(B-V) > 0.25$, meeting our red QSO definition. 
Another 35 sources have $E(B-V) > 0.25$, based on their spectral fits, but their KX colors are outside the selection criteria of Eqn \ref{eqn:kx}.  Most of these sources are at low redshift ($z<0.3$) and have $E(B-V) < 0.3$, lying in the tail of the reddening distribution. Visual inspection of their spectra confirm that the host-galaxy subtracted AGN spectrum does not resemble the AGN-dominated red systems we are pursuing and they are not included in the red QSO counts.
However, five of these sources are at higher redshifts ($z>0.4$) and their spectra are well fit by a reddened QSO template. Their colors are outside of the KX-selection box either due to strong emission lines affecting their photometry or variability between the 2MASS and SDSS epochs yielding artificially blue colors. 

We also matched the SDSS QSOs to the F2M red quasar sample of \citet{Glikman12} and identified 11 additional sources. Two of these had $E(B-V) < 0.25$ based on the SDSS spectrum alone, but when combined with their near-infrared spectrum from \citet{Glikman12}, the fit yielded $E(B-V) > 0.25$ which supersedes the SDSS-derived value.  The other nine F2M sources were already recognized as red quasars through our various methods above. 
Therefore, we add the 19 SDSS QSOs in the KX selection box plus the five $z>0.4$ SDSS QSOs with $E(B-V) > 0.25$ and two F2M quasars with SDSS spectra to our sample of 37 red QSOs, as they obey the same selection criteria, including the infrared color conditions of Eqn \ref{eqn:diag}, as the W2M sources listed in Table \ref{tab:candidates}. The final number of red QSOs that we identify in this survey is thus 63 (37+19+5+2), leaving 1049 blue QSOs. 

We note that this number of red QSOs is a lower limit, since near infrared spectroscopy can reveal broad lines in objects that show narrow emission their optical spectra (e.g., F2M1313+1453). 
We also miss red QSOs whose flux is dimmed below our $K<14.7$ mag limit when reddened with $E(B-V) > 0.25$, but whose unreddened counterparts are present in the blue QSO sample.

\section{Luminosity-Restricted sample}\label{sec:luminosity}

The host-galaxy subtraction (\S \ref{sec:gandalf}) allowed us to compute the reddening experienced by the QSOs without contamination from stellar light. We used the distribution of those reddenings (Figure \ref{fig:gandalf}, right) to define a sample of reddened QSOs with $E(B-V)>0.25$. The resultant $E(B-V)$ values can then be used compute de-reddened absolute $K$-band magnitudes, $M_K$. At the same time, since all the QSOs have a measured {\em WISE} $W4$ (22$\mu$m) magnitude -- a wavelength at which extinction is negligible -- we can check whether $W4$ absolute magnitude and de-reddened absolute $K$-band magnitude are consistent. 

\begin{figure*}
\figurenum{13}
\plottwo{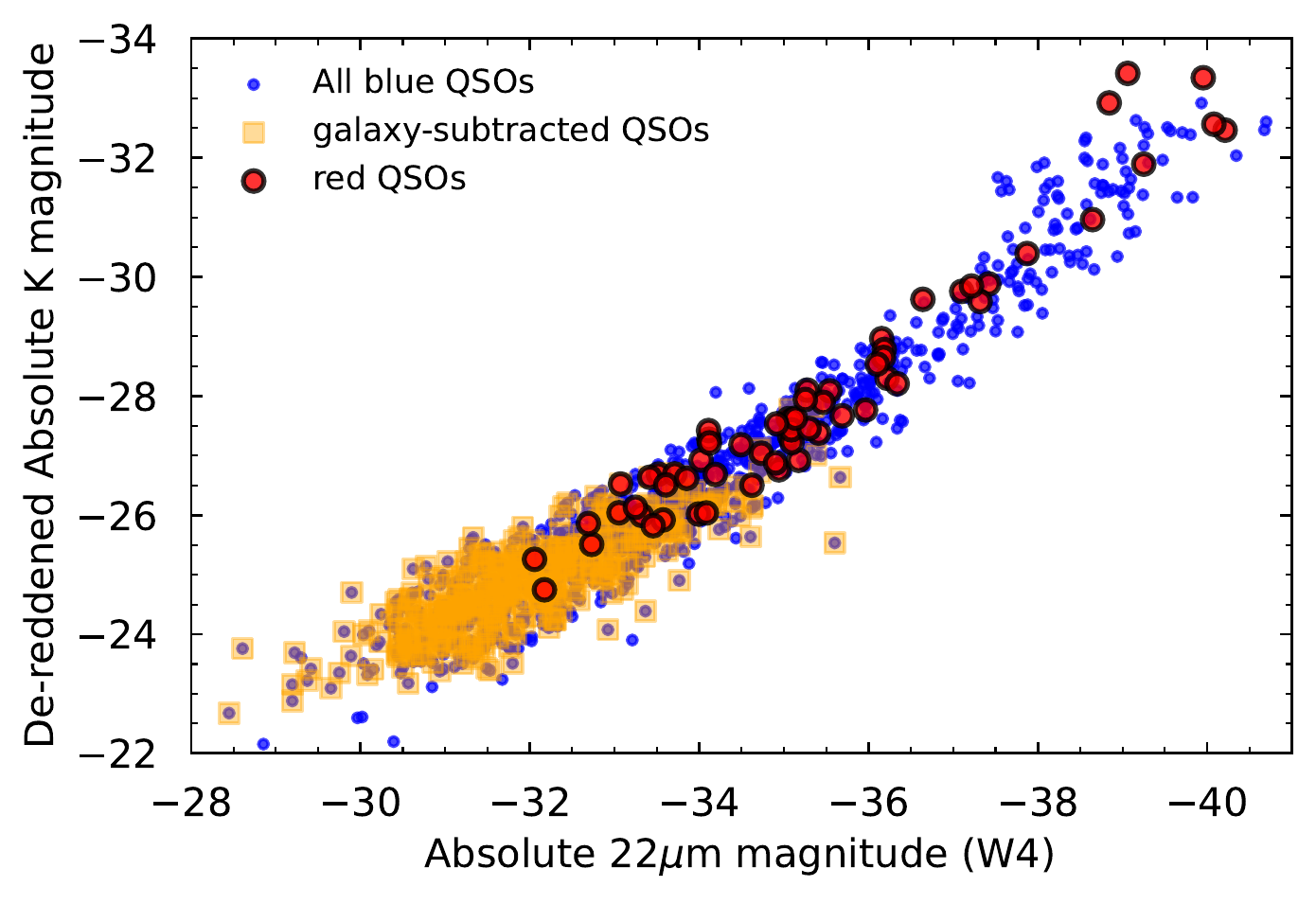}{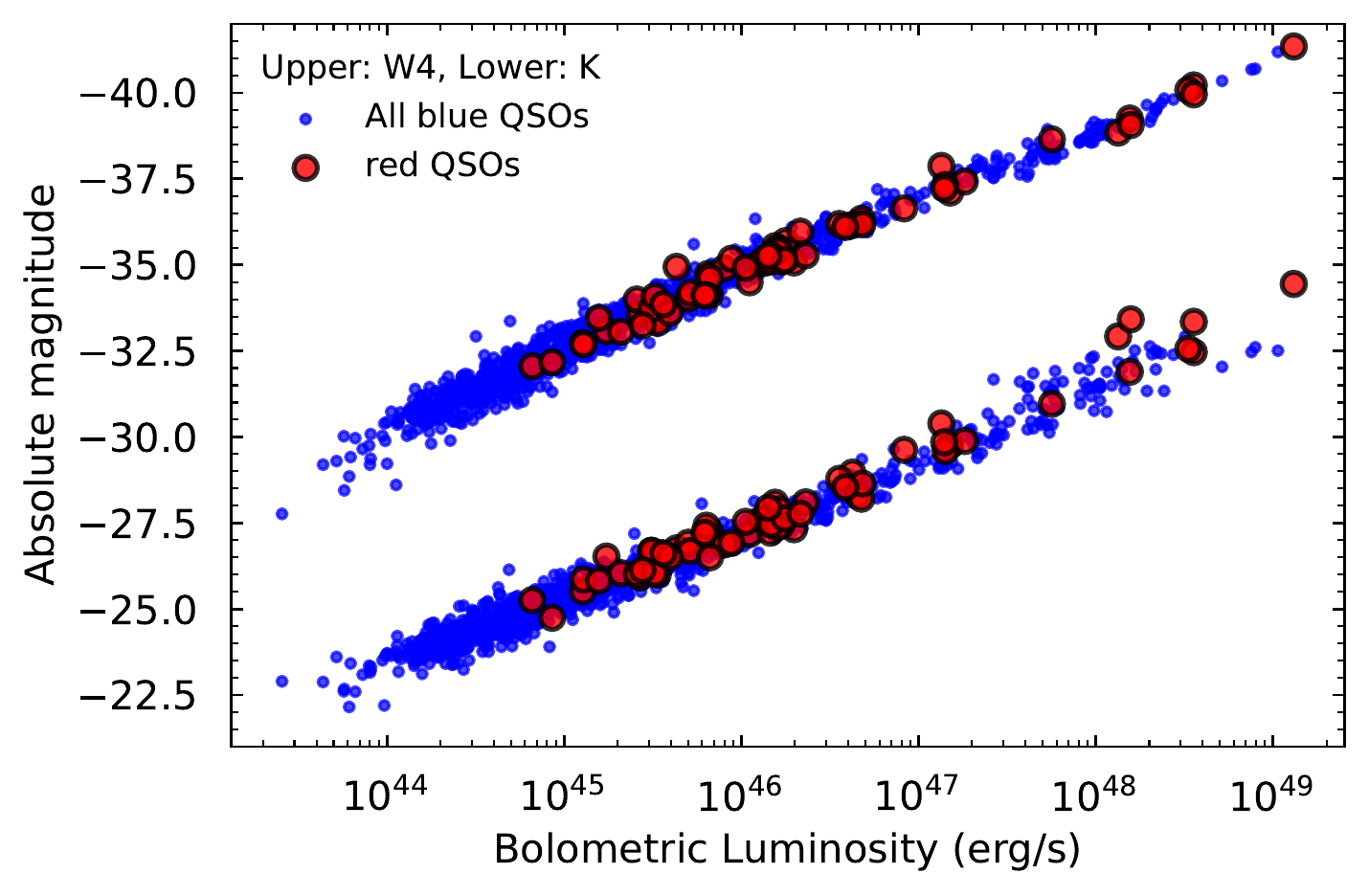}
\caption{ 
{\em Left --} De-reddened absolute $K$-band magnitude from 2MASS versus absolute 22$\mu$m ($W4$) magnitude, computed in the observed frame, for the QSO sample.
A direct relationship is seen across the luminosity range, except at the highest luminosities where the $K$-band magnitudes are slightly enhanced compared with $W4$. 
Overplotted are the sources whose host galaxy light was subtracted to determine the QSO reddnenings (\S \ref{sec:gandalf}; orange squares). The red circles are the red QSOs whose absolute $K$-band magnitude have been de-reddened. The red QSOs lie on the same relation with no apparent offset. 
{\em Right --} Absolute magnitude in $W4$ (upper locus) and de-reddened $K$ (lower locus) magnitudes versus the QSO bolometric luminosity. Both relations correlate well, although there is more scatter in the $K$-band relation at higher luminosities. 
These agreements provide reassurance that the reddening estimates in Section \ref{sec:reddening} are reliable.
} \label{fig:K_W4_Lbol}
\end{figure*}

In the left panel of Figure \ref{fig:K_W4_Lbol}, we plot the de-reddened, absolute $K$ magnitude versus absolute $W4$ (22$\mu$m) magnitude, in the observed frame, with no reddening correction. The unreddened QSOs are shown with blue circles, with the QSOs that had host-galaxy subtraction overplotted with orange squares. 
The orange squares lie atop the blue points with no apparent offset, which provides reassurance that the reddenings derived after host galaxy removal are reliable (i.e., had we {\em correccted} their $K$-band magnitudes by the erroneous $E(B-V)$ values assumed from the poor spectral fits, the orange points would have been shifted toward higher luminosities). The red circles are the red QSOs whose $K$-magnitudes are de-reddened and they lie along the same relation as all the other QSOs, providing further assurance that the reddening analysis of our sample is reliable.  

In the right panel of Figure \ref{fig:K_W4_Lbol}, we plot the $W4$ and de-reddened $K$-band absolute magnitudes versus the bolometric luminosity, which we compute by interpolating between the infrared fluxes in the {\em WISE} bands, corrected for reddening in the red QSOs, to measure the rest-frame $6\mu$m luminosity and apply a bolometric correction factor of 7.82 using the QSO spectral energy distribution (SED) from \citet{Richards06}. The $W4$ magnitudes are plotted in the upper relation and the de-reddened $K$-band magnitudes are shown in the lower relation. Both relations correlate well, though there is some scatter in the high luminosity $K$-band measurements. 

Although our use of {\em WISE} colors to select QSOs avoids the color bias experienced by optically selected QSO samples that miss most reddened QSOs, the imposition of the $K<14.7$ mag limit introduces a luminosity bias because, even at 2.2$\mu$m, moderately reddened QSOs will be dimmed below the flux limit and therefore missed unless they are intrinsically more luminous. 
The bias has a strong redshift dependence because, at higher redshifts, 2.2$\mu$m represents rest-frame optical emission, which is more sensitive to the effects of dust-extinction. 
Therefore, to conduct a valid comparison between blue and red populations, we must compare objects with similar intrinsic absolute magnitude thresholds at all redshifts. 

\begin{figure}
\figurenum{14}
\plotone{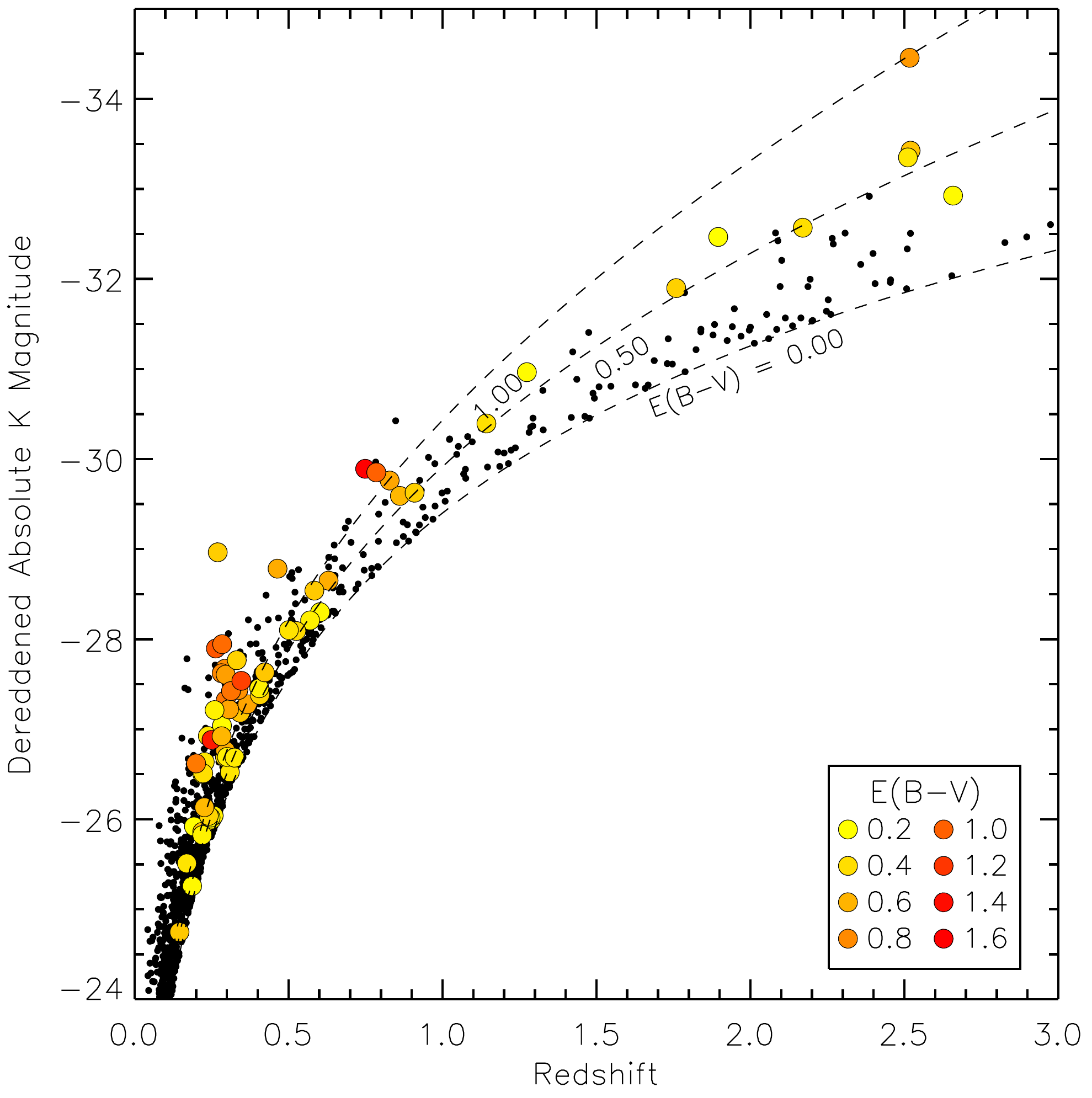}
\caption{ 
Absolute $K$-band magnitude from 2MASS, where the red QSOs have been de-reddened using $E(B-V)$ determined in \S \ref{sec:reddening} and color coded to reflect the original value, defined in the legend. The dashed lines indicate the survey limit of $K < 14.7$, and for increasing amounts of extinction. The small dots are the blue QSOs.
} \label{fig:lum_z}
\end{figure}

We plot the absolute $K$-band magnitudes for the sample in Figure \ref{fig:lum_z}, where the red QSOs are de-reddened and colored by their $E(B-V)$ values versus redshift. The dashed lines indicate the flux limit of $K<14.7$ mag with no reddening and with increasing amounts of extinction. Here, the luminosity bias becomes obvious with an absence of red QSOs near the flux limit especially toward increasing redshifts.   
This trend is well-explained by the fact that higher-redshift sources with even moderate amounts of reddening must be more luminous in order to pass the selection criteria.  At lower redshifts, the effect of reddening is weaker and we identify heavily reddened sources with intrinsic luminosities consistent with their blue QSO counterparts. 

A similar analysis on the F2M red quasars revealed that they are more luminous than their unobscured counterparts, when corrected for extinction \citep{Glikman07,Glikman12}. In the W2M sample, we see that at lower redshifts ($z<1$) the red and blue QSOs occupy similar luminosities, while at higher redshifts $z>1$ there appear to be more high luminosity red QSOs compared with blue QSOs. 

\begin{figure}
\figurenum{15}
\plotone{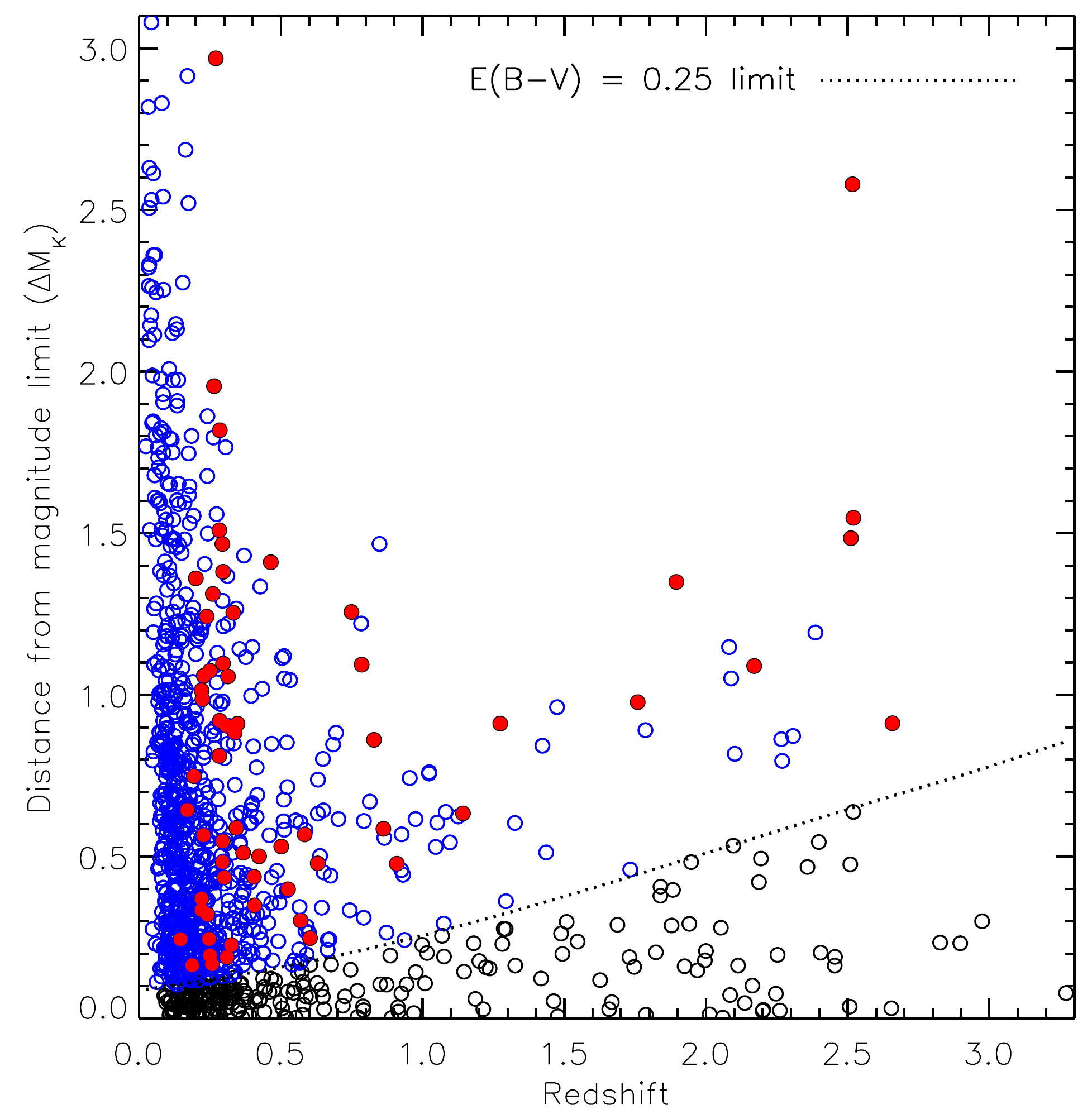}
\caption{The magnitude difference between the $K=14.7$ magnitude limit of the W2M survey minus the de-reddened absolute $K$ magnitude for red and blue QSOs showing a luminosity bias imposed by the flux limit on reddened QSOs. De-reddened red QSOs are plotted with red filled circles. Blue QSOs are shown with open symbols. The dotted black line indicates the magnitude limit when $E(B-V)= 0.25$ such that blue QSOs above the line (blue circles) have comparable absolute magnitudes to red QSOs.}\label{fig:kdiff}
\end{figure}

In order to define a blue QSO subsample that has a similar intrinsic luminosity limit, we plot in Figure \ref{fig:kdiff} the difference between the flux limit (dashed line labeled $E(B-V) = 0$ in Figure \ref{fig:lum_z}) and the de-reddened $K$-band magnitude.  Open circles are blue QSOs while filled red circles are red QSOs.  
We also plot the flux limit for sources with $E(B-V)=0.25$ with a dotted line and use this as a cut to separate out the lower luminosity blue QSOs that have no red QSO counterparts.
Imposing this luminosity cut leaves 798 blue QSOs which we plot as blue open circles in Figure \ref{fig:kdiff} (for a total of 861 QSOs, when including the 63 red QSOs). We use this sample to compare the fraction of red and blue QSOs and their radio properties in the sections that follow. 
We list in the last three columns of Table \ref{tab:w2m} the de-reddened absolute $K$-band magnitude, $M_K$, the bolometric luminosity, $L_{\rm bol}$, and a boolean indication of whether the object is part of the luminosity restricted subsample.  

We note that two of the most luminous infrared sources seen at $z\simeq 2.5$ are both gravitationally lensed QSOs of the rare quad variety.  The blue QSO is the well-studied Cloverleaf Quasar H1413+117 \citep{Kayser90} while the red QSO was newly-discovered in the current survey and is fully analyzed in (Glikman et al. 2018b).  
We exclude these sources from the statistical analyses that follow.

\begin{figure*}
\epsscale{1}
\figurenum{16}
\plottwo{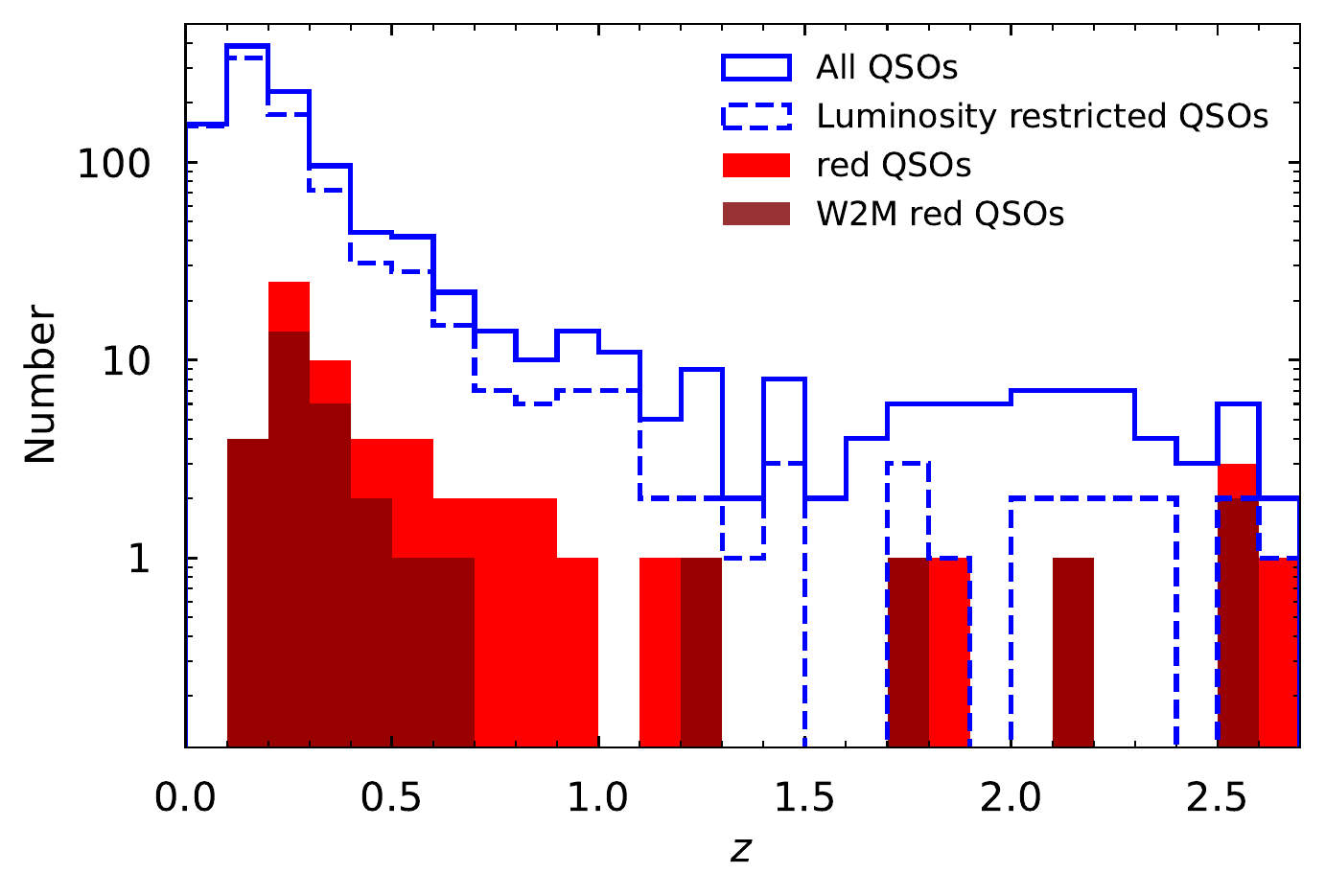}{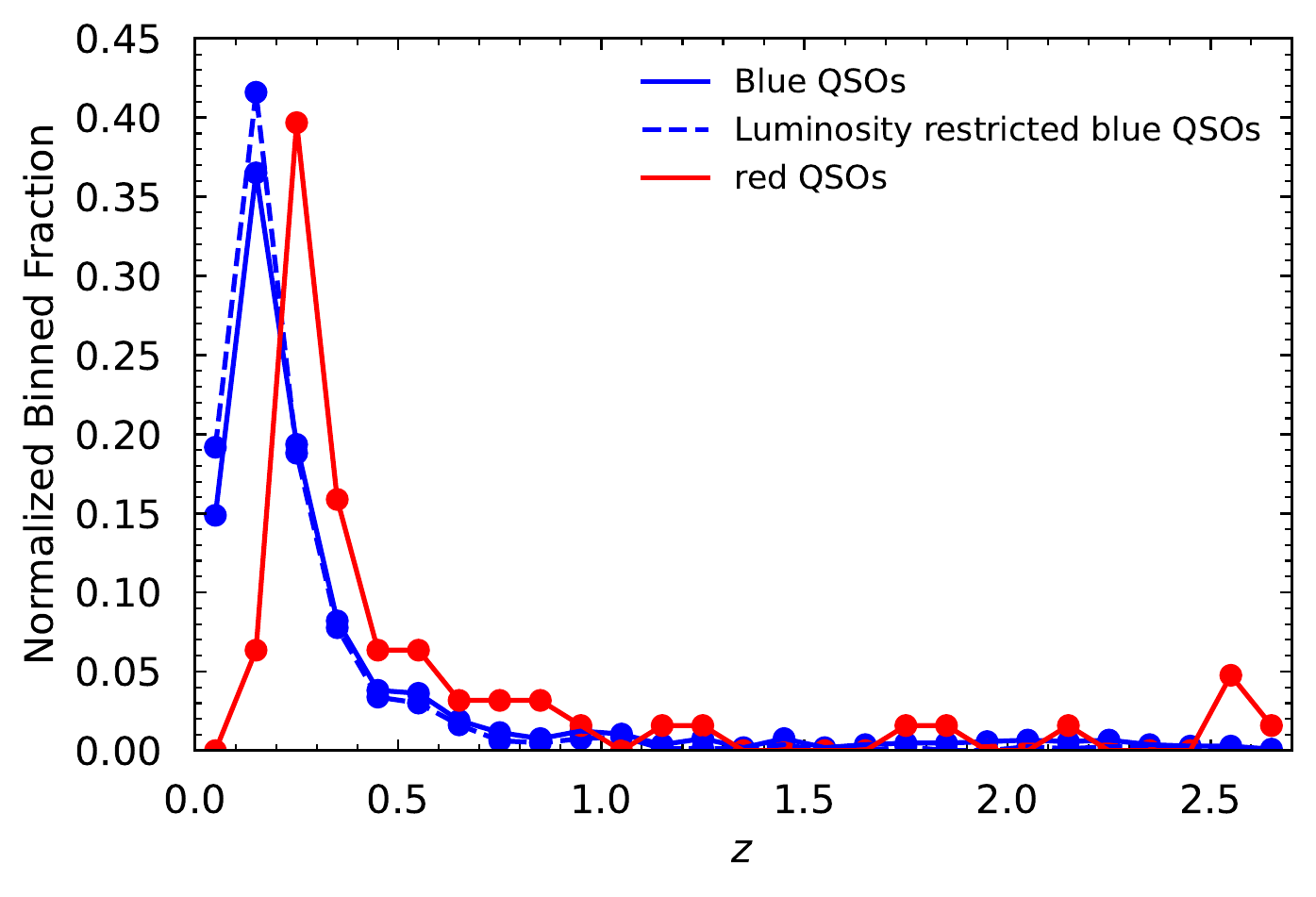}
\caption{{\em Left --} QSO redshift distribution plotted on a logarithmic scale. The solid blue line represents the full set of 1112 QSOs, while the dashed blue line shows the 861 luminosity-restricted sources. The filled red histogram overplots the 63 red QSOs while the dark-red shading represents the 37 newly identified W2M red QSOs (\S \ref{sec:ebv_w2m}). 
{\em Right --} Redshift distribution of the QSO subsamples normalized by the total number of objects in each sample, plotted on a linear scale. Here, the solid and dashed blue lines are the blue QSOs and luminosity-restricted blue QSOs, respectively. The red QSOs are shown in red. The redshift distributions of the blue QSOs are similar for the full and luminosity-restricted samples, while the red QSOs are shifted such that they have a higher relative incidence at higher redshifts. 
}\label{fig:zdist}
\end{figure*}

Figure \ref{fig:zdist} shows the redshift distribution of the Type 1 SDSS QSOs (in blue) along with the newly added W2M red QSOs (in red). The left panel shows a histogram of all sources on a logarithmic scale to better view both the blue and red populations. On the right we normalize the blue and red histograms by the total number of QSOs in each subsample so that their redshift distributions could be better compared. 
There is a dearth of red QSOs at low redshifts ($z<0.1$). A similar observation was made in \citet{Glikman18} for the pilot sample of luminous infrared selected red and blue QSOs in Stripe 82 where it was shown that Type 2 (narrow line) QSOs dominated the obscured QSO population at low redshifts, while the numbers of red QSOs increased at higher redshifts. 
This behavior is also seen in X-ray selected red QSOs, suggesting an evolutionary explanation \citep{LaMassa17}. 
On the other hand, The absence of $z<0.1$ red QSOs may be a selection effect, as these will be lower luminosity AGN such that, when further reddened, will have their mid-infrared fluxes contaminated by host galaxy light, potentially shifting them out of the selection box.
With this in mind, we also consider comparisons between red and blue QSOs restricted to $z>0.1$, which amounts to 645 blue sources (for a total of 708 QSOs, when including the 63 red QSOs).
Although the red QSO sample is otherwise predominantly low redshift ($z\lesssim 1.0$), which is largely due to the shallow $K$-band flux limit imposed on our selection, higher-redshift red QSOs appear to be more represented among the red QSOs than in the blue population. This too could point to an evolutionary explanation. 

\begin{figure*}
\figurenum{17}
\plottwo{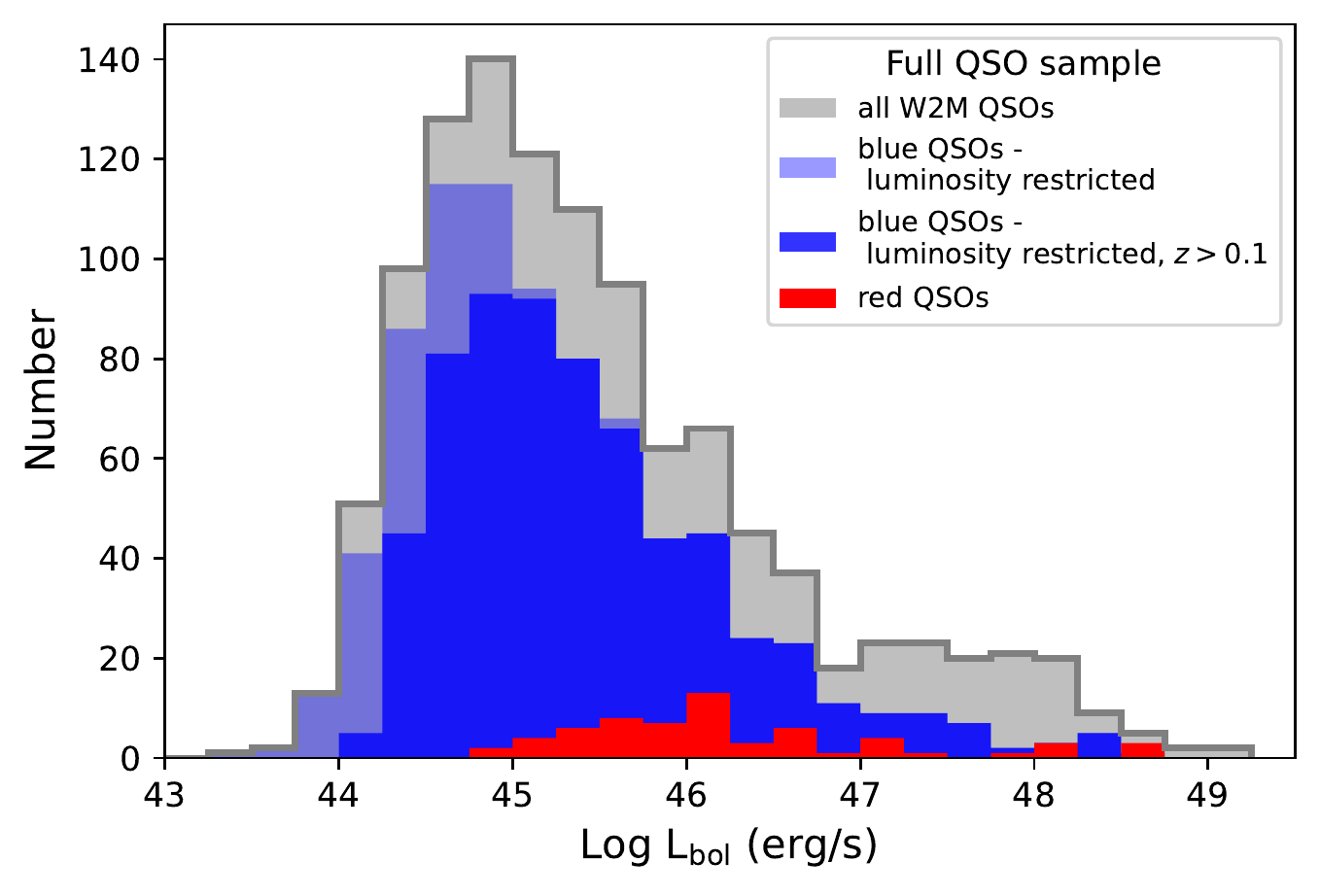}{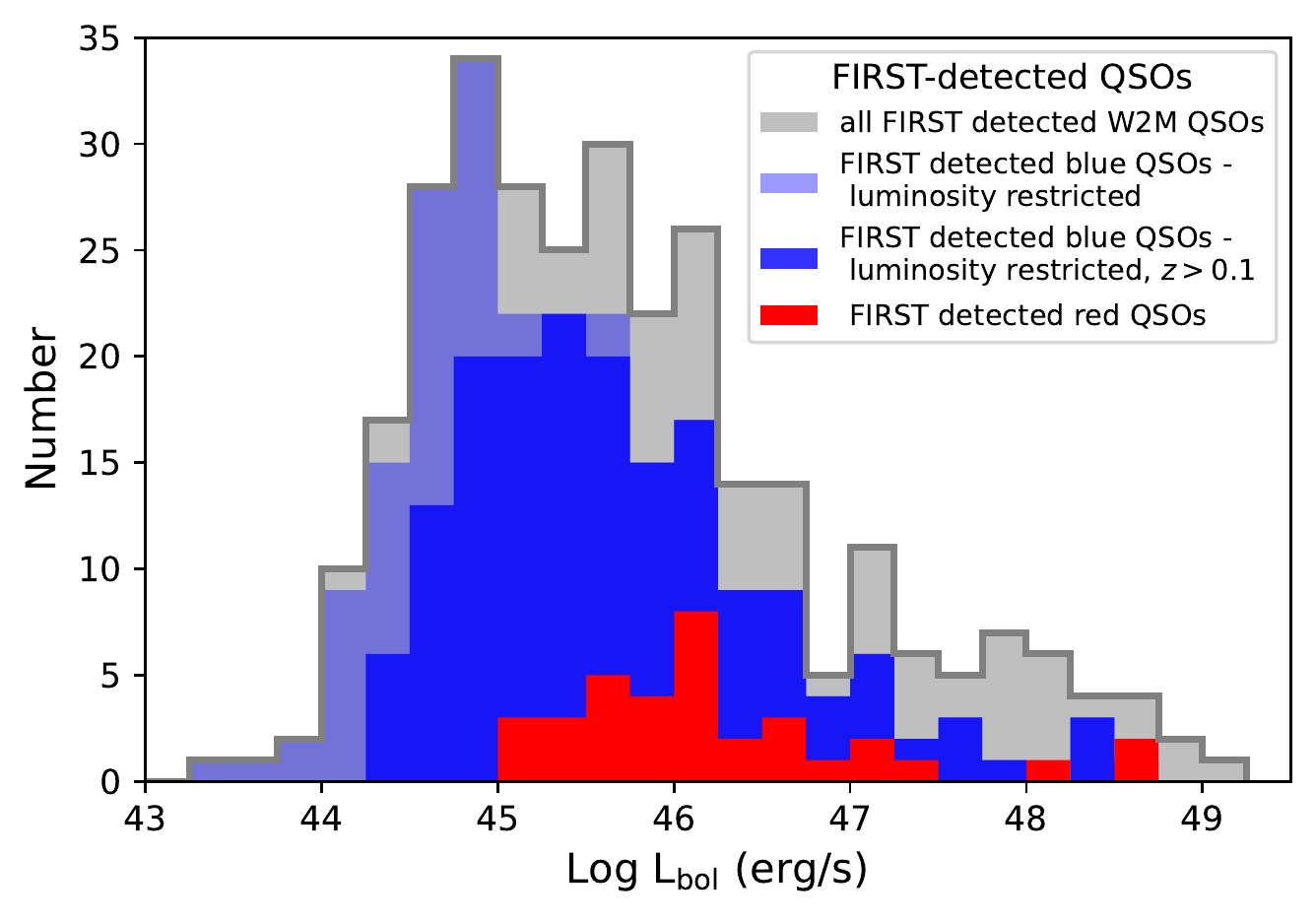}
\caption{Bolometric luminosity distribution of W2M QSOs. {\em Left --} The full QSO sample. {\em Right -- } The FIRST-detected subsample.  The gray filled histogram shows all W2M sources. The light blue histogram represents the luminosity restricted subsample of blue QSOs, and the dark blue shows blue QSOs with $z>0.1$.  The red histogram represents the red QSOs.  
} \label{fig:lum_hist}
\end{figure*}

Figure \ref{fig:lum_hist} shows a histogram of the bolometric luminosities for the W2M QSOs, divided into various subsamples. The left panel shows all the the W2M QSOs, while the right panel shows only the FIRST-detected sources. In both panels, the gray shaded histogram represents all the QSOs within that category, and reflects the full luminosity range accessible to our selection criteria. The blue histogram shows the full luminosity restricted blue QSO sample, and the dark blue shaded subset represents only QSOs with $z>0.1$. 
We use these luminosity-restricted subsamples to compare red and blue QSOs in the sections that follow.
However, we caution that because we are unable to create perfectly luminosity-matched samples, owing to the fact that we do not know how many red QSOs we are missing as they drop below our selection threshold, some of our results could be explained by differences in the luminosity functions of the two samples.

We note that this luminosity restricting effort is the minimum needed to establish comparable populations. The blue QSOs excluded by this process removes objects that would fall below the flux limit if reddened by $E(B-V) = 0.25$. However, the red QSO sample is still incomplete in ways that cannot be easily corrected without knowing the distribution of $E(B-V)$ as a function of luminosity and redshift. The lower luminosity bins of the red QSO histogram are therefore incomplete and those sources can only be recovered with a deeper survey. 

\section{Results}

We have constructed a carefully-selected, luminosity-restricted sample of 63 red and 798 blue Type 1 QSOs based on their mid and near-infrared properties. All of these sources are spectroscopically confirmed either through publicly available spectra from SDSS or from supplementary spectroscopy. With this sample of QSOs in hand, we study their radio properties and demographics in the sections below.

\subsection{Radio Properties}\label{sec:radio}

A recent study by \citet{Klindt19} investigated the fraction of FIRST-detections among blue and red QSOs in SDSS, defined according to how their observed $g-i$ colors compared to the median QSO $g-i$ at a given redshift \citep[this is effectively the `relative color' defined in][]{Richards03} and found that red QSOs have a FIRST-detection fraction that is $\sim 3$ times higher than blue QSOs, across redshifts.
A follow-up study by \citet{Fawcett20} used stacking of the radio images of red and blue QSOs, identified in a similar way but that lacked a radio detection, and found that the integrated flux density of the median red QSO is 30\% higher than the median blue QSO's integrated flux density. 
 
The SDSS spectroscopic survey is not sensitive to heavily reddened QSOs, therefore the sample defined as `red' by \citet{Klindt19} is dominated by objects with $E(B-V)\sim 0.05 - 0.2$ and contains very few sources with $E(B-V) > 0.25$.  
With the W2M sample, we have taken a more conservative approach in separating reddened QSOs from blue QSOs, having shown that most QSOs have a natural distribution of spectral slopes that, because of their power-law shape, can mimic the presence of dust (or `negative dust' for very blue objects) when a template QSO spectrum is fitted to a host-galaxy-subtracted AGN spectrum. 

If the differences in radio properties seen in \citet{Klindt19} and \citet{Fawcett20} are best-explained by dust-reddened QSOs being fundamentally distinct from `normal' QSOs then those differences should be more pronounced when comparing radio properties between blue and red QSOs in the W2M sample. 

\subsubsection{radio detected fraction}\label{sec:first_frac}

We matched the luminosity-restricted W2M sample to two radio surveys that overlap our fields. 
There were 249 matches within 1\farcs5 to the FIRST survey which reaches a 5-$\sigma$ sensitivity of 1 mJy at 1.4 GHz and has an angular resolution of 5\arcsec. 
We found 186 matches to VLASS, which reaches a median rms of 120 $\mu$Jy at $2-4$ GHz with a spatial resolution of 2\farcs5, using the catalog constructed by \citet{Gordon21}.

\begin{figure*}
\figurenum{18}
\plottwo{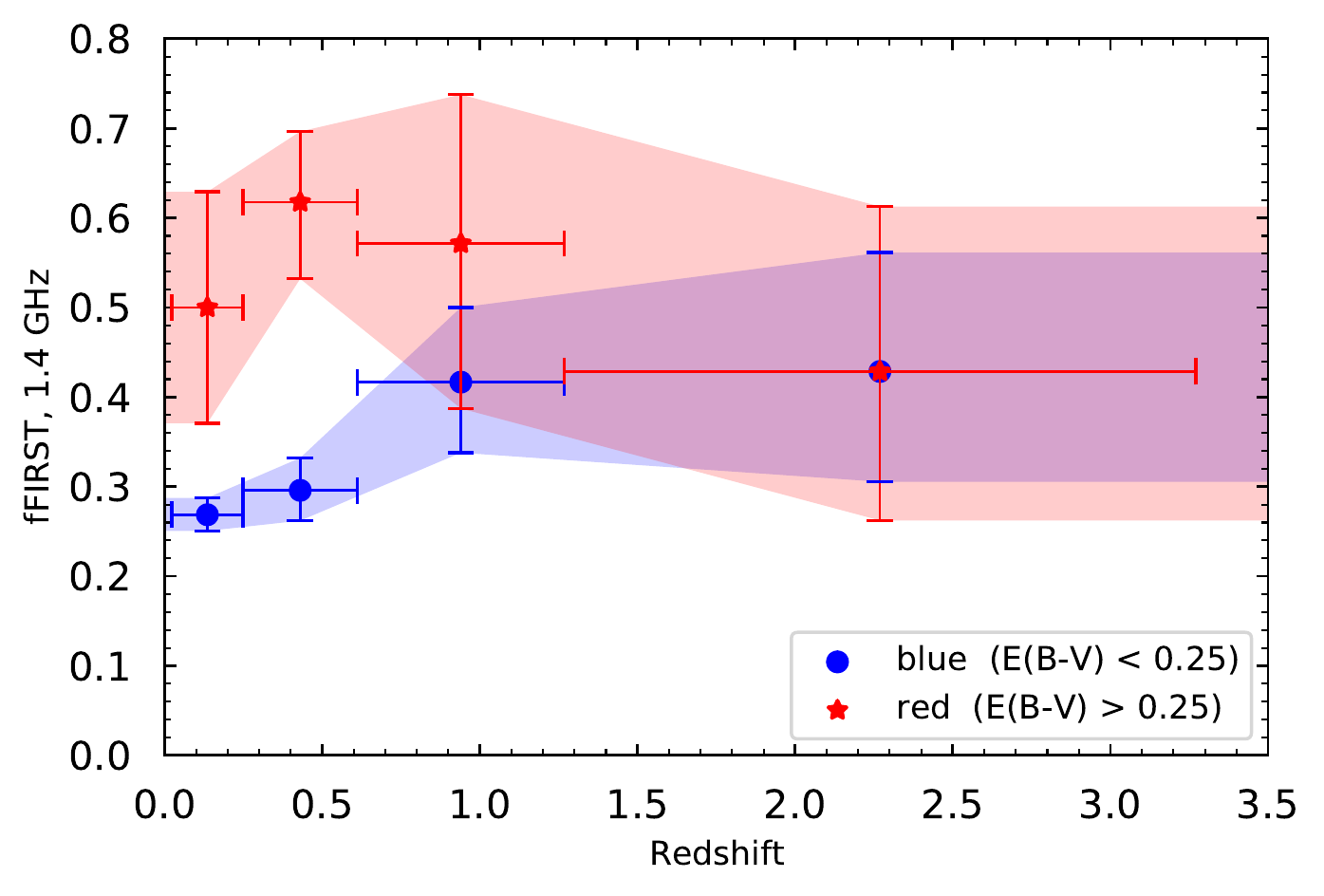}{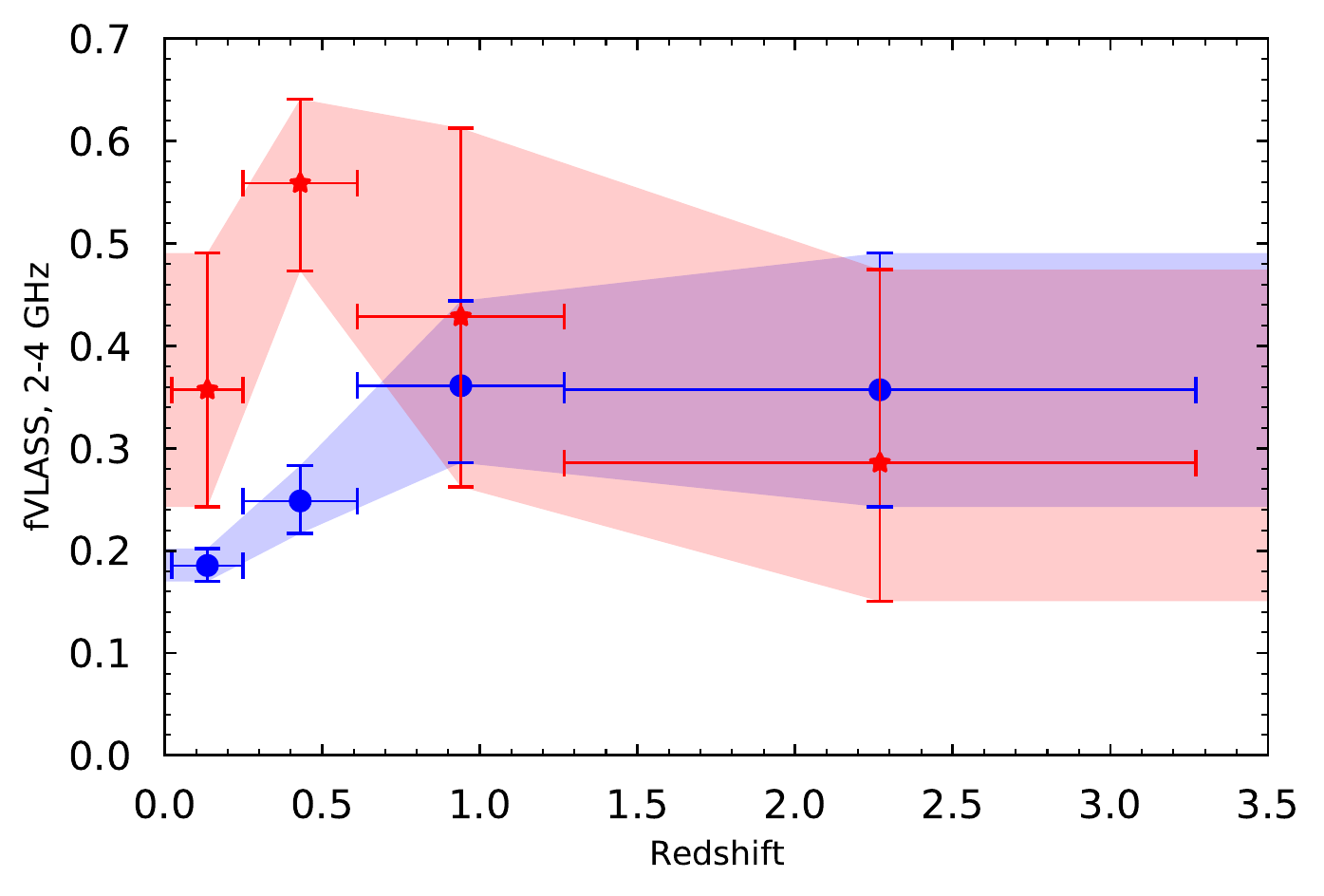}
\caption{The fraction of FIRST-detected ({\em left}) and VLASS-detected ({\em right}) QSOs as a function of redshift for the red QSOs (red symbols) and  blue QSOs (dark blue symbols).
Each redshift bin spans 2.85 Gyr in lookback time. 
At low redshifts, the red QSOs have a significantly higher fraction compared to all the sources we deem unreddened ($E(B-V) < 0.25$). However, at higher redshifts, the radio fraction of red QSOs declines and cannot be distinguished from the fraction of blue QSOs due to the small sample sizes in those bins. 
These trends are similar in both VLASS and FIRST. 
The error bars here are computed using binomial proportion confidence interval using a \citet{Wilson27} interval for small number counts.
} \label{fig:first_frac}
\end{figure*}

Figure \ref{fig:first_frac}, left, shows the fraction of FIRST-detected sources among the red and blue QSO subsets as a function of redshift.  
We divided the redshift range of our sample into four bins each representing 2.85 Gyrs in lookback time corresponding to redshift limits of $z = 0.248, 0.612, 1.267, 3.271$. 
We confirm the finding from \citet{Klindt19} that red QSOs have a significantly higher fraction of FIRST-detections and that the fraction increases toward lower redshifts. 
However, we find a higher overall fraction for both red and blue QSOs;  \citet{Klindt19} find $\sim 7\%$ and $\sim 17\%$ for blue and red QSOs, compared to $\sim 28\%$ (226/798) and $\sim 56\%$ (35/62) for blue and red QSOs in the W2M sample, respectively. 
The right panel shows the same for VLASS, largely corroborating the FIRST behavior.
We attribute this overall increase in the detection fraction to be due to the different flux limits of the two samples, as it is known that more luminous QSOs in general have a higher radio-loudness fraction \citep{Lacy01}.  
While the red QSOs have a higher detection fraction in the lower redshift bins, at higher redshifts the difference between the red and blue fractions cannot be distinguished due to the small numbers of sources in those bins. 
The breakdown of FIRST- and VLASS-detected sources for the blue and red subsamples is provided in Table \ref{tab:radio_frac}.  

\begin{table*}[!htb]
	\tablenum{4}
	\caption{Radio detection fraction in redshift bins for blue and red QSOs, as shown in Figure \ref{fig:first_frac}. }
	\begin{ruledtabular}
		\begin{tabular}{l c c c c c}
			Sample & $N_{\rm all}$ & $N (0.023 \le z < 0.248) $ & $N (0.248 \le z < 0.612)$ & $N (0.612 \le z < 1.267)$ & $N (1.267 \le z < 3.271)$ \\
			\hline 
			Parent QSO & 858\tablenotemark{a} & 577 & 185 & 37 & 79 \\
			\hline
			\multicolumn{6}{c}{Color-selected QSO samples} \\
			\hline
			blue QSOs & 798 & 577 & 169 & 36 & 14 \\
			red QSOs & 62 & 14 & 34 & 7 & 7\\ 

			FIRST blue & 226 & 155 & 50 & 15 & 6 \\
			FIRST red & 35 & 7 & 21 & 4 & 3\\

			VLASS blue & 167 & 107 & 42 & 13 & 5 \\
			VLASS red & 29 & 5 & 19 & 3 & 2\\
			\hline
			& \multicolumn{5}{c}{Radio-detected fraction (\%)}\\
			\hline
			& & \multicolumn{2}{c}{FIRST} & \multicolumn{2}{c}{VLASS} \\
			Redshift  & & blue  &  red  &  blue  &  red  \\
			\hline
			$0.023 \le z < 0.248 $ & & $26.9^{+1.9}_{-1.8}$     & $50.0\pm12.9$              & $18.5^{+1.7}_{-1.6}$    & $35.7^{+13.4}_{-11.5}$ \\
			$0.248 \le z < 0.612 $ & & $29.6^{+3.6}_{-3.4}$     & $61.7^{+7.9}_{-8.6}$     & $24.8^{+3.5}_{-3.2}$    & $55.9^{+8.2}_{-8.6}$ \\
			$0.612 \le z < 1.267 $ & & $41.7^{+8.3}_{-7.9}$     & $57.1^{+16.6}_{-18.4}$ & $36.1^{+8.3}_{-7.5}$    & $42.9^{+18.4}_{-16.6}$ \\
			$1.267 \le z < 3.271 $ & & $42.9^{+13.3}_{-12.3}$ & $42.9^{+18.4}_{-16.6}$ & $35.7^{+13.4}_{-11.5}$ & $28.6^{+18.9}_{-13.5}$ \\	
		\end{tabular}
	\end{ruledtabular}\label{tab:radio_frac}
	\tablenotetext{a}{We remove one gravitational lens and two sources lacking coverage in FIRST from the total sample of 861.}
\end{table*}

\subsubsection{radio morphologies}\label{sec:radio_morph}

A major result from \citet{Klindt19} was that when categorized by their FIRST morphology among FIRST-detected sources, a much larger fraction of red QSOs had a compact appearance compared to blue QSOs (7\% vs 2\% of the entire subsample, respectively) while their fractions of extended sources are approximately the same. 
A similar finding was reported in \citet{Fawcett20}.

\begin{figure*}
\figurenum{19}
\plottwo{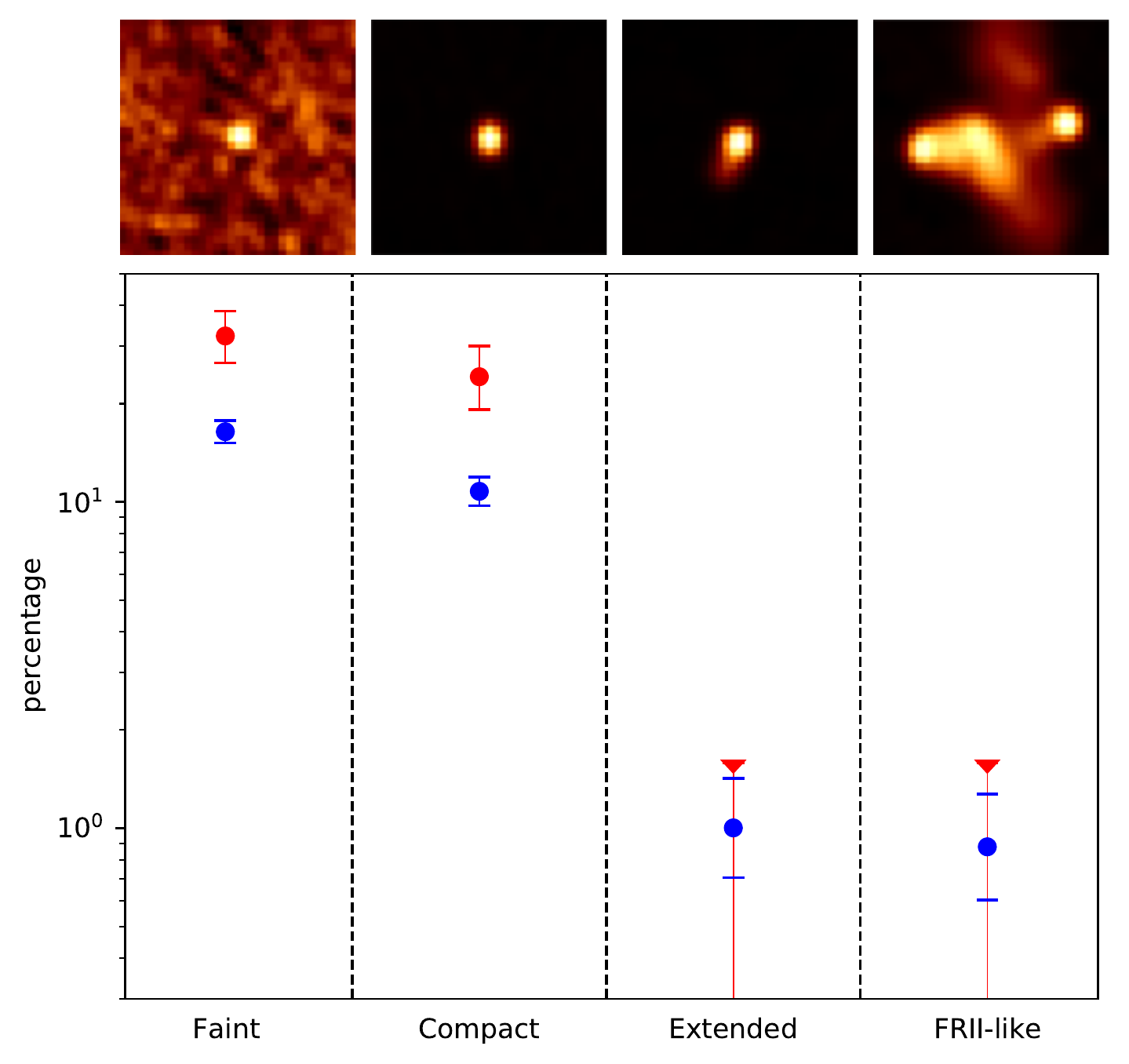}{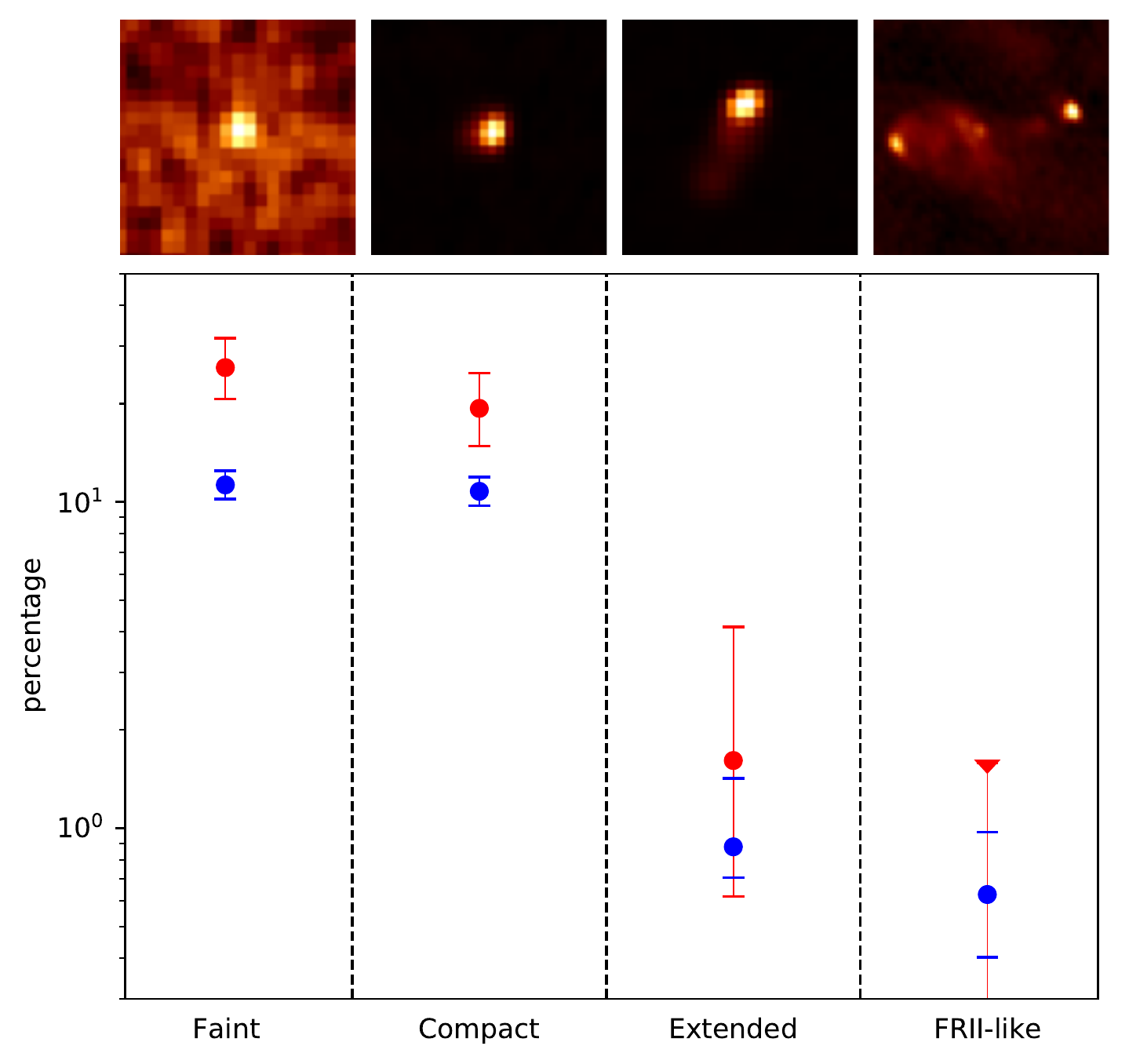}
\caption{The percentage of red and blue QSO samples with different radio morphologies, as defined by \citet{Klindt19}. {\em Left --}  FIRST-detected QSOs in our survey. {\em Right --} VLASS-detected QSOs.  The four morphological categories are illustrated by examples at the top of the figure, where the Faint category is distinguished from the Compact category by $F_{\rm pk, FIRST} < 3$ mJy. The fraction of red QSOs in the `Faint' and `Compact' categories is approximately twice as high as for the blue QSOs in the FIRST-detected sample and more than ten times higher in the VLASS detected sample. Due to the overall small number of red QSOs (62) we detect no (or few) `Extended', or `FRII-like' sources.}
\label{fig:radio_morph}
\end{figure*}

We examined the FIRST morphologies of our red and blue samples following the same morphological classifications that \citet{Klindt19} employed (Faint, Compact, Extended, FRII-like). The left panel of Figure \ref{fig:radio_morph} shows the fraction of red and blue QSOs with FIRST detections in the different morphological categories. 
We see a similar trend with a higher detection fraction of faint compact red QSOs compared to blue QSOs (32\% and 16\%, respectively).  

We note that \citet{Klindt19} find that red QSOs have a compact morphology fraction that is $\sim 3.5$ times higher than for blue QSOs, whereas the W2M red QSOs have a compact morphology fraction that is only $\sim 2$ times higher than for blue QSOs. 
This may be due to the fact that the red QSOs in our sample are significantly redder, and that some of the sources that we define as blue, with $E(B-V)\sim 0.05 - 0.2$, would be categorized as red in the \citet{Klindt19} sample. 
In addition, there may be luminosity-based effects as there are more radio-loud blue QSOs overall in the W2M sample. 

We repeat this exercise in the right panel of Figure \ref{fig:radio_morph} with the VLASS images, which sample the sources at a higher frequency than FIRST.  With both surveys, we separate the `Faint' from `Compact' categories using a flux cutoff of $S_{\rm pk} < 3$ mJy to maintain consistency with \citet{Klindt19}. 
Table \ref{tab:morphology} lists the percentages for each morphology class in the red and blue QSO samples for FIRST and VLASS.




\begin{deluxetable}{ccccc}




\tablecaption{Percentage of QSOs by morphology  \label{tab:morphology}}

\tablenum{5}

\tablehead{\colhead{Sample} & \colhead{Faint} & \colhead{Compact} & \colhead{Extended} & \colhead{FR II-like} \\ 
\colhead{} & \colhead{(\%)} & \colhead{(\%)} & \colhead{(\%)} & \colhead{(\%)} } 

\startdata
\cutinhead{FIRST}
blue & $16.4_{-1.3}^{+1.4}$ & $10.8_{-1.0}^{+1.1}$ & $1.0_{-0.3}^{+0.4}$ & $0.9_{-0.3}^{+0.4}$ \\
red & $32.3_{-5.6}^{+6.1}$ & $24.2_{-5.0}^{+5.8}$ & $<1.6$ & $<1.6$ \\
\cutinhead{VLASS}
blue & $11.3_{-1.1}^{+1.2}$ & $10.8_{-1.0}^{+1.1}$ & $0.9_{-0.3}^{+0.4}$ & $0.6_{-0.2}^{+0.3}$ \\
red & $25.8_{-5.1}^{+5.9}$ & $19.4_{-4.5}^{+5.5}$ & $1.6_{-1.0}^{+2.5}$ & $<1.6$ \\
\enddata




\end{deluxetable}

\subsubsection{median radio properties via stacking}\label{sec:rad_stack}

The majority of the QSOs in our sample are undetected in neither FIRST nor VLASS. We therefore employ the method of image stacking to study the median radio properties of the red and blue QSOs in our sample following \citet{White07}. Image stacking involves creating a three-dimensional image cube made up of image cutouts centered on the individual QSO positions and collapsing it onto a two-dimensional image with each pixel having the median value along the stacked axis. 
The stacking process involves adding signal from sources well within the noise while reducing the background rms.  The brightness of the resultant stacked image represents the median flux density of the input sample.  
The measured flux density of stacked images of sub-threshold FIRST sources experience a flux density bias \citep[dubbed ``snapshot bias'' in][]{White07} whereby the stacked source has a $\sim 30\%$ deficit in its measured flux.  Sources that are detected with a peak flux density above 0.75 mJy also have a deficit of 0.25 mJy beam$^{-1}$, known as the CLEAN bias. 
As a result, \citet{White07} carefully calibrated the peak flux densities derived from stacking and established a bias correction formula: $S_{p,{\rm corr}} = {\rm min}(1.40S_p,~S_p +0.25~{\rm mJy})$. 

When performing the stacking with the VLASS data, we include only a single epoch observation for each QSO so that each source has a uniform weighting. We use the first epoch data if available and only use a second epoch observation if not. 
Because the broad frequency range of the VLASS data ($2-4$ GHz) results in minimal side-lobes, which are the target of the CLEAN algorithm, they do not suffer from CLEAN bias the way FIRST sources do \citep[][]{Rau16}.  
To investigate whether a ``snapshot'' bias correction is needed, we stacked the positions of 12 unresolved COSMOS sources, observed with the same $2-4$ GHz band with the VLA, whose fluxes were between 0.2 and 0.5 mJy, with an average of 0.306 mJy \citep{Smolcic17}.  
 We measure a peak flux density to their VLASS image stack of 0.289$\pm$0.40 mJy, which is consistent within the errors. 
However, given that this test was only conducted on a single stack of a small number of sources, there may be a correction that is up to $\sim20\%$ of the stacked value if the uncertainty to the stacked flux is considered. 

As part of the analysis of FIRST image stacking, \citet{White07} found that redder QSOs, as parametrized by their $g-i$ relative color, have higher median radio fluxes. Compared to the SDSS composite, objects redder by 0.8 mag have radio fluxes that are $\gtrsim 3$ times higher. The median radio flux densities of QSOs bluer than the mean are flat and do not change with color.  

\begin{figure}
\figurenum{20}
\begin{center}
\includegraphics[scale=0.28]{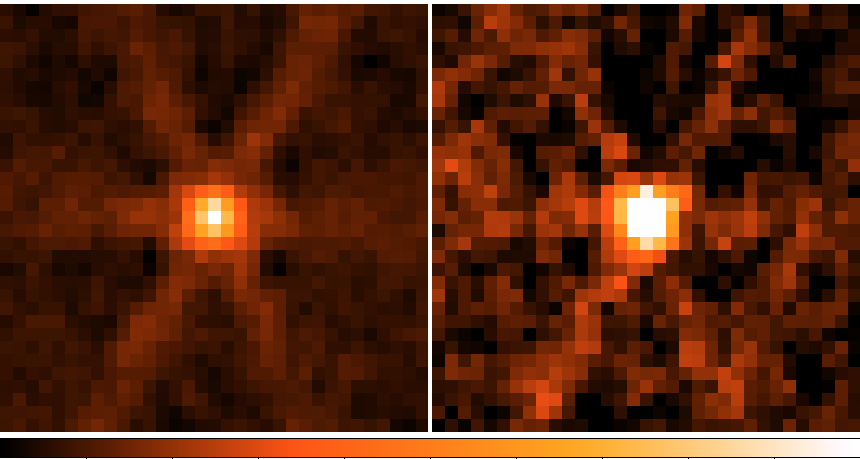}\\
\includegraphics[scale=0.28]{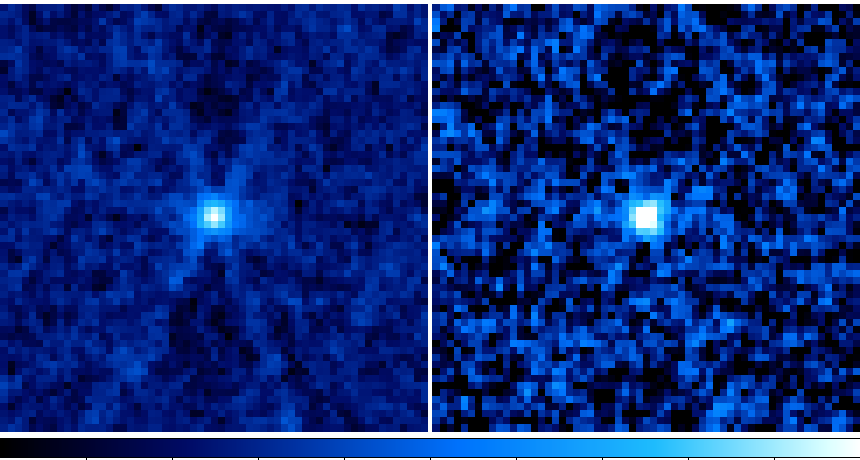}
\caption{Stacked radio images of blue (left) and red (right) QSOs in our survey. {\em Top -- } FIRST. Images are 33 pixels on a side (59.4\arcsec) and are displayed with a linear color scale. {\em Bottom -- } VLASS. Images are 61 pixels on a side (61\arcsec) and are displayed with a asinh color scale. In both cases, the brightness limits are fixed to the min/max of the blue QSO stack so that the brighter average flux density of the red QSO is apparent.}
\label{fig:red_blue_stack}
\end{center}
\end{figure}

We stacked the FIRST images of the 62 red and 796 blue QSOs\footnote{Excluding the two gravitational lenses, as noted in \S \ref{sec:luminosity}.} and examined their median radio image properties. The top row of Figure \ref{fig:red_blue_stack} shows the resultant blue (left) and red (right) median stacked images.  The scaling of the color bar is matched for both images, reflecting the result that red QSOs have a higher median flux density at 1.4 GHz.  After fitting a two dimensional Gaussian profile to the image and applying the bias correction to the peak flux density, we find that the blue QSOs have $S_{p,{\rm blue}} = 0.390\pm0.008$ mJy and red QSOs have $S_{p,{\rm red}} = 1.183\pm0.024$ mJy. This is a significant difference in median flux density, and is consistent with the results from the previous sections showing a higher incidence of FIRST detections among the red QSOs.  \citet{Fawcett20} also found a higher flux density for red QSOs but with significantly lower luminosities and dominated by sources with much smaller $E(B-V)$ values compared with the W2M red QSOs. They find that the red QSO stack is 35\% brighter than the blue QSO stack.  

Similarly, we performed image stacking using cutouts from the VLASS survey and show the stacked red and blue QSO images in the bottom row of Figure \ref{fig:red_blue_stack}.  With a median peak flux density of $S_{p,{\rm red}} = 0.698\pm0.022$ mJy, the red QSOs have a higher flux density than the blue QSOs, $S_{p,{\rm blue}} = 0.301\pm0.006$ mJy, at higher frequencies as well.

\begin{figure}
\figurenum{21}
\begin{center}
\includegraphics[scale=0.28]{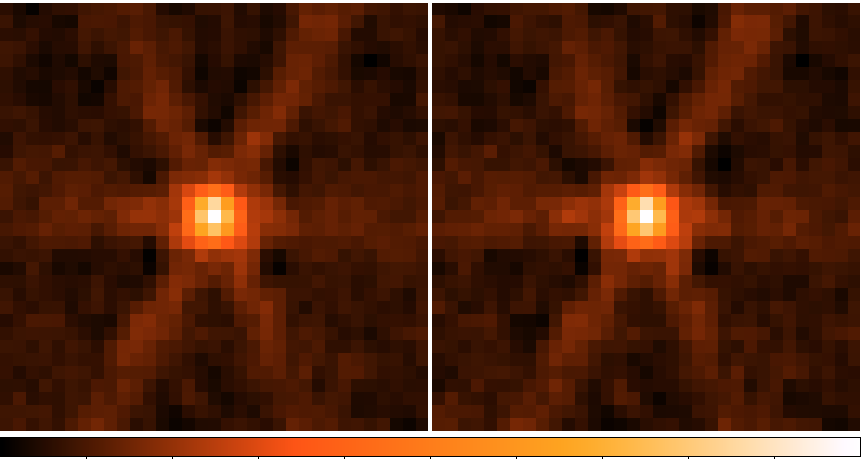}\\
\includegraphics[scale=0.28]{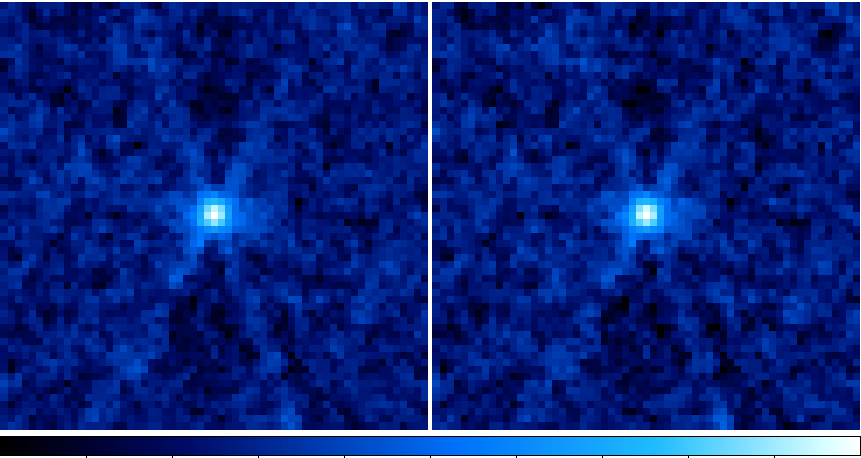}
\caption{Stacked radio images of blue QSOs obeying the box (left; Eqn. \ref{eqn:gator_sel}) and diagonal (right; Eqn. \ref{eqn:diag}) selected QSOs in our survey. {\em Top -- } FIRST. The images are 33 pixels on a side (59.4\arcsec) and are displayed with a linear color scale. {\em Bottom -- } VLASS. Images are 61 pixels on a side (61\arcsec) and are displayed with a asinh color scale. 
The stacked images are near-identical.  In both cases, the brightness limits are fixed to the min/max of the box blue QSO stack on the left.}
\label{fig:box_diag_stack}
\end{center}
\end{figure}

To test for biases among the blue QSOs in our sample, we conducted additional median stacks of several subsets. 
Given that the `box' selection contains sources with significant host galaxy emission in their optical spectra, we constructed a median stack of only QSOs that obey the `diagonal' selection to compare with the blue QSO stack in Figure \ref{fig:red_blue_stack}. Figure \ref{fig:box_diag_stack} shows the resultant stacks, FIRST in the top row and VLASS in the bottom row, with the blue QSOs from the `box' selection (Eqn. \ref{eqn:gator_sel}) shown on the left (same as in Figure \ref{fig:red_blue_stack}) and the `diagonal' selection (Eqn. \ref{eqn:diag}) shown on the right. The bias-corrected peak flux densities in FIRST are $S_{p,{\rm blue}} = 0.390\pm0.008$ mJy and $S_{p,{\rm blue,diag}}= 0.402\pm0.008$ mJy, and for VLASS are $S_{p,{\rm blue}} = 0.301\pm0.006$ mJy and $S_{p,{\rm blue,diag}}= 0.323\pm0.007$ mJy, both of which are near identical. 
This provides reassurance that we can compare the red QSOs, which are shifted toward higher $W1-W2$ colors, to blue QSOs that extend to lower $W1-W2$ colors without biasing our analysis of their radio properties.

Since all of the red QSOs are at $z>0.1$, we also stacked the luminosity-restricted blue QSOs with $z>0.1$ to ensure that the 153 low-redshift and low-luminosity blue QSOs are not skewing the median radio flux density to lower values. We find that the median radio flux densities of the $z>0.1$ blue QSOs have $S_{p,{\rm blue,highz}}= 0.373\pm0.007$ mJy in FIRST and $S_{p,{\rm blue,highz}}= 0.293\pm0.007$ mJy in VLASS which are both consistent with -- and even slightly fainter than -- the full stacked sample. 

Considering that the red QSOs are skewed toward higher luminosities we also stacked only blue QSOs with $L_{\rm bol} > 10^{45}$ erg s$^{-1}$ which overlaps the histogram of dereddened $K$-band magnitudes for the red QSOs (see Figure \ref{fig:lum_hist}) to ensure that the lower-lumniosity blue QSOs are not skewing the median radio flux densities. This subsample contains 424 such sources (considering only those that are luminosity-restricted). Here too, we find that the median radio flux densities are largely consistent with the full stacked sample. The median FIRST flux density of the high luminosity blue QSOs is $S_{p,{\rm blue,highlum}}= 0.435\pm0.009$ mJy. The median VLASS flux density is $S_{p,{\rm blue,highlum}}= 0.336\pm0.009$ mJy. Both values are slightly higher than the full blue QSO sample stack, but still significantly lower than the red QSO value.

\begin{figure}
\figurenum{22}
\begin{center}
\includegraphics[scale=0.284]{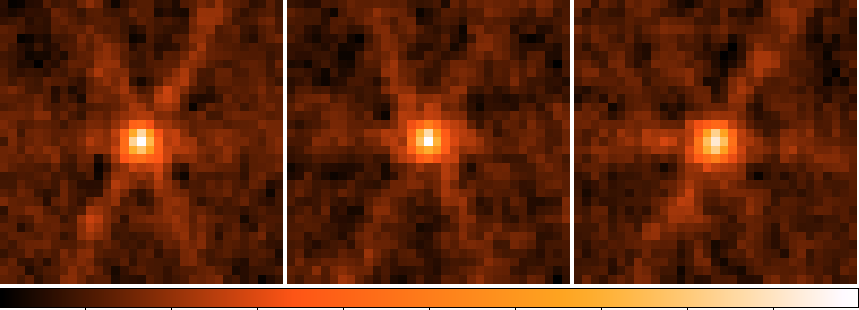}
\includegraphics[scale=0.284]{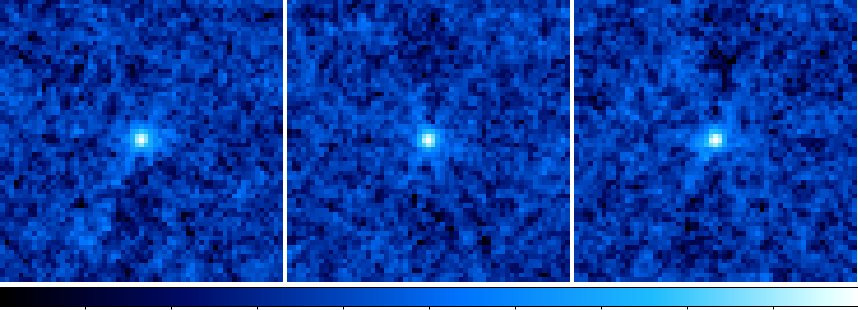}
\caption{Stacked radio images of blue QSOs broken into three bins spanning the blue side, peak, and red side of the $E(B-V)$ distribution seen in Figure \ref{fig:ebv_hist}, right, which we interpret as a proxy for the intrinsic distribution of QSO continuum slopes rather than dust reddening.  
{\em Top -- } FIRST. The images are 33 pixels on a side (59.4\arcsec) and are displayed with a linear color scale. {\em Bottom -- } VLASS.  Images are 61 pixels on a side (61\arcsec) and are displayed with a asinh color scale. 
In both cases, the brightness limits are fixed to the min/max of the blue QSO stack in the middle image.
The stacked images are very similar suggesting that unreddened QSOs do not have significant intrinsic variations in their radio properties with the shape of their optical spectrum. }
\label{fig:ebv_bin_stack}
\end{center}
\end{figure}

Finally, we broke the blue sample into three bins of $E(B-V)$ to explore whether QSOs that are not dust-reddened, but whose intrinsic continua vary to give slightly flatter and steeper slopes, might exhibit different median radio properties. To test this, we divided all 796 blue QSOs into three equal sized bins\footnote{Here, as noted in \S \ref{sec:reddening}, $E(B-V)$ is merely a proxy for the intrinsic spectral shape of the QSO continuum.}. The bins span  $E(B-V) < -0.05$ for the first bin (Bin 1), $-0.05 \le E(B-V) < 0.07$ for the second bin (Bin 2), encompassing the peak of the distribution, and $0.070 \le E(B-V)$ (Bin 3).  Figure \ref{fig:ebv_bin_stack} shows the median stacked images of these three subsets, with FIRST at the top and VLASS on the bottom.  
The bias-corrected FIRST peak flux densities for the subsets are $S_{p,{\rm Bin1}} = 0.404\pm0.012$ mJy, $S_{p,{\rm Bin2}} = 0.375\pm0.011$ mJy, $S_{p,{\rm Bin3}} = 0.398\pm0.012$ mJy, and $S_{p,{\rm Bin1}} = 0.268\pm0.011$ mJy, $S_{p,{\rm Bin2}} = 0.316\pm0.011$ mJy, $S_{p,{\rm Bin3}} = 0.314\pm0.011$ mJy  in VLASS.  
Regardless of intrinsic color, the overall the median radio flux densities are largely very similar and significantly lower than the average red QSO flux density. 

In addition, there are differences among the average morphologies of the various QSO subsets based on the widths along the x and y axes derived from the two-dimensional Gaussian fits to the blue and red stacked images. 
We find that the red QSOs are more compact than the blue QSOs, with FWHM$_{\rm red}$ = 3.1 pixels and FWHM$_{\rm blue} \sim$4.1 pixels\footnote{The size of a pixel in the FIRST images is 1\farcs 8.}.  
Similarly, in VLASS, the red quasars have FWHM$_{\rm red} = 3.4$ pixels versus FWHM$_{\rm blue} \sim$  4.0 pixels\footnote{The size of a pixel in the VLASS images is 1\arcsec.}. 
In both cases, the sources are slightly resolved (FIRST resolution is 2.8 pixels, VLASS resolution is 3 pixels). 
We note, that stacking will cause some morphological distortion by spreading out the PSF depending on where the source's true peak exists within the central image pixel.  
Nonetheless, we find consistency with the earlier findings (\S \ref{sec:radio_morph}) that, among the radio-detected sources, red quasars have more compact morphologies. 

Table \ref{tab:stacking} summarizes all the stacking results from the various sub-samples and their derived quantities. 
Our main takeaway is that red QSOs display significantly enhanced radio emission compared with the un-reddened sample, particularly when red QSOs are defined as having $E(B-V)>0.25$, beyond the normal spread of spectral shapes. 




\begin{deluxetable*}{l|ccc|ccc}




\tablecaption{Statistics of stacking results \label{tab:stacking}}

\tablenum{6}

\tablehead{\colhead{QSO } & \colhead{Num} & \colhead{$F_{pk}$} & \colhead{FWHM\tablenotemark{a}} & \colhead{Num} & \colhead{$F_{pk}$} & \colhead{FWHM\tablenotemark{a}} \\ 
\colhead{Subset} & \colhead{} & \colhead{(mJy)} & \colhead{(pixels)} & \colhead{} & \colhead{(mJy)} & \colhead{(pixels)} \\
\hline
\colhead{} & \multicolumn{3}{c}{FIRST} & \multicolumn{3}{c}{VLASS\tablenotemark{b}} } 
\startdata
Red            &   62    &  $1.183\pm0.024$  &  3.1  &   62   &  $0.698\pm0.022$  &  3.4 \\
Blue           &  796   &  $0.390\pm0.008$  &  4.1  &  790  &  $0.301\pm0.006$  &  4.1 \\
Blue Diag   &  640  &  $0.402\pm0.008$  &  4.0  &  636   &  $0.323\pm0.007$  &  3.9 \\
\hline
Blue, $z>0.1$            & 643 & $0.373\pm0.007$ & 4.1 & 639 & $0.293\pm0.007$ & 4.0 \\
Blue, $L_{\rm bol}  > 10^{45}$ & 424 & $0.435\pm0.009$ & 4.1 & 421 & $0.336\pm0.009$ & 3.9 \\
\hline
Bin 1          &  259   &  $0.404\pm0.012$  &  4.0  &  258   &  $0.268\pm0.011$  &  4.6 \\
Bin 2          &  278   &  $0.375\pm0.011$  &  4.2  &  275   &  $0.316\pm0.011$  &  3.8 \\
Bin 3          &  259   &  $0.398\pm0.012$  &  4.1  &  257   &  $0.314\pm0.011$  & 4.0  \\
\enddata

\tablenotetext{a}{A single FWHM is reported as the average FWHM along the x and y directions.}
\tablenotetext{b}{An additional $20\%$ uncertainty may be added to the VLASS stacks due to possible ``snapshot bias'' (see \S \ref{sec:rad_stack}). }
\tablecomments{The color-separated bins in the bottom three rows are defined as Bin 1: $E(B-V) < -0.05$, Bin 2: $-0.05 \le E(B-V) < 0.07$, Bin 3: $0.070 \le E(B-V)$.\\
The differences between the reported FIRST and VLASS numbers in columns (2) and (6) are due to some objects not having coverage in the respective radio imaging survey. }


\end{deluxetable*}

\subsection{a note on incompleteness}

The red W2M QSOs obey the KX color cuts (i.e., red box in Figure \ref{fig:kx_colors}) and are nearly spectroscopically complete with either a spectrum from SDSS or spectra that we obtained in the near-IR and optical. The blue QSO sample is not spectroscopically complete. Among the sources obeying the {\em WISE} color selection outside the KX color cuts, 747 sources lacked a spectrum in SDSS. These sources could be a mix of QSOs, galaxies, and perhaps a few stars. 
Might the exclusion of the QSOs missed from this set of objects affect the different mean radio properties that we observe? 

Given that the radio properties of blue QSOs do not change with $W1-W2$ color, and that an object is more likely to be a galaxy below the diagonal line cut (Eqn. \ref{eqn:diag}), we consider the 262 objects with no SDSS spectrum above that line. We found that 69 of these objects had a FIRST match within 2\arcsec, which is 26\%, consistent with the 28\% found for confirmed blue QSOs. Even under the most extreme assumption that all 69 sources are QSOs, while the remaining objects are not, the FIRST-detected fraction would rise to 31\%, which is still far below the 52\% found for red QSOs. 
A visual examination of the SDSS images of these spectrum-free sources reveals that a majority appear to be extended galaxies. We therefore conclude that the spectroscopic incompleteness of the blue QSO sample does not bias our conclusions about the differences between the radio properties of red and blue QSOs. 

\section{Discussion}

We have defined a sample of QSOs selected in a radio-independent manner such that difference in their radio properties ought to reflect intrinsic differences between the two populations. 
We took care to correct for host galaxy light that reddens the spectral shape of otherwise unobscured QSOs to define a clear distinction between dust-reddened QSOs and unreddened, blue QSOs with an intrinsically redder optical continuum. 
We then identified a luminosity-restricted subsample to further minimize the effects of reddening combined with the survey's flux limit. 
Below, we discuss the implications of the differences seen in the radio properties of these red and blue QSOs. We also estimate the fraction of red QSOs in this radio-independent population, given the differences in the radio properties of red and blue QSOs uncovered in this work. 

\subsection{Radio emission in red quasars}

\begin{figure}
\figurenum{23}
\plotone{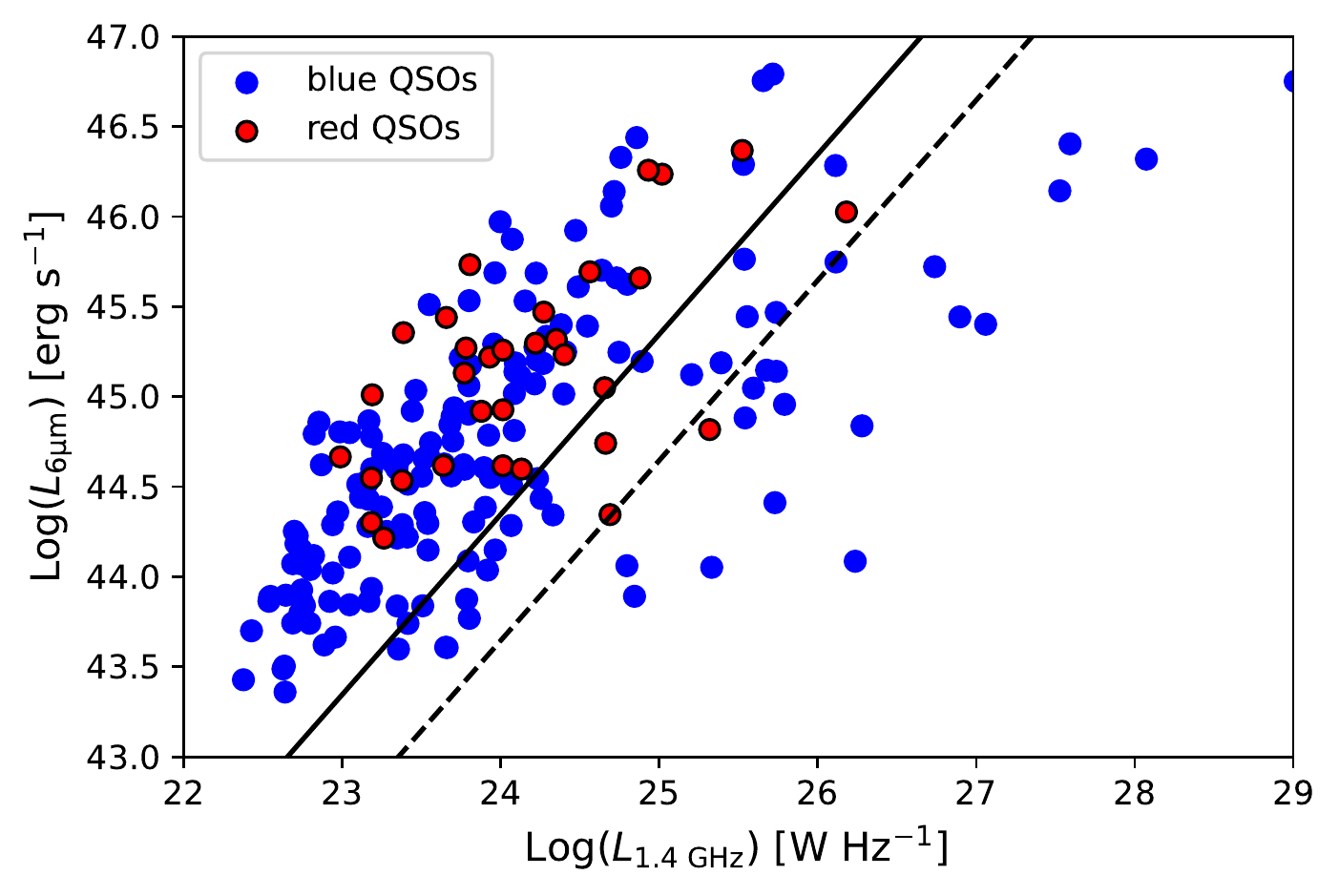}
\caption{Rest-frame infrared luminosity at 6 $\mu$m vs. radio luminosity at 1.4 GHz for the luminosity-restricted blue and red QSOs at $z>0.1$ with FIRST detections. The diagonal black line represents $\mathcal{R} = -4.6$, which is the curtoff for radio-loud QSOs, seen to the right. The dashed black line is $\mathcal{R} = -3.5$ which includes radio-intermediate sources. }\label{fig:radio_IR}
\end{figure}

Although most of the QSOs in our sample are undetected in FIRST and VLASS, we investigate the radio emission of the sources with detections to understand the radio luminosities and radio-loudnesses of the blue and red populations. Because radio-loudness is defined as the ratio of radio to optical emission, the presence of reddening and extinction at optical wavelengths will result in red QSOs being artificially apparent as radio-loud. This was addressed in \citet{Glikman12} by de-reddening the optical flux based on $E(B-V)$ (e.g., \S \ref{sec:reddening}).  However, \citet{Klindt19} and \citet{Fawcett20} define radio-loudness using the QSO luminosity at 6 $\mu$m, 
\begin{equation}
\mathcal{R} = \log_{10} \frac{1.4\times 10^{16} L_{1.4 {\rm GHz}}}{L_{6 \mu{\rm m}}},
\end{equation}
which is less sensitive to dust extinction but still probes the QSO continuum.  In this formulation, the radio-quiet/radio-loud divide occurs at $\mathcal{R} = -4.6$. Figure \ref{fig:radio_IR} shows the 6 $\mu$m luminosity versus the 1.4 GHz  radio luminosity for the QSOs with FIRST detections that are part of the luminosity-restricted subsample and at $z>0.1$. The solid black line represents $\mathcal{R} = -4.6$ and demonstrates that all but four of the red QSOs are radio quiet.  A similar investigation is performed in \citet{Fawcett20} who find a larger representation of radio-loud red QSOs.

The median radio properties of red and blue QSOs are significantly distinct, with red QSOs emitting $\sim 2-3$ times more flux, at both FIRST (1.4 GHz) and VLASS frequencies (3.0 GHz). 
Assuming a power-law shape to the radio spectrum, $S_\nu \propto \nu^\alpha$, we can compare the ratio of flux densities for the stacked red and blue QSOs to investigate differences in their spectral slopes.   
Using the median flux values in FIRST and VLASS, reported in Table \ref{tab:stacking}, we compute an estimated spectral index,
\begin{equation}
\alpha = \frac{\log{(S_{\rm FIRST}/S_{\rm VLASS} )} }{ \log{(1.4{\rm~GHz}/3.0{\rm~GHz})} },
\end{equation}
with an uncertainty, $\alpha_{\rm err}$, derived using standard error propagation. 
We find that the red quasars have a median index of $\alpha_{\rm red} = -0.70\pm0.05$ while the blue quasars have significantly flatter spectra, with $\alpha_{\rm blue} = -0.34\pm0.04$.  

It is interesting to note that the median spectral index for the red QSOs is the spectral index typically used for applying $k$-corrections to radio luminosities \citep[e.g.][]{Kimball11}.  
However, given the lack of robust calibration of a `snapshot bias' correction for VLASS, these values are best used in comparison between the two populations rather than as a true value. However, the `snapshot bias' correction involves scaling the flux upward, which in both the red and blue cases would result in a more negative (steeper) slope.

Indeed, a similar trend was seen in the F2M quasars with contemporaneous VLA observations of 44 F2M sources at 1.4 GHz and 8.3 GHz \citep[i.e., 20 cm and 3.6 cm;][]{Glikman07}. 
These were compared to 214 FIRST Bright Quasar Survey \citep[FBQS;][]{Gregg96} sources, which is the blue quasar equivalent of F2M, observed with the same VLA configuration \citep{Lacy01}.  
When broken up into bright ($S_{\rm pk, FIRST} > 10$ mJy) and faint ($S_{\rm pk, FIRST} < 10$ mJy) subsamples, the F2M red quasars were found to have a similar spread of spectral indices in the bright sample, but the faint red quasars had steeper spectral slopes with the a median spectral index of $\alpha = -0.68$ and most sources having $\alpha$ between  $-0.5$ and $-1.1$\footnote{Among the 11 F2M red quasars recovered here, only two had spectral indices measured in \citet{Glikman07}. F2M1004+1229 has $\alpha = 0.03$ and, with a FIRST flux density of 12.3 mJy, belongs in the bright sample. F2M2216$-$0054 has $\alpha = -1.08$ and, with a FIRST flux density of 1.3 mJy, is faint. With only two sources that span the range of spectral indices, we are unable to generalize more broadly. }. Given that only 7 W2M red QSOs have $S_{\rm pk, FIRST} > 10$ mJy and that the spectral index estimated from the stacked images is similar to the median spectral index for the F2M red quasars with $S_{\rm pk, FIRST} < 10$ mJy, we are likely probing a similar population of faint but enhanced radio sources that are somehow associated with a dusty environment that is reddening their optical to near-infrared spectra. 

This apparent connection between enhanced radio emission and reddening suggests a different physical mechanism driving the radio emission of the red and blue populations. 
It may be tempting to explain the flatter median spectral index for the blue QSOs as due to orientation effects from relativistic jets.  
To check for the influence of core-dominated emission from radio-loud QSOs, whose beamed emission may flatten the slope, we stacked only QSOs with $\mathcal{R} \le -3.5$. These objects appear to the left of the dashed line in Figure \ref{fig:radio_IR} and excludes the radio-loudest systems (18 blue and 1 red). We found $\alpha_{\rm red} = -0.69\pm0.05$ and $\alpha_{\rm blue} = -0.36\pm0.04$, which are unchanged from the full sample, within the uncertainties.
Furthermore, given that the vast majority of the QSOs are blue, it is unlikely that they are all viewed within the small angle needed to witness relativistic beaming effects. 

Given that our sample is dominated by relatively low redshift sources, we consider the possibility that star formation is a significant contributor to the radio emission. \citet{Kimball11} constructed a radio-luminosity function for luminous blue QSOs, including radio-quiet sources, which has a shape characterized by two-components; star formation dominates in sources with $\log(L_{6 \rm GHz} {\rm [W~Hz^{-1}]}) \lesssim 22.5$. In Figure \ref{fig:radio_IR}, we see that among the FIRST-detected sources, there are very few QSOs near this luminosity threshold at 1.4 GHz. Furthermore, the red QSOs are not found below $\log(L_{1.4 \rm GHz} {\rm [W~Hz^{-1}]}) \sim 23$. Since these detected red QSOs account for  $\gtrsim 50\%$ of the red QSO sample, we cannot attribute most of the radio emission to star-formation processes. Other studies of radio emission from radio quiet QSOs have also argued against star-formation as the dominant source of emission in radio-quiet QSOs \citep[e.g.,][]{White17,Laor19} and even in red QSOs, \citet{Fawcett20} argue that the enhanced radio emission is likely due to AGN activity. 
We note that these conclusions are based on indirect analyses and more targeted studies, such as high spatial resolution, multi-frequency radio imaging of the red QSOs would directly test whether the radio emission is AGN dominated.

Recently, employing high-resolution radio imaging (0\farcs2 at $\sim1.5$ GHz) of a sample of red and blue QSOs chosen from the \citet{Klindt19} study, \citet{Rosario21} found unresolved radio cores in the majority of both groups arising from regions smaller than 2 kpc in size. However, the red QSOs did have a significantly higher fraction of extended or multi-component radio emission compared with the blue QSOs. The authors propose that dusty winds are both reddening the QSOs and driving shocks that generate radio emission. This interpretation is corroborated by \citet{Calistro-rivera21} who find an excess of near-infrared emission in the mean SEDs of red QSOs also drawn from the \citet{Klindt19} study. The authors interpret this excess emission as arising from hot outflowing dust, i.e., a dusty wind \citep[see also ][]{Zakamska14}.

We note that if we consider just the {\em excess} radio emission in the red QSO population, we find a spectral index of $\alpha \simeq -0.9$, which is even steeper.  
\citet{Laor19} find a strong correlation between $\alpha$ and Eddington ratio ($L/L_{\rm Edd}$) for a sample of luminous radio-quiet QSOs such that the steeper the slope, the higher the accretion rate. The F2M red quasars are known to have Eddington ratios that are significantly higher than comparable blue quasars \citep[e.g.,][]{Urrutia12,Kim15} and the same may be true for the W2M red QSOs. \citet{Laor19} interpret the steep spectral slope in radio-quiet high-accretion QSOs as possibly due to AGN-driven winds generating outflows and associated shocked gas that results in synchrotron radiation emitted in the radio. This is consistent with the presence of broad absorption line systems seen in the F2M red quasars \citep{Urrutia09,Glikman12}. The highest redshift W2M QSOs also show evidence for outflows, either in absorption or emission, when an optical (i.e., rest-frame UV) spectrum exists. 
These interpretations are also consistent with the dusty wind scenario proposed by \citet{Klindt19}, \citet{Rosario21}, and \citet{{Zakamska14}}. 

\subsection{The fraction of red QSOs}\label{sec:fred}

If the difference between blue and red QSOs is not due to orientation with respect to our viewing angle, as is suggested by the radio results, then we can directly compare the two populations to find a true fraction of each sub group within the Type 1 QSO population (i.e., we can control for viewing angle, as illustrated in Figure \ref{fig:cartoon}, while exploring dust-reddening as the variable). With this assumption, we can compute the fraction of red quasars to be, 
\begin{equation}
\frac{N_{red~QSO}}{N_{red~QSO} + N_{blue~QSO}}. \label{eqn:frac}
\end{equation}

However, as Figure \ref{fig:lum_hist} shows, the fraction of red QSOs appears to have a strong luminosity dependence which must be considered when determining a red QSO fraction.  
In both panels, red QSOs make up a large percentage of all QSOs at high luminosities. On the lower luminosity end blue QSOs dominate. 

\begin{figure}
\figurenum{24}
\plotone{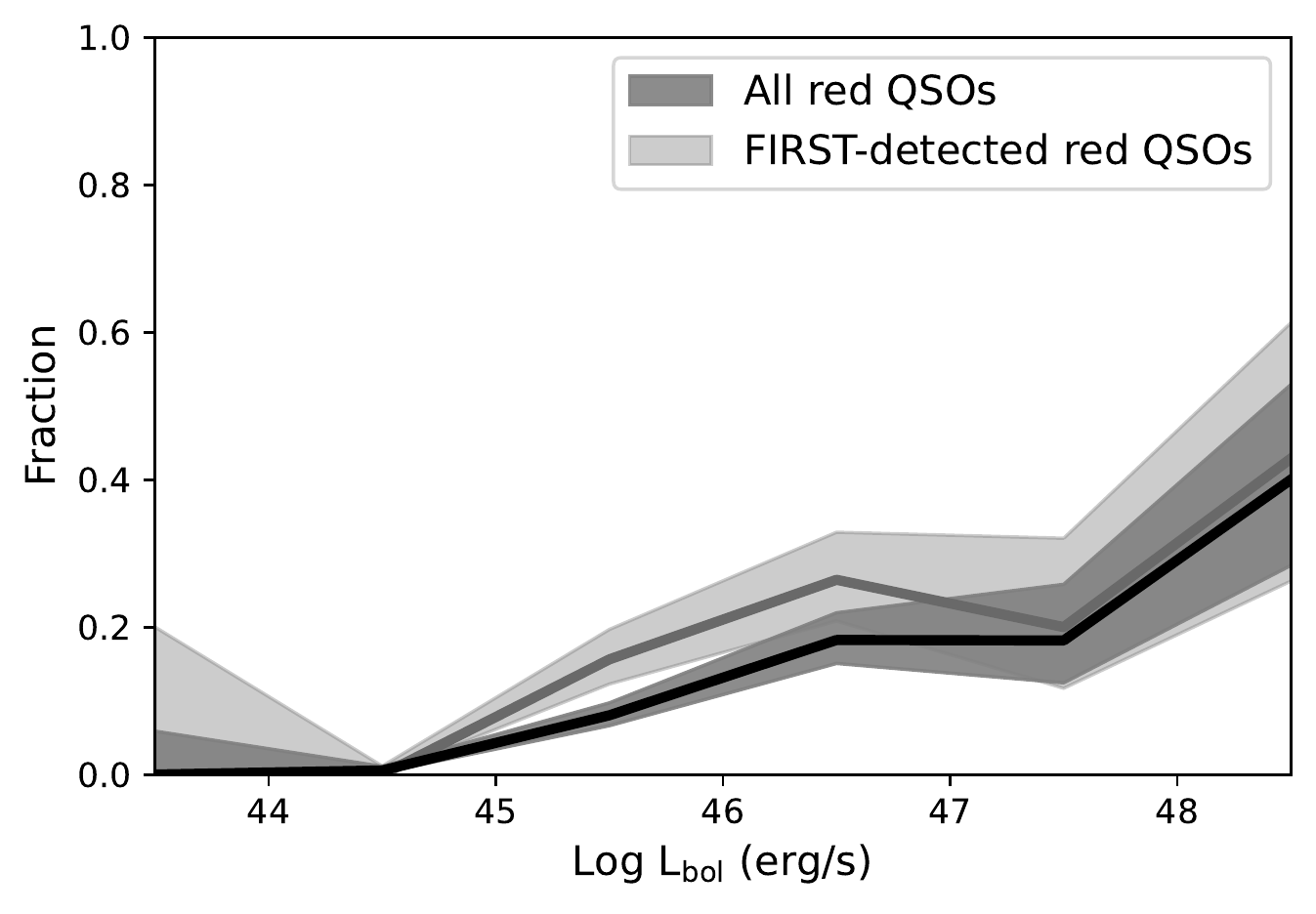}
\caption{Red QSO fraction as a function of bolometric luminosity, determined from the histograms in Figure \ref{fig:lum_hist} and obeying Equation \ref{eqn:frac}. The data were smoothed to 1 dex bins and uncertainty intervals (shaded gray areas) were computed using binomial proportion confidence interval for small number counts \citep{Wilson27}. The black line and dark shaded area represent the full W2M sample, while the gray line and light gray shading represent the FIRST-detected subsample. At the highest luminosities, we see that red QSOs dominate the overall QSO population, regardless of their radio properties. At decreasing luminosities, the radio-detected sources have a higher fraction compared the full sample of QSOs. These results are consistent with previous findings that red QSOs make up $\sim 20\%$ (and up to $40\%$) of all QSOs, at least at the highest luminosities. }\label{fig:rqso_frac}
\end{figure}

Figure \ref{fig:rqso_frac} shows the fraction of red QSOs found in our survey as a function of bolometric luminosity. We determine this fraction by using the red QSO histogram and the luminosity-restricted blue QSO histogram, with a binning of 1 dex to smooth out fluctuations, and taking a ratio following Equation \ref{eqn:frac}.  There is a strong luminosity effect, showing that red QSOs dominate the overall QSO populations regardless of radio properties. The radio detected red QSO fraction is higher at decreasing luminosities. However, in both cases, red QSOs make up at least 20\% and up to 40\% of the overall QSO population at the highest luminosities. 
These fractions are lower limits, since we miss more heavily reddened QSOs, while accounting for the blue QSOs, at the same luminosities. 

We note that these fractions are consistent with the fraction of red quasars estimated in \citet{Glikman12} for the F2M sample, which extends to fainter magnitudes ($K\lesssim15.5$ mag, the 2MASS limit).  Figure 17 of that paper shows that red quasars dominate at the highest de-reddened absolute $K$-band magnitudes of a similar range.  
\citet{LaMassa17} see similar behavior for X-ray-selected red QSOs, finding that red QSOs make up a larger fraction of all X-ray-selected QSOs when corrected for absorption; above $L_{X} = 10^{44}$ erg s$^{-1}$, red QSOs make up $\sim 20\%$ of all quasars in that luminosity regime. 
In \citet{Glikman18}, the red QSO sample identified over Stripe 82 is combined with deeper, mid-infrared red QSO surveys conducted over smaller areas to enable a luminosity function calculation and the comparison of red QSO space density versus blue and Type 2 QSOs (which are more heavily obscured, likely due to orientation).  That study finds that red QSOs make up $\sim 30 - 40\%$ of the overall QSO population at the highest luminosities, above $\nu L_{5\mu{\rm m}} = 10^{45.5}$ erg s$^{-1}$, which this work corroborates. 

However, the limitations of the relatively shallow $K<14.7$ mag limit of the W2M survey limits our ability to disentangle the effects of redshift, luminosity, reddening, and radio emission and provides motivation for a deeper, mid-infrared-selected QSO study. 
In fact, \citet{Glikman13} showed that the space density of FIRST-selected quasars rises more steeply than for blue quasars when approaching a deeper flux limit of $K=17$ mag. 
Work is currently underway to expand the W2M QSO sample to a fainter $K$-band magnitude limit. 

\subsection{Mergers, radio emission, and red quasars as an evolutionary phase}

The distinct differences between the radio properties of red and blue QSOs complicates our ability to address the role of mergers in the co-evolutionary picture for QSOs and their host galaxies. 
It is unclear whether the dusty winds proposed to explain the enhanced radio and reddening properties of red QSOs are associated with the high merger fraction ($>80\%$) seen in the host galaxies of F2M red quasars \citep{Urrutia08,Glikman15}, which are radio selected. 
Many of the F2M red quasar properties are consistent with a `blowout' phase. 
The existence of broad lines rules out the source of reddening being due to orientation along a line-of-sight that intersects with a dusty torus. 
Their spectra show an unusually high fraction of low-ionization broad absorption line sources \citep[LoBALs;][]{Urrutia09,Glikman12}, high accretion rates compared to unreddened quasars \citep{Urrutia12,Kim15}, and absorption-corrected bolometric luminosities that are higher than blue quasars at similar redshifts \citep{Glikman12,Treister12}. 

However, the results for red and obscured QSOs at lower radio luminosities are a less clear-cut. 
\citet{Zakamska19} analyzed {\em HST} images of ERQs, which are hyper-luminous broad-lined QSOs at $z = 2-3$. These objects are infrared bright, heavily obscured in X-rays, exhibit outflows in their emission line profiles, and are possibly accreting at super-Eddington rates. The radio properties of ERQs are interpreted as also arising from accretion-driven winds. However, these objects, which are similar to the F2M red quasars in many ways, do not show a significant merger fraction ($\sim20\%$). Their main distinction is in their radio properties, with the F2M red quasars being an order of magnitude more luminous at rest-frame 1.4 GHz \citep[$10^{41.9}$ erg s$^{-1}$ versus $10^{40.9}$ erg s$^{-1}$;][]{Zakamska19}. 

Interestingly, \citet{Chiaberge15} studied a large sample of radio galaxies from the 3C catalog between $1<z<2.5$ with {\em HST} and found a remarkably high merger rate ($>$90\%), while radio-quiet Type 2 analogs at the same redshifts have a merger fraction consistent with inactive galaxies.
They conclude that major mergers not only trigger star formation and SMBH growth as the models predict, but are also responsible for launching jets. 
It is possible, then, that the enhanced radio emission in the W2M red QSOs can be due to a combination of phenomena, where winds are a ubiquitous main driver of low-level radio emission and jets, possibly associated with mergers, contribute to the radio emission in the more radio-luminous sources. 

The main driver for dust-driven winds may then be bolometric luminosity or accretion rate. Dust-driven winds are invoked to explain the properties of the most luminous Type 2 QSOs at $z\lesssim1$ \citep{Zakamska14}. 
Gas-rich mergers are particularly effective at providing an abundant fuel source to the SMBH, resulting in a higher luminosity AGN possibly explaining them being a significant presence among the radio-selected red quasars.
Given that the `blowout' phase derives its meaning in a merger scenario, it is important to distinguish between different feedback phenomena before interpreting the red QSO fraction (e.g., Figure \ref{fig:rqso_frac}) as a phase duration. 
One way to test this would be to study the host morphologies of W2M QSOs with {\em HST} to determine whether mergers are a universal phenomenon among all red QSOs or whether radio-selection is biased toward merging systems. This analysis is currently underway with an {\em HST} imaging program. 

\section{Conclusions}

We have identified a sample of 1,112 QSOs selected according to their mid-infrared colors with a near-infrared flux limit of $K<14.7$ mag over 2,213 deg$^2$. This selection method identifies blue and red QSOs with minimal contamination and reddening bias and without the need for a radio selection criterion.  We performed a careful analysis of the QSO spectra, removing host galaxy light which artificially mimics reddening by dust in some otherwise blue QSOs.  We also defined a luminosity-restricted subsample in the $K$-band, after correcting for reddening. This enabled us to create intrinsically blue (798) and red (63) QSO subsamples whose properties we studied and compared. 

We investigated the fraction of sources detected in two radio surveys, FIRST (1.4 GHz) and VLASS ($2-4$ GHz), and we employed radio stacking to study the flux densities of sources undetected in these surveys.  We found that red QSOs are significantly more likely to be detected at both 1.4 GHz and 3.0 GHz and are more likely to appear compact in morphology. We also found that red QSOs have brighter median radio flux densities compared with blue QSOs. 
These results are consistent with recent work by \citet{Klindt19} and \citet{Fawcett20} who find similar radio enhancement at 1.4 GHz for SDSS quasars that have redder colors compared with their blue counterparts. We note that, compared with red QSOs in the SDSS sample, the W2M red QSOs reach higher $E(B-V)$ values and show a more pronounced distinction in the stacked radio flux ratios. Considering both frequencies, we find that red QSOs have steeper median radio spectra compared with blue QSOs (i.e., red QSOs have higher FIRST to VLASS flux ratios than blue QSOs). 
We speculate that a dusty AGN-driven wind can account for both the unique radio and reddening properties of red QSOs, as has been noted elsewhere \citep{Zakamska14,Zakamska19,Klindt19,Fawcett20,Calistro-rivera21,Rosario21}.

The red QSOs in this study are among the more luminous QSOs, especially at high redshift ($z>1.5$), though our survey is not sensitive to the same luminosity distributions sampled for the red and blue QSOs.
We also note an absence of red QSOs at $z<0.1$, which is consistent with evolutionary behavior seen in previous work \citep[e.g.,][]{Glikman18,LaMassa17}. 
We therefore investigated the fraction of red and blue QSOs as a function of de-reddened absolute $K$-band magnitude in a de-reddened-luminosity-restricted and redshift-matched subsample. We find that red QSOs dominate the QSO population at the highest luminosities, remaining a significant fraction of the QSO population at $\log(L_{\rm bol} [{\rm erg~s^{-1}}]) > 46$ with the radio-detected red QSOs having a $\sim 40\%$ higher fraction. 

The results of this study suggest that previous conclusions about the fraction of red quasars, determined from radio-selected samples, is too high to be extended to the overall QSO population, and a radio-independent selection is essential for understanding the nature of dust-reddened QSOs. 
The fact that red QSOs appear to be a predominantly high luminosity phenomenon with distinct radio properties showing enhanced emission, strongly implies that red QSOs are not simply an apparent orientation effect but are rather a distinct population that can shed light on supermassive black-hole growth and the quasar phenomenon. 

\acknowledgments

We thank the anonymous referee for a careful reading of the text and for suggesting revisions and analyses that greatly improved the presentation of the paper. 
EG acknowledges the generous support of the Cottrell Scholar Award through the Research Corporation for Science Advancement.
SGD and MJG acknowledge a partial support from the NASA grant 16-ADAP16-0232 and the NSF grants AST-1749235, AST-1815034, and AST-2108402.

This research made use of the cross-match service provided by CDS, Strasbourg.

Research at Lick Observatory is partially supported by a generous gift from Google.

This publication makes use of data products from the Wide-field Infrared Survey Explorer, which is a joint project of the University of California, Los Angeles, and the Jet Propulsion Laboratory/California Institute of Technology, funded by the National Aeronautics and Space Administration.

This publication makes use of data products from the Two Micron All Sky Survey, which is a joint project of the University of Massachusetts and the Infrared Processing and Analysis Center/California Institute of Technology, funded by the National Aeronautics and Space Administration and the National Science Foundation.

Funding for SDSS-III has been provided by the Alfred P. Sloan Foundation, the Participating Institutions, the National Science Foundation, and the U.S. Department of Energy Office of Science. The SDSS-III web site is http://www.sdss3.org/.

SDSS-III is managed by the Astrophysical Research Consortium for the Participating Institutions of the SDSS-III Collaboration including the University of Arizona, the Brazilian Participation Group, Brookhaven National Laboratory, Carnegie Mellon University, University of Florida, the French Participation Group, the German Participation Group, Harvard University, the Instituto de Astrofisica de Canarias, the Michigan State/Notre Dame/JINA Participation Group, Johns Hopkins University, Lawrence Berkeley National Laboratory, Max Planck Institute for Astrophysics, Max Planck Institute for Extraterrestrial Physics, New Mexico State University, New York University, Ohio State University, Pennsylvania State University, University of Portsmouth, Princeton University, the Spanish Participation Group, University of Tokyo, University of Utah, Vanderbilt University, University of Virginia, University of Washington, and Yale University.

\facility{Palomar, IRTF, APO, LBT, Lick}
\software{Astropy \citep{astropy:2013, astropy:2018}, IRAF \citep{Tody86,Tody93}, Spextool \citep{Cushing04}, TOPCAT \citep{Taylor05}}

\bibliographystyle{apj}
\bibliography{w2m.bbl}

\end{document}